\documentclass[ 11pt, a4paper]{article}
               
\usepackage[T1]{fontenc}
\usepackage{mathptmx}
\usepackage{amsmath,amsfonts,amssymb}
\usepackage{graphicx}\graphicspath{{Graphics/}}
\usepackage[onehalfspacing]{setspace}
\usepackage{ragged2e} 
\usepackage{textcomp}
\usepackage{caption}
\usepackage[square,comma,sort&compress]{natbib}
\usepackage[bottom]{footmisc}
 
\usepackage[toc,page]{appendix}
\usepackage{bm}
\usepackage{float}
\usepackage{booktabs,multirow}
\usepackage[table]{xcolor}
\usepackage{fancyhdr}
\usepackage{pifont}
\usepackage{marvosym}
\usepackage{hyperref}
 \hypersetup{
     bookmarks=true,         
     colorlinks=true,        
     citecolor=blue,         
     linkcolor=black,
    } 
\usepackage{cleveref}
\usepackage{fancybox}
\usepackage{epsfig}
\usepackage{epstopdf}
\usepackage[top=1in, bottom=1in, left=1.25in, right=1in]{geometry} 
\usepackage{subcaption}
\usepackage{color}

\newcommand{\mi}{\boldsymbol{-} \mathrel{\mkern -16mu} \boldsymbol{-}}

\definecolor{cream}{RGB}{222,217,201}
\definecolor{forestgreen}{rgb}{0.13, 0.55, 0.13}
\definecolor{brown}{rgb}{0.24, 0.17, 0.12}
\definecolor{magenta}{rgb}{1.0, 0.0, 1.0}
\definecolor{forestgreen}{rgb}{0.13, 0.55, 0.13}
\definecolor{brown}{rgb}{0.24, 0.17, 0.12}
\definecolor{cream}{RGB}{222,217,201}
\definecolor{forestgreen}{rgb}{0.13, 0.55, 0.13}
\definecolor{brown}{rgb}{0.24, 0.17, 0.12}

\newcommand{\rr}{\textcolor{red}}
\newcommand{\bb}{\textcolor{blue}}

\title{Deformation of a biconcave-discoid capsule in extensional flow and electric field}
\author{Sudip Das, Shivraj D. Deshmukh and Rochish M Thaokar\\
Department of Chemical Engineering, Indian Institute of Technology Bombay,
Mumbai,\\ 400076, India
}

\def\l[{{[\![}}
\def\r]{{]\!]}}

\def\be{\begin{equation}}
\def\ee{\end{equation}}
\def\.{\cdot}

\date{}

\begin{document}
\maketitle



%
%

\begin{abstract}
Biconcave-discoid (empirical shape of a red blood cell) capsule finds numerous applications in the field of bio-fluid dynamics and rheology, apart from understanding the behavior of red blood cells (RBC) in blood flow. A detailed analysis is, therefore, carried out to understand the effects of the uniaxial extensional flow and electric field on the deformation of a RBC and biconcave-discoid polymeric capsule in the axisymmetric regime. The transient deformation is computed numerically using axisymmetric boundary integral method for Stokes flow considering the Skalak membrane constitutive law as the model for the area incompressible RBC/biconcave-discoid capsule membrane. A remarkable biconcave-discoid to prolate spheroid transition is observed when the elastic energy is overcome by the viscous or Maxwell electric stresses.  Moreover, the significance of membrane stresses developed during the deformation and at steady state and different modes of deformation are presented. This study should be useful in designing an artificial system involving biological cells under an electric field.
\end{abstract}
\section{Introduction}
The study of the interfacial rheological response of a red blood cell (RBC) membrane is of considerable significance given their relevance in the field of biofluid dynamics and rheology. The membrane rheology of RBC has been used to ascertain the health of biological cells in several studies using microfluidic devices~\citep{henon14, yaginuma13, Lee09}. From a mechanics point of view, the red blood cells can be considered to be comprised of a Newtonian solution of hemoglobin dispersed in water and  encapsulated by a lipid bilayer membrane and associated protein skeleton (spectrin network)~\citep{evans80}. The lipid bilayer has several embedded, but, mobile proteins. The RBC membrane incompressibility and interfacial viscous behavior are attributed to the lipid bilayer and the proteins, respectively. The bilayer membrane is supported by a rigid protein (Spectrin) network that endows an elastic behavior~\citep{mohandas94}.  This complex structure makes the cell membrane an area-preserving (incompressible) two-dimensional medium and therefore, it resists change in the area by generating inhomogeneous, but, anisotropic tension.  Moreover, unlike lipid bilayer membranes which offer negligible resistance to shear, the red blood cells  offer minimal but significant resistance to shear deformation. Further, the resistance to bending~\citep{poz90, pozmodel03,kwak01} is substantial in the regions of high curvatures.

Another important area of interest is the understanding of the rheological behavior as well as the response to deforming forces for polymeric biconcave-discoid capsules \citep{veronika14}, which have been recently suggested as very promising drug-delivery carriers. Inspired by the shapes of the carriers in biological processes such as human red blood cells as well as bio-organisms in nature such as  bacteria~\citep{shupeng13,She14}, particles of varying shapes have been suggested for the efficient drug delivery.  Improved bio-functionality has been reported with hydrogel microparticles~\citep{Merkel11} and capsules~\cite{veronika14} similar in size and shape with human red blood cell. It is reported that shape of the carrier/capsule is an important parameter in biological processes, influencing the enhanced cellular uptake of the loaded particle~\citep{Champion07,Venkataraman11}. Nonspherical capsules have an ability to migrate laterally towards the wall of the vascular network in laminar flow~\citep{sei09,Decuzzi08}. Although few experimental studies have been reported for near RBC-shaped microcapsules, they lack theoretical analysis to comprehend their rheological behavior. This is essentially due to their complicated shapes, thereby, rendering difficulty in obtaining closed formed analytical solutions. Thus, numerical solutions are often necessary to describe their deformation.

When a red blood cell is subjected to a straining flow, it initially deforms like a liquid drop. However, unlike a liquid drop, it does not deform continuously due to the incompressibility of the red blood cell membrane. With an increase in the intensity of the straining flow, the membrane tension increases. Typically, the tension beyond $\sim10\ mN/m$ leads to the rupture of the membrane, resulting in hemolysis~\citep{evans76}. Although shear flow is commonly encountered in biological systems, one could often encounter extensional flows, especially at bifurcations of blood vessels. Moreover, extensional flow provides an excellent framework to investigate soft particles, mainly due to its axisymmetric deformation field. The computational~\citep{yen15} and experimental~\citep{yen15,Lee09} analyses suggest that the stress caused by the extensional flow plays a significant role in the hemolysis of red blood cell as compared to the shear stress. Therefore, to understand the sustainability of a synthetic biconcave-discoid capsule in maximum stress condition encountered in applications, the analysis in extensional flow is essential.

\citet{poz90} studied the dynamics of a red blood cell in uniaxial extensional flow considering a spherical shape with high oblate perturbation by second degree Legendre mode as the model geometry~\citep{poz90}. In this study, an initial oblate shaped cell is positioned axisymmetrically, and its dynamics is solved by the boundary element method. Along with membrane elasticity, area incompressibility constraint was employed, whereas the bending rigidity of the membrane was neglected. It was observed that oblate perturbed initial shapes evolved to prolate spheroid shapes through different transitional shapes depending upon the reduced shear elasticity.  In another analysis of the deformation of a spherical capsule in uniaxial extensional flow~\citet{kwak01} reported that the bending rigidity of the membrane prevents the formation of sharp pointed tips at large deformation, and results in nearly cylindrical shapes with rounded caps.      

An electric field can induce deformation~\citep{Chang1985}, change in orientation~\citep{Friend75}, dielectrophoresis~\citep{cruz98}. It is also reported that electric field can cause the fusion and lysis of biological cells~\citep{Scheurich1980}. \citet{sackmann88} measured the shear elastic moduli of the erythrocyte membrane by carrying out deformation study in a high-frequency electric field~\citep{sackmann88}. When the electric field is turned off, a highly elongated erythrocyte returns to the actual shape of the cell, supporting the proposed shape memory theory of erythrocyte~\citep{bagchi17}. The reversible deformation is noticed at a low-intensity AC electric field, and the hemolysis of the cells are observed at a high-intensity field~\citep{sackmann88,gass91,krueger97}. Moreover, the deformation also depends upon the conductivity of internal and external fluids~\citep{Sukhorukov98,Kononenko00,kononenko02a}. Apart from electrodeformation, electrorotation and its dependency on conductivities of the fluids as well as the frequency of the applied AC electric field is observed. Attempts have been made to calculate Maxwell stresses at the interface of an erythrocyte~\citep{sebastian06}. An improved understanding of the mechanics of RBC has enabled development of better medical diagnostic devices~\citep{sackmann88,Kononenko00,kononeko02,Thom06,Thom09,Du14}. A similar analysis can also be employed to measure the mechanical properties of the synthetic biconcave-discoid capsule.

Even though the change in shape and lysis of a red blood cell in an electric field was observed long back~\citep{amiram68}, the underlying physics is still not well understood. Most of the studies consider erythrocyte as a spheroidal~\citep{Chang1985, Ashe1988,Joshi2002} or oblate~\citep{poz90} shape, although it is a biconcave-discoid under normal quiescent conditions.  Thus, it is important to carry out a consistent electrohydrodynamic analysis on the exact geometry of the red blood cell.

In the current work, the aim is to investigate the effects of extensional flow and electric field on the deformation of a discocyte (empirical RBC shape) and biconcave-discoid polymer capsule in the axisymmetric regime. The transient deformation is computed numerically using boundary element method for Stokes flow and considering Skalak law as the model for the elastic membrane. A high value of Skalak model coefficient ($C$) makes the enclosing interface area conserving. The objective here is to assess the significance of membrane stresses developed during the transient deformation and at steady-state and to understand the modes of deformation. Additionally, this analysis should provide the distribution of tensions along the arc length of the membrane, which could suggest the location where a biconcave-discoid capsule and a RBC can undergo breakup. Moreover, the possibility of a biconcave-discoid $\rightarrow$ Prolate transition under extensional flow and an electric field is explored. 
 
\section{Boundary Integral Formulations}
\begin{figure}
\centering
 \includegraphics[width=0.6\linewidth]{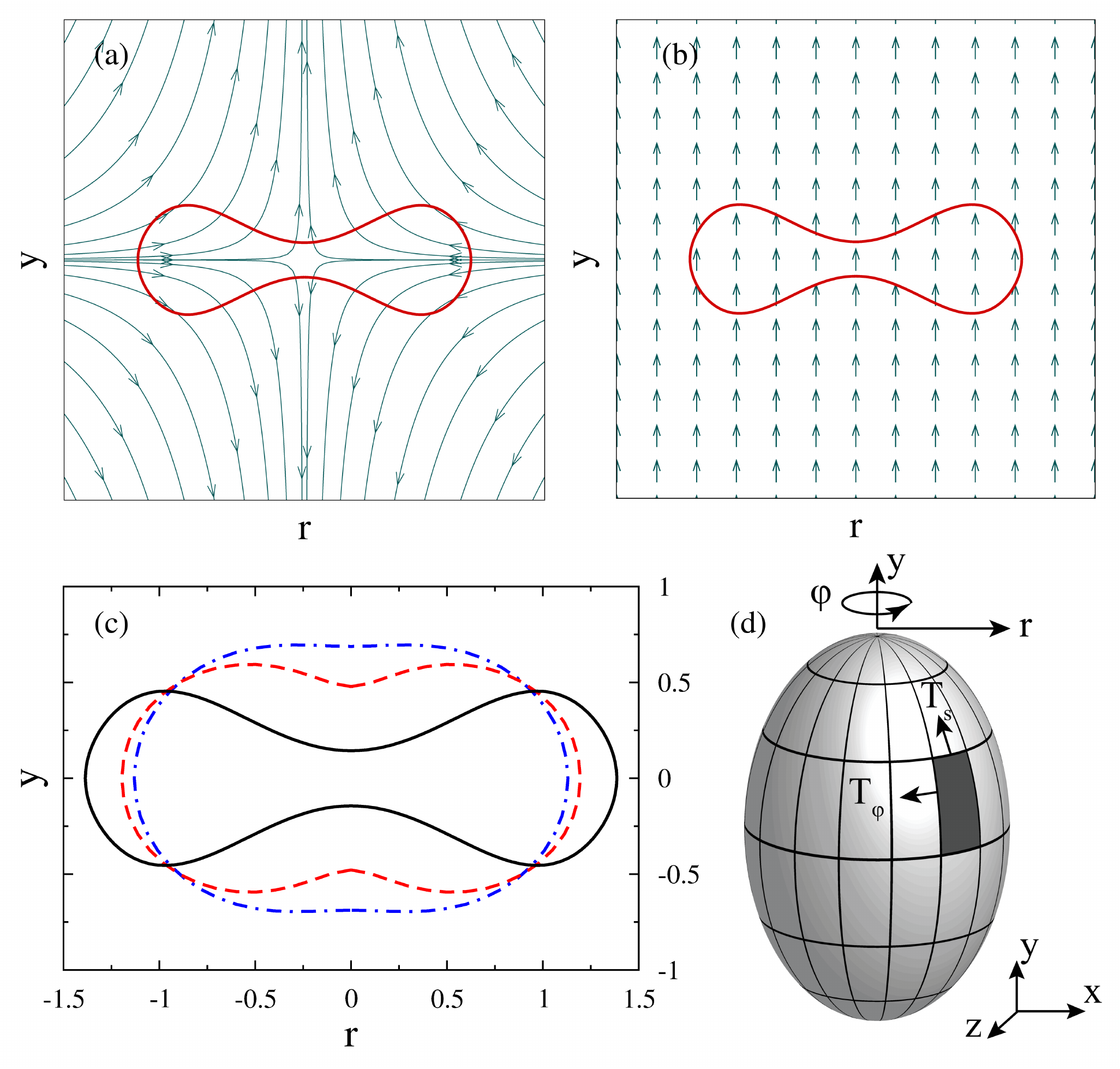}
\caption{Schematic representation of a capsule in (a) extensional flow (curves with arrows represent streamlines) and (b) uniform electric field (arrows represent field lines). (c) Comparison of the shape of a red blood cell ($\pmb{\mi}$) with the oblate spheroids perturbed with second degree Legendre mode (\rr{$\pmb{\bf --}$} for $\epsilon=-0.5$ and $\bb{\pmb{\bf-\cdot-}}$ for $\epsilon=-0.3$). (d) Schematic presentation of a prolate spheroid representing azimuthal ($T_\phi$) and meridional ($T_s$) elastic tensions. Center of the spheroid is at the origin of the Cartesian coordinate system.}
\label{fig:schematic2}
\end{figure}

The axisymmetric deformation of a biconcave-discoid elastic capsule and RBC (\emph{commonly called capsule in this article}) under a uniaxial extensional flow at vanishing Reynolds number is investigated. In the Cartesian coordinate system, the axisymmetric extensional flow is given by:
\begin{equation}\label{eq:freestream}
 \tilde{{\bf u}}^\infty=e\left[
\begin{array}{ccc}
					2 & 0 & 0 \\ 
					0 &-1 & 0 \\
					0 & 0 & -1\\
			\end{array}  \right]\tilde{\bf x},
\end{equation}
where  $e$ is the axisymmetric uniaxial extensional strain rate and $\tilde{\bf x}$ is the position vector. In the Cartesian coordinate system, the position vector is given by  $\tilde{\bf x}=\tilde{y} {\bf e}_y+\tilde{x}{\bf e}_x+ \tilde{z}{\bf e}_z$, where ${\bf e}_y$, ${\bf e}_x$, and ${\bf e}_z$ are the unit vectors in the $y$, $x$ and  $z$-directions, respectively. The axis of symmetry of the capsule is assumed to coincide with the $y$-axis, such that its centroid is at the origin (see Fig.~\ref{fig:schematic2}). In this article, all the notations with and without a tilde (\ $\tilde{}$\ ) represent dimensional and their nondimensional counterpart, respectively.

The shape of a capsule is considered as that of a healthy human Red Blood Cell (RBC, erythrocyte). Although the RBCs have been modeled as oblate spheroids~\citep{poz90}, there is a significant difference between the shape of a biconcave-discoid and  an oblate spheroid. Fig.~\ref{fig:schematic2}c shows that an oblate spheroid considered by~\citet{poz90} are quite different from the commonly conceived  biconcave-discoid  shape of a healthy human  RBC. Equation~\ref{rbcshapeeqn} describes the shape of a RBC possessing a well-known biconcave-discoid shape under normal physiological conditions~\citep{evans72},

\begin{equation} \label{rbcshapeeqn}
 \tilde{y}=a\frac{\alpha}{2}\left(1-\frac{\tilde{r}}{\alpha}^2\right)\left(0.207161 + 2.002558 \frac{\tilde{r}}{\alpha}^2 - 1.122762 \frac{\tilde{r}}{\alpha}^4 \right),
\end{equation}
where $\alpha=1.385818$ is the ratio between the maximum radius of the biconcave-discoid in the transverse plane of symmetry and equivalent radius $a$. Here, $\tilde y$  and $\tilde r=\sqrt{\tilde x^2+\tilde z^2}$ are the surface coordinates of the RBC in the cylindrical coordinate system, where ${\tilde r}$ defines the radial coordinate. The thickness of the RBC membrane is $\sim 10\ nm$ which is much smaller  than the RBC's equatorial diameter~($\sim 8\ \mu m$)~\citep{Helfrich1974}.

Both the external and the internal fluids are Newtonian, and the viscosities of the external and internal fluids are denoted by $\mu_e$ and  $\lambda\mu_e$, respectively, where $\lambda$ is the viscosity ratio. The mechanical properties of the membrane are elasticity, $E_s$ and bending rigidity, $\kappa_b$. The membrane of the capsule is considered to be purely elastic, and the viscous resistance is neglected. In the analysis of deformation in extensional flow and electric field, the length is scaled by $a$, the equivalent radius of the capsule. For the analysis under extensional flow, the time, fluid velocity, stress (force/area), pressure and tensions are scaled by $\mu_ea/E_s$, $E_s/\mu_e$, $E_s/a$, $E_s/a$ and $E_s$, respectively. 

In another case, an investigation is carried out for the axisymmetric deformation of a biconcave-discoid capsule subjected to a uniform DC electric field represented by $ \tilde{{\bf E}}^{inf}_{DC}=\tilde{E}_0^{DC} {\bf e}_y$ where $\tilde{E}_0^{DC}$ is the electric field strength. Moreover, to investigate the electrohydrodynamic response of a RBC in an AC electric field, $\tilde{{\bf E}}^{inf}_{AC}=\tilde{E_0}^{AC}\cos(\tilde{\omega} \tilde{t}) {\bf e}_y$ with the frequency $\tilde{\omega}$ is applied.  Here, $\tilde{E_0}^{AC}$ is the strength of the applied electric field, and $\tilde{t}$ is the time.

For the internal and external fluid media, the dielectric constants and electrical conductivities are $\epsilon_{i,e}$ and  $\sigma_{i,e}$, respectively. Subscripts $i$ and $e$ identify parameters for internal and the external fluids, respectively. The ratio of conductivities of the internal and external fluids is denoted by $\sigma_r=\sigma_i/\sigma_e$ and the ratio of dielectric constants is denoted by $\epsilon_r=\epsilon_i/\epsilon_e$. The electrical properties of the capsule membrane are capacitance, $ C_m$, and conductance, $ G_m$. In the analysis of the dynamics of a biconcave-discoid capsule in DC electric field, the scalings are motivated by the electrostatics model and are discussed later.

For simplicity of computations, the interface is considered to be impermeable, thereby preventing osmotic flows across the membrane.  The effect of bending is also considered in the present analysis, which was ignored in earlier work by~\citet{zhou95}. The ratio of the elastic to bending force is given by $\frac{\Delta X/h^2 L}{C^2}$, where $h$ is the thickness of the membrane, $L$ is the typical size, and $\Delta X$ is the deformation. The bending force can be neglected only if the curvature $C<\sqrt{\frac{\Delta X}{h^2 L}}$. All the non-hydrodynamic and hydrodynamic forces, responsible for the deformation of a biconcave-discoid capsule and RBC, are discussed in the following sections.

\subsection{Elastic forces}\label{sec:elasticforce}
The interface of a capsule is considered to be  thin and elastic as described by the Skalak model~\citep{skalak73}. According to the Skalak law, the non-dimensional membrane elastic tension in the principal direction $i$ ($s$ or $\phi$, see Fig.~\ref{fig:schematic2}d) is expressed as a function of principal stretch ratios ($\lambda_{i,j}$) for a strain hardening membrane as
\begin{equation}\label{eq:skalaknd}
{\tilde T}_{i}^{SK} = \frac{G^{SK}}{\lambda_i\lambda_j}\left[\lambda_i^2(\lambda_i^2-1)+C(\lambda_i\lambda_j)^2\{(\lambda_i\lambda_j)^2-1\}\right]. 
\end{equation}
Membrane tension in the principal direction $j$ can be expressed by interchanging indices in the constitutive relation (eq.~\ref{eq:skalaknd}). In this work, $i$ is the meridional and $j$ is the azimuthal principal direction and the corresponding tensions are considered as $T_{s}$ and $T_{\phi}$. The principal stretch ratios are defined as $\lambda_s=d\tilde{s}/d\tilde{S}$ and $\lambda_\phi=\tilde{r}/\tilde{R}$, where $s$ and $r$ are the measure of arc length and the radius of the deformed capsule and the upper case alphabets represent their counterparts in the stress-free shape, respectively. The first and second terms on the right-hand side of eq.~\ref{eq:skalaknd} are the measures of the shear deformation associated with the modulus $G^{SK}$ and the area dilatation with the modulus $C G^{SK}$, respectively~\citep{skalak73}. The area dilatation parameter $C$ regulates the extent of change of surface area, a large value of $C$ restricts the membrane dilatation.  In the small deformation limit, $G^{SK}$ is related to the surface Young modulus, $E_s$,~\citep{barthesbiesel02} as $E_s=2G^{SK}(1+2C)/(1+C)$. 

The membrane elastic traction (force/area), governed by the combined contribution of the elastic tensions (force/length) obtained from the constitutive law, is given by $\Delta{{\bf f}^{el}}=\tau_n^{el} {\bf n}+\tau_t^{el} {\bf t}$. The components of the elastic stresses can be obtained as
\begin{subequations}
\label{eq:elaststress}
 \begin{eqnarray}
 \tau_n^{el} &=& -(K_sT_{s}+K_{\phi}T_{\phi})\\
 \tau_t^{el} &=& \frac{d T_{s}}{ds}+\frac{1}{r}\frac{dr}{ds}(T_{s}-T_{\phi})
 \end{eqnarray}
 \end{subequations}
where $K_s=\left|\frac{d{\bf t}}{ds}\right|$ and $K_\phi=\frac{n_r}{r}$ are the curvatures of the meridional surface in the principal directions $s$ and $\phi$, respectively. Components of unit normal and tangent vectors at the interface can be calculated as $t_y=n_r=\frac{dy}{ds}$ and $t_r=-n_y=\frac{dr}{ds}$. 
\subsection{Resistance to bending}
Even though thin elastic membranes offer very low resistance to bending, it can have a significant contribution to the restoring forces especially in the regions of high curvature. The overall nondimensional bending force (force/area) is given by 
\begin{equation}
 \Delta{\bf f}^b =\hat{\kappa}_b\left[2\Delta_s H+(2H-c_0)(2H^2-2K_G+c_0H)\right] {\bf n},
\end{equation}
where, $\hat{\kappa}_b=\kappa_b/a^2E_s $ is the nondimensional bending rigidity, $H=\frac{1}{2}(k_s+k_\phi)$ is the mean curvature, $K_G=k_sk_\phi$ is the Gaussian curvatures, and $c_0$ is the spontaneous curvature~\citep{vlahovska09,bagchi11}. 
In the axisymmetric cylindrical coordinate system, the Laplace Beltrami of the mean curvature is given as~\citep{hu14}
\begin{equation}
 \Delta_sH=\nabla_s\cdot(\nabla_sH)=\frac{1}{r\mid{\bf x}_s\mid}\frac{\partial}{\partial s}\left(\frac{r}{\mid{\bf x}_s\mid}\frac{\partial H}{\partial s}\right),
\end{equation}
where $\mid{\bf x}_s\mid=\sqrt{\left(\frac{\partial r}{\partial s}\right)^2+\left(\frac{\partial y}{\partial s}\right)^2}$. A spectral method is used to calculate the higher order derivatives of mean curvature with respect to the arc length, especially for improved accuracy in estimating bending force~\citep{trefethen94}. For the analyses of the deformation of a capsule in both the extensional flow and electric field, a fixed value of $\hat{\kappa}_b=0.001$ is considered.

\subsection{Electrostatics}\label{sec:electricforce}
For the analysis of the dynamics of deformation of a biconcave-discoid capsule in externally applied DC electric field, the charge relaxation time of the outer fluid, $\tilde{t}_e={\epsilon_e\epsilon_0}/{\sigma_e}$ is considered as the scaling for time, where $\epsilon_0$ is the permittivity of the free space. Other timescales of relevance are hydrodynamic response time $\tilde{t}_H=\mu_ea/E_s$, Maxwell-Wagner relaxation time $\tilde{t}_{MW}=\epsilon_0(\epsilon_i+2\epsilon_e)/(\sigma_i+2\sigma_e)$ and membrane charging time $\tilde{t}_{cap}=aC_m(1/\sigma_i+1/2\sigma_e)$. Nondimensional counterparts of these timescales are  $t_e=1$, $t_H=\tilde{t_H}/\tilde{t}_e$, $t_{MW}=\tilde{t}_{MW}/\tilde{t}_e=(2+\epsilon_r)/(2 +\sigma_r)$ and $t_{cap}=\tilde{t}_{cap}/\tilde{t}_e=\hat{C}_m(1/2+1/\sigma_r)$~\citep{grosse92}. Fluid velocity ($\tilde{{\bf u}}$),  electric field ($\tilde{{\bf E}}$), potential ($\tilde V$) and stress are scaled by $E_s/\mu_e$, $E_0$, $E_0 a$ and $E_s/a$, respectively. 

The electric potential inside and outside the biconcave-discoid capsule satisfy the Laplace equation, $\nabla^2 V_{i,e}=0$. The electric potential can be obtained, by solving the Laplace equation and using Green's theorem, as
\begin{subequations}
 \label{eq:ponieeqns}
 \begin{eqnarray}
 \frac{1}{2}V_i({\bf x}_0) &=& \int_s \left[G^E({\bf x},{\bf x}_0){\bf \nabla}V_i({\bf x})\cdot {\bf n}({\bf x})-V_i({\bf x}){\bf n}({\bf x})\cdot {\bf \nabla}G^E({\bf x},{\bf x}_0)\right] ds({\bf x})\label{eq:poti},\\ 
 \frac{1}{2}V_e({\bf x}_0) &=& V^\infty ({\bf x}_0)\nonumber \\
 &&-\int_s \left[G^E({\bf x},{\bf x}_0){\bf \nabla}V_e({\bf x})\cdot {\bf n}({\bf x})-V_e({\bf x}){\bf n}({\bf x})\cdot {\bf \nabla}G^E({\bf x},{\bf x}_0)\right] ds({\bf x})\label{eq:pote},
 \end{eqnarray}
\end{subequations}
where $G^E({\bf x},{\bf x}_0)=\frac{1}{4\pi |\hat{{\bf x}}|}$ is the Green function for the Laplace equation~\citep{mcconnell15sm}. $\hat{{\bf x}}={\bf x}-{\bf x}_0$, where ${\bf x}_0$ and ${\bf x}$ are the source and observation points, respectively. 

The discontinuity in electrical potential at the interface is termed as the transmembrane potential ($V_m=V_i-V_e$), and it can be calculated solving eqs.~\ref{eq:poti} and \ref{eq:pote} along with the electrical current continuity across the membrane
\begin{equation}\label{currentcont}
 \sigma_r E_{n,i}+\epsilon_r\frac{dE_{n,i}}{dt}=E_{n,e}+\frac{dE_{n,e}}{dt}=\hat{C}_m \frac{d V_m}{dt}+\hat{G}_m V_m,
\end{equation}
where $E_{n,i,e}$ are normal electric fields inside and outside of the membrane. The nondimensional capacitance and conductance of the membrane are $\hat{C}_m={aC_m}/{\epsilon_e\epsilon_0}$ and $ \hat{G}_m={aG_m}/{\sigma_e}$.
 
 The Maxwell electric stress tensor in a fluid is defined as $\tilde{{\bf  \tau}}^E=\epsilon \epsilon_0 (\tilde{{\bf  E}}\tilde{{\bf  E}}-\frac{1}{2} \tilde{E}^2{\bf I})$, where ${\bf I}$ is the identity tensor. The net non-dimensional electric traction at the interface is given by, $\Delta {\bf f^E}={\bf n}\cdot ({\bf \tau}^E_e-{\bf \tau}^E_i)=\tau^{E}_n{\bf n}+\tau^{E}_t{\bf t}$, where the components of electric stresses are
\begin{subequations}
\label{eq:estress}
 \begin{eqnarray}
 &&\tau_n^E = \frac{1}{2}\left[(E_{n,e}^2-E_{t,e}^2)-\epsilon_r(E_{n,i}^2-E_{t,i}^2)\right],\\
 &&\tau_t^E = E_{n,e}E_{t,e}-\epsilon_rE_{n,i}E_{t,i}.
\end{eqnarray}
\end{subequations}

\subsection{Hydrodynamics}
 Typical characteristic length of a capsule is $\sim5\mu m$; therefore the flow inside and outside of the capsule can be described by the Stokes equations
\begin{equation}
{\bf \nabla} .{\bf u}=0, {\bf \nabla} .{\bf \tau^H}=0,
\end{equation}
where ${\bf u}$ is the fluid velocity and ${\bf \tau^H}$ is the viscous stress. Viscous stresses for the internal and external fluid media are given by
\begin{subequations}
\label{eq:hstress}
\begin{eqnarray}
  &&{\bf \tau}^H_i = -P{\bf I}+\lambda({\bf \nabla} {\bf u}+{\bf \nabla} {\bf u}^T)\\
  &&{\bf \tau}^H_e = -P{\bf I}+({\bf \nabla} {\bf u}+{\bf \nabla} {\bf u}^T),
\end{eqnarray}
\end{subequations}
respectively, where $P$ is the pressure. 

At large distances from the capsule membrane ($\vert {\bf x}\vert \rightarrow \infty$), ${\bf u}\rightarrow {\bf u}^\infty$, where ${\bf u}^\infty$ is assumed to be the undisturbed flow velocity of the exterior fluid and is given by the dimensional form of applied velocity in eq.~\ref{eq:freestream}. The solution of the above equations give rise to an integral equation for the interfacial velocity~\cite{rallison78} in a non-dimensional form as,

\begin{equation}\label{eq:veleqn}
{\bf u}({\bf x}_0)=\frac{2}{1+\lambda}{\bf u}^\infty ({\bf x}_0)-\frac{1}{1+\lambda}\frac{1}{4\pi}\int_s\Delta{{\bf f}}({\bf x})\cdot{\bf G}({\bf x},{\bf x}_0)ds({\bf x})
+\frac{1}{4\pi}\frac{1-\lambda}{1+\lambda}\int_s {\bf u}({\bf x})\cdot{\bf Q}({\bf x},{\bf x}_0)\cdot{\bf n}({\bf x})ds({\bf x}), 
\end{equation}
where ${\bf G}({\bf x},{\bf x}_0)$ and ${\bf Q}({\bf x},{\bf x}_0)$ are Green's functions for velocity and stress, respectively and $\Delta{\bf f}$ is the unbalanced non-hydrodynamic traction at the interface. For an unbounded three dimensional flow, explicit expressions for these tensors are,
\begin{equation}
 {\bf G}({\bf x},{\bf x}_0)=\frac{{\bf I}}{|\hat{{\bf x}}|}+\frac{\hat{{\bf x}}\hat{{\bf x}}}{{|\hat{{\bf x}}|}^3},\quad {\bf Q}({\bf x},{\bf x}_0)=-6\frac{\hat{{\bf x}}\hat{{\bf x}}\hat{{\bf x}}}{{|\hat{{\bf x}}|}^5}.
\end{equation}
The general boundary integral equation (eq.~\ref{eq:veleqn}) is solved for the velocity of the interface of an axisymmetric deformable capsule in an axisymmetric flow and applied electric field. For the analysis of the dynamics of a biconcave-discoid capsule and a RBC subjected to a uniform electric field, the stagnant external fluid medium is considered, therefore ${\bf u}^{\infty}=0$. 

Finally, the shape can be evolved to ${\bf x}(t+\Delta t)$, using the kinematic condition given by
 \begin{equation}\label{eq:kin}
  {\bf x}(t+\Delta t)={\bf x}(t)+k_f{\bf u} ({\bf x})\Delta t,
 \end{equation}
 where ${\bf x}(t)$ is the shape of the capsule at the current time, $t$, and  $\Delta t$ is the time step considered for the boundary integral simulation. In eq.~\ref{eq:kin}, the kinematic factor $k_f$ depends upon the scaling of the variables and is described in the corresponding sections.
 
 \subsection{Pressure calculation}
Internal and external pressure of the capsule can be calculated using the boundary integral equations, given as
\begin{subequations}
\label{eq:pressure}
\begin{eqnarray}
 P_i({\bf x}_0) &=&-\frac{1}{8\pi}\int_s\Delta {\bf f}({\bf x})\cdot {\bf p}({\bf x},{\bf x}_0)dS({\bf x})+\lambda\frac{1-\lambda}{8\pi}\int_s {\bf u}({\bf x})\cdot{\bf \pi}({\bf x},{\bf x}_0)\cdot{\bf n}({\bf x})dS({\bf x})\label{eq:pri},\\
 P_e({\bf x}_0) &=&-\frac{1}{8\pi}\int_s\Delta {\bf f}({\bf x})\cdot {\bf p}({\bf x},{\bf x}_0)dS({\bf x})+\frac{1-\lambda}{8\pi}\int_s {\bf u}({\bf x})\cdot{\bf \pi}({\bf x},{\bf x}_0)\cdot{\bf n}({\bf x})dS({\bf x})\label{eq:pre},
 \end{eqnarray}
\end{subequations}
where ${\bf p}({\bf x},{\bf x}_0)$ and ${\bf \pi}({\bf x},{\bf x}_0)$ are the free space Green's functions for pressure~\citep{pozrikidis92,lac07}. 
\begin{equation}
 {\bf p}({\bf x},{\bf x}_0)=2\frac{\hat{ {\bf  x}}}{\hat{ {\bf  x}}^3}, \quad {\bf \pi}({\bf x},{\bf x}_0)=4\left(-\frac{{\bf I}}{|\hat{{\bf x}}|^3}+3\frac{\hat{ {\bf  x}}\hat{{\bf  x}}}{|\hat{{\bf x}}|^5}\right)
\end{equation}

\section{Results and Discussion}
\subsection{Dynamics of capsule in extensional flow} 
The scaling of eq.~\ref{eq:freestream}, gives rise to a nondimensional quantity $Ca_f={\mu e a}/E_s$, which is termed as the flow capillary number, and it determines the dynamics of the deformation of a capsule. The hydrodynamic force at the interface is balanced by the elastic and bending forces, i.e., $\Delta {\bf f}^H=-(\Delta {\bf f}^{el}+\Delta {\bf f}^b)$. It is assumed that the kinematic factor $k_f=1$ and the Skalak membrane parameter $C$ is assumed to be $1$ and $10$ for two separate studies. 

\subsubsection{Consideration of Small dilatation parameter: C=1}
\begin{figure}
\centering
 \includegraphics[width=0.45\linewidth]{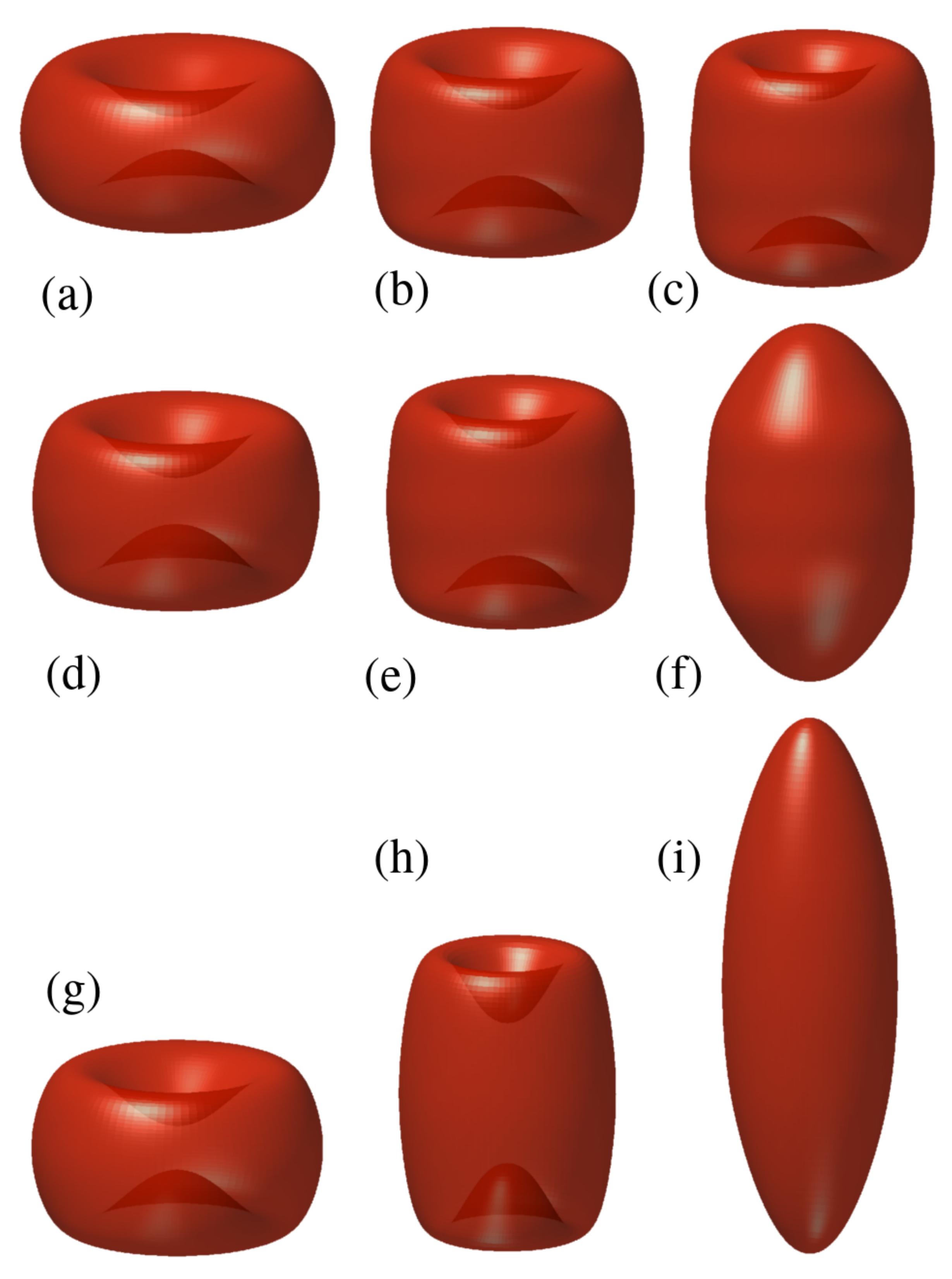}
 \caption{During the evolution of a capsule, observed shapes at (a) $t=10$, (b) $t=40$, (c) $t=\infty$ for $Ca_f=0.015$, (d) $t=20$, (e) $t=70$ (f) $t=\infty$ for $Ca_f=0.017$ and (g) $t=2$, (h) $t=7$, (i) $t=\infty$ for $Ca_f=0.1$, considering the dilatation parameter $C=1$. }
 \label{fig:c1}
\end{figure}

The analysis of the dynamics of deformation of a capsule in extensional flow is presented for a small dilatation parameter (C=1 in this case). It is observed that a considerable change in area ($\sim 13\%$) is admitted when the capsule is subjected to a strong extensional flow. The capillary numbers are chosen in such a way that they demonstrate three possible modes of deformation. Fig.~\ref{fig:c1}a-c show that at a small value of destabilizing force $(Ca_f=0.015)$, the deformation is small and biconcavity is maintained. The capsule slowly deforms such that the concavities at the poles are affected the most~(see Fig.~\ref{fig:c1}a and b), Subsequently the flow reduces the depression (in a way opening them up) and eventually reaches to a steady state shape as shown in Fig.~\ref{fig:c1}c.  A capsule does not open up into a spheroid until a threshold capillary number is reached, (in this case, $Ca_f=0.017$). At such a threshold capillary number, $(Ca_f=0.017)$ the capsule shows a remarkable shape transition wherein the biconcave cavities~(see Fig.~\ref{fig:c1}d and e) open up and transform into a steady spheroid shape~(see Fig.~\ref{fig:c1}f). At a still higher capillary number $(Ca_f=0.1)$, the dynamics is faster, and a very similar behavior is observed~(see Fig.~\ref{fig:c1}g-i), such that the final spheroid shape is a highly stretched one~(see Fig.~\ref{fig:c1}i).

\begin{figure}
 \includegraphics[width=0.95\linewidth]{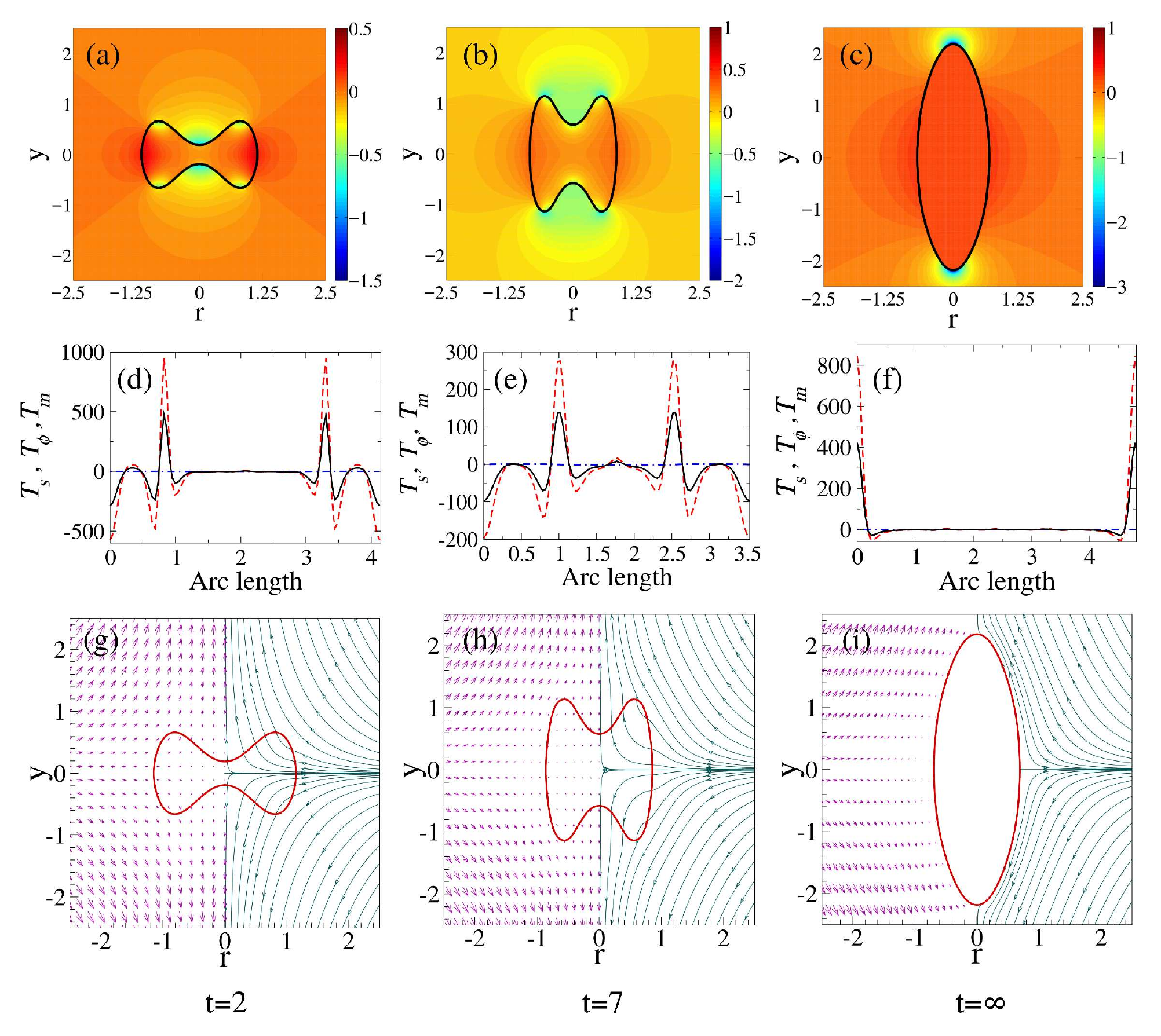}
 \caption{At $Ca_f=0.1$ and $C=1$, (a, b, c) Pressure profile, (d, e, f) variation of meridional, $T_s$ (\textcolor{red}{$\pmb{--}$}), azimuthal, $T_\phi$ (\textcolor{blue}{$\pmb{-\cdot-}$}) and mean, $T_m$ (\textcolor{black}{$\mi$}) elastic tensions along the arc length (g, h, i) streamlines (shown only at right half), velocity profile (relative magnitude of the flow is proportional to the length of the short arrows shown at left half). First, second and third columns of figures are corresponding to the times $t=2$, $7$ and $\infty$, respectively.}
 \label{fig:force}
\end{figure}

To understand the mechanism of transition from a biconcave shape to a spheroid, the pressure profile and elastic tension over the half arc length are plotted in Fig.~\ref{fig:force}a-c and Fig.~\ref{fig:force}d-f, respectively. The applied free stream uniaxial extensional flow does not have an associated pressure profile because the velocity is linear. When a rigid spherical particle is placed in such a flow, it generates a velocity field to bring the velocity of the applied flow to zero on the particle surface. This generates a stress on the particle from the equator to the poles, which the particle transmits to the fluid (from poles to equator) thereby leading to a stresslet disturbance velocity field, the stresslet being a pusher (kind). Since in the case of a rigid spherical particle, the particle brings the fluid to rest at its surface, the fluid has to generate a positive pressure at the equator to resist the incoming flow and a negative pressure at the poles to suck fluid into the poles. When the surface/interface is not rigid, e.g., a drop, it can give way under this pressure and builds curvature. In a drop, this can result in a dimple at the equator and a bump at the pole (the drop will try to adjust its curvature to satisfy young Laplace at the interface) leading to a prolate shape. Similar arguments will mean that the shape of a drop will be oblate in a biaxial extensional flow. 

In a capsule, the meridional and azimuthal tensions are present which could be anisotropic and compressive, unlike the isotropic tension in a liquid-liquid interface (a drop).While a spherical drop cannot maintain its shape in extensional flow, a capsule can maintain a near-spherical shape by generating a pressure inside that is intermediate between that at the equator and the poles on the outer side. This can be attained by a compressive membrane tension (predominantly azimuthal) at the equator  and a tensile membrane tension at the poles. In the case of a biconcave-discoid capsule, a similar distribution of membrane tension (compressive at the equator and tensile at the poles) can support an extensional flow, except the pressure inside now is lesser than that at the poles.

When the capillary number is increased beyond the threshold, at short times ($t=2$), the pressure is lower at the poles, thereby generating an internal flow from the equator to the poles. This leads to the biconcave cavities opening up. The pressure generated at the equator is so strong that the resulting internal pressure  at the poles continues to drive the fluid to the poles (t=7). This fluid flow (see Fig.~\ref{fig:force}g-i) , eventually, leads to the opening up of the poles and transformation of the biconcave-discoid shape into a prolate spheroid (see Fig.~\ref{fig:c1}i). 

The elastic tensions are predominantly compressive, except at the shoulders. At the poles, as the curvature is negative and the inside pressure is higher than at outside (see Fig.~\ref{fig:force}a, b), the stresses have to be compressive (see Fig.~\ref{fig:force}d, e). The tension at the shoulders, though, is tensile, leading to a higher pressure inside the capsule, which further assists the shoulders to disappear by pushing the fluid into the poles. 

For the deformed spheroid at the steady state, at the equator, the internal pressure is uniform (see Fig.~\ref{fig:force}c); therefore, the flow inside disappears (see Fig.~\ref{fig:force}i). The pressure inside is nearly identical to the external pressure, due to the very small curvature at the equator. Thereby, the mean elastic tension at the equator disappears (see Fig.~\ref{fig:force}f). At the poles, the external pressure is lower compared to the internal pressure (see Fig.~\ref{fig:force}c), and since the curvature is positive, a high tensile mean elastic tension (see Fig.~\ref{fig:force}f) is generated. 

\subsubsection{Large dilatation parameter: C=10}
\begin{figure}
\centering
\includegraphics[width=0.45\linewidth]{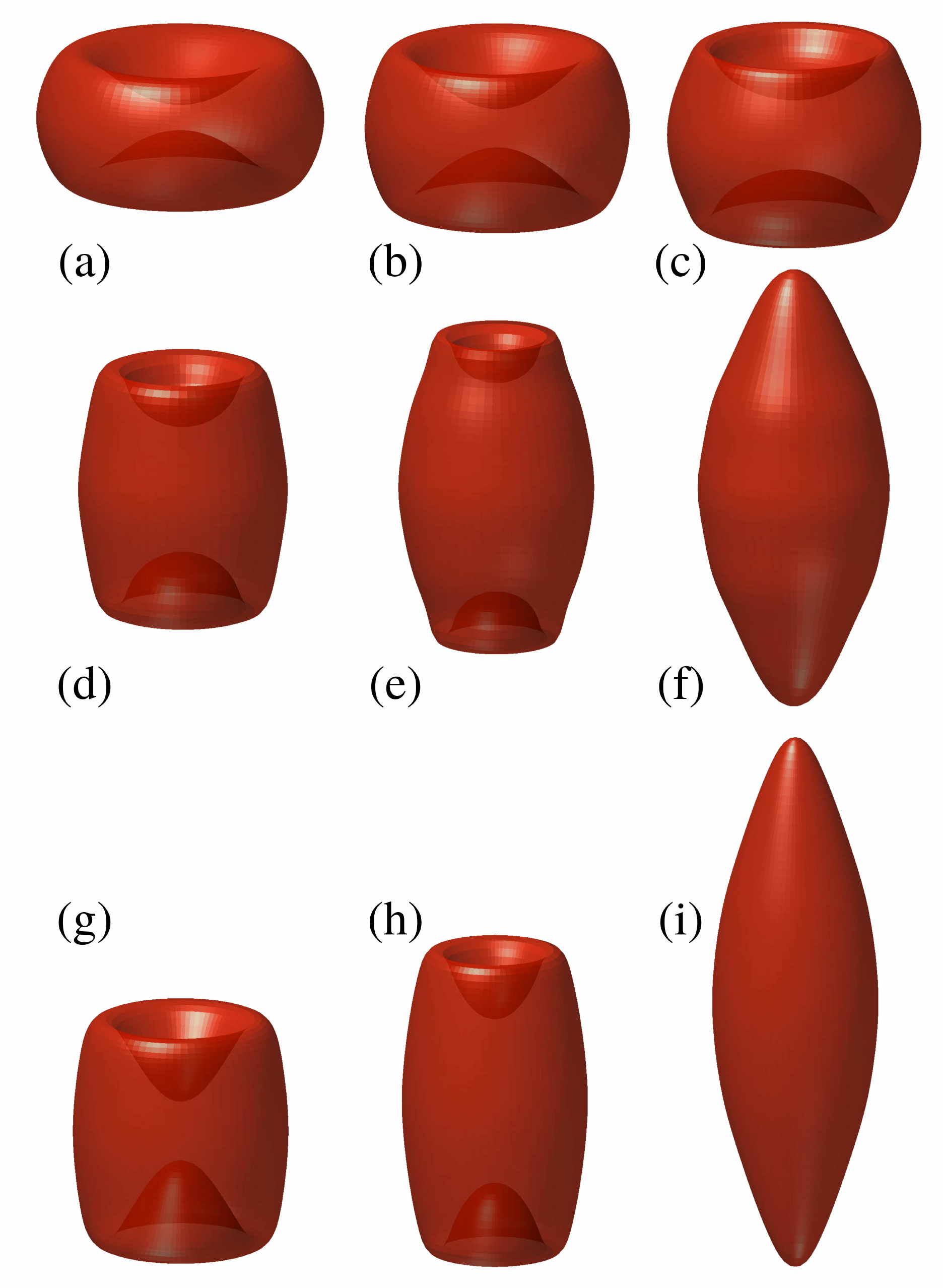}
\caption{During the evolution of a capsule, observed shapes at (a) $t=10$, (b) $t=25$, (c) $t=\infty$ for $Ca_f=0.015$, (d) $t=30$, (e) $t=75$ (f) $t=\infty$ for $Ca_f=0.04$ and (g) $t=5$, (h) $t=10$, (i) $t=\infty$ for $Ca_f=0.1$, considering the dilatation parameter $C=10$.}
\label{fig:c10}
\end{figure}
For the deformation of a Skalak capsule, a large dilatation parameter restricts the change in surface area, although at the cost of numerical stiffness. Therefore, the analysis is restricted to a maximum value of dilatation parameter $C=10$ allowing $\sim 10\%$ change in surface area. At a small capillary number, $Ca_f=0.015$, a behavior similar to Fig.~\ref{fig:c1}a-c (for $Ca_f=0.015$ at $C=1$) in the deformation of a  capsule is observed. However significant differences exist between the shapes shown in Fig.~\ref{fig:c1}d-i and Fig.~\ref{fig:c10}d-i.

The threshold value of the capillary number to evolve into a spheroid is comparatively higher, $Ca_f=0.04$ (compared to $Ca_f=0.017$ at $C=1$). From Fig.~\ref{fig:c10}d-f, it can be observed that, unlike Fig.~\ref{fig:c1}d and e, a capsule does not take a cylindrical shape (there is a bulge at the equator). In this case, the final steady-state shape is highly stretched (compared to Fig.~\ref{fig:c1}f), due to high $Ca_f$. However, for $Ca_f=0.1$ and $C=10$, a capsule deforms through intermediate cylindrical shapes (see Fig.~\ref{fig:c10}g and h) to a highly stretched steady state spheroid with pointed tips (see Fig.~\ref{fig:c10}i). 

\subsection{Dynamics of biconcave-discoid capsule in DC electric field} \label{sec:efield}
The analysis of deformation of an elastic biconcave-discoid capsule in DC electric field is carried out in the absence of free stream fluid velocity,  ${\bf u}^\infty=0$. Therefore the hydrodynamic traction is balanced by the elastic traction, electric traction and the traction due to bending rigidity, i.e.,  
\begin{equation}\label{eq:fbel}
  \Delta {\bf f}^H=-(\Delta{\bf f}^{el}+Ca_e \Delta{\bf f}^E+\hat{\kappa}_b \Delta{\bf f}^b),
\end{equation}
where $Ca_e=\epsilon_e\epsilon_0aE_0^2/E_s$ is the electric capillary number. The scaling of variables results in the kinematic factor as the hydrodynamic response time, i.e., $k_f=1/t_H$.  Considering a typical set of parameters $E_s\sim 0.1\ N/m$, $\mu_e\sim 0.89\ mPa\cdot s$, $\sigma_e\sim 10\ mS/m$ and $\epsilon_e\sim 80$, for the dynamics of a capsule with equivalent radius $a\sim 5\mu m$, the hydrodynamic response time can be calculated to be unity, i.e., $t_H=t_e=1$. For this analysis, the consideration of $t_H=t_e=1$ implies that the hydrodynamic response of the capsule is over the same time scale as the evolution of the electric stresses.  Also, the membrane of the capsule is considered to be purely capacitive with $\hat{C}_m=50$ and $\hat{G}_m=0$. 

\subsubsection{More conducting external fluid medium: $\sigma_r=0.1$} 
\begin{figure}
\centering
\includegraphics[width=0.95\linewidth]{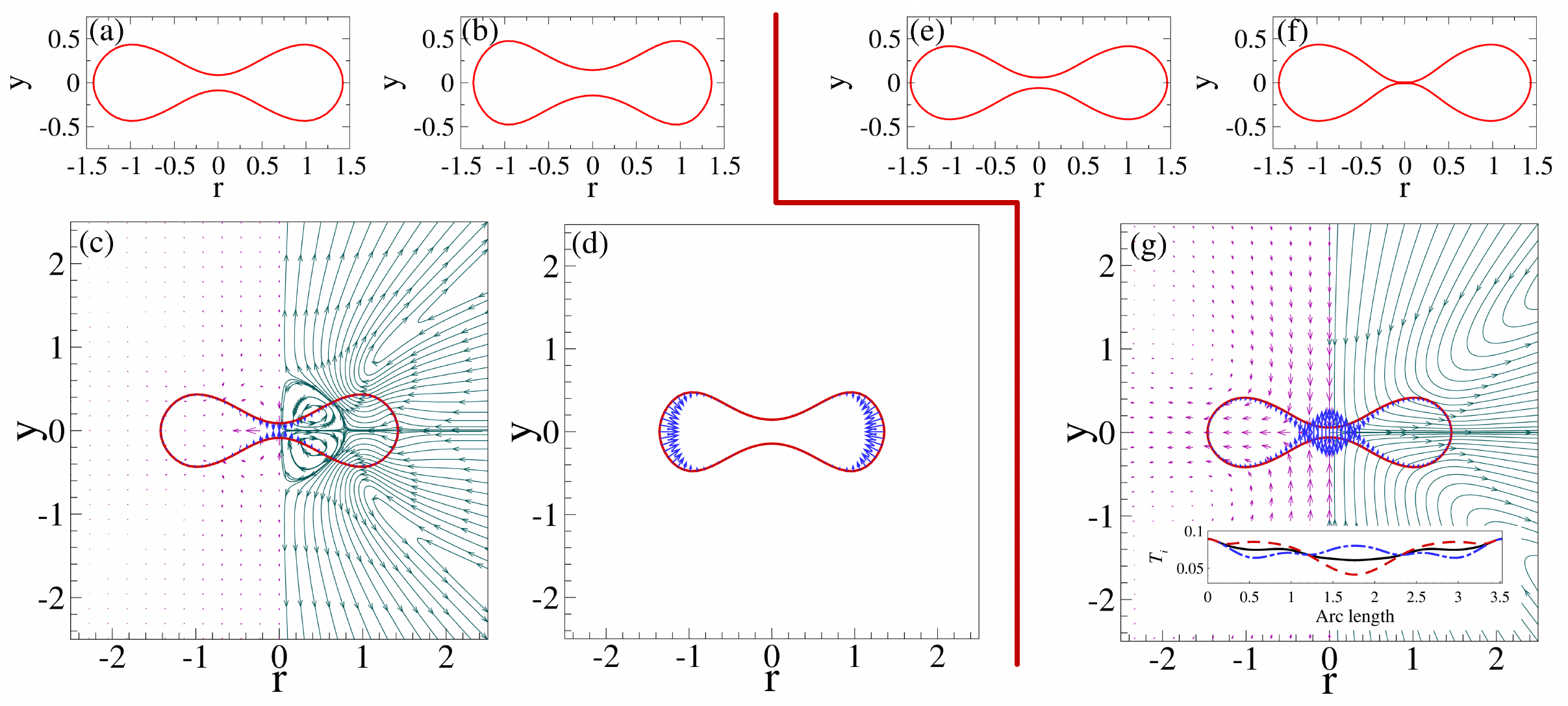}
\caption{During the evolution of a biconcave-discoid capsule, observed shapes at (a) $t=26$, (b) $t=\infty$ at $Ca_e=0.02$ and $\sigma_r=0.1$, corresponding streamlines (shown only at right half), velocity profile (relative magnitude of the flow is proportional to the length of the short arrows shown at left half), electric stress distribution (shown by short arrows from the interface) are shown in (c) and (d), respectively. (e) $t=15$ and (f) $t=65$ represent the observed shapes during deformation at $Ca_e=0.05$ and $\sigma_r=0.1$. Streamlines, velocity profile, electric stress distribution are shown in (g) corresponding to the shape shown in (e). Variation of meridional, $T_s$ (\textcolor{red}{$\pmb{\pmb{--}}$}), azimuthal, $T_\phi$ (\textcolor{blue}{$\pmb{\pmb{-\cdot-}}$}) and mean, $T_m$ (\textcolor{black}{$\pmb{\mi}$}) membrane tensions along the arc length are shown in the inset of (g).}
\label{fig:el0p1}
\end{figure}

\begin{figure}
\centering
\includegraphics[width=0.95\linewidth]{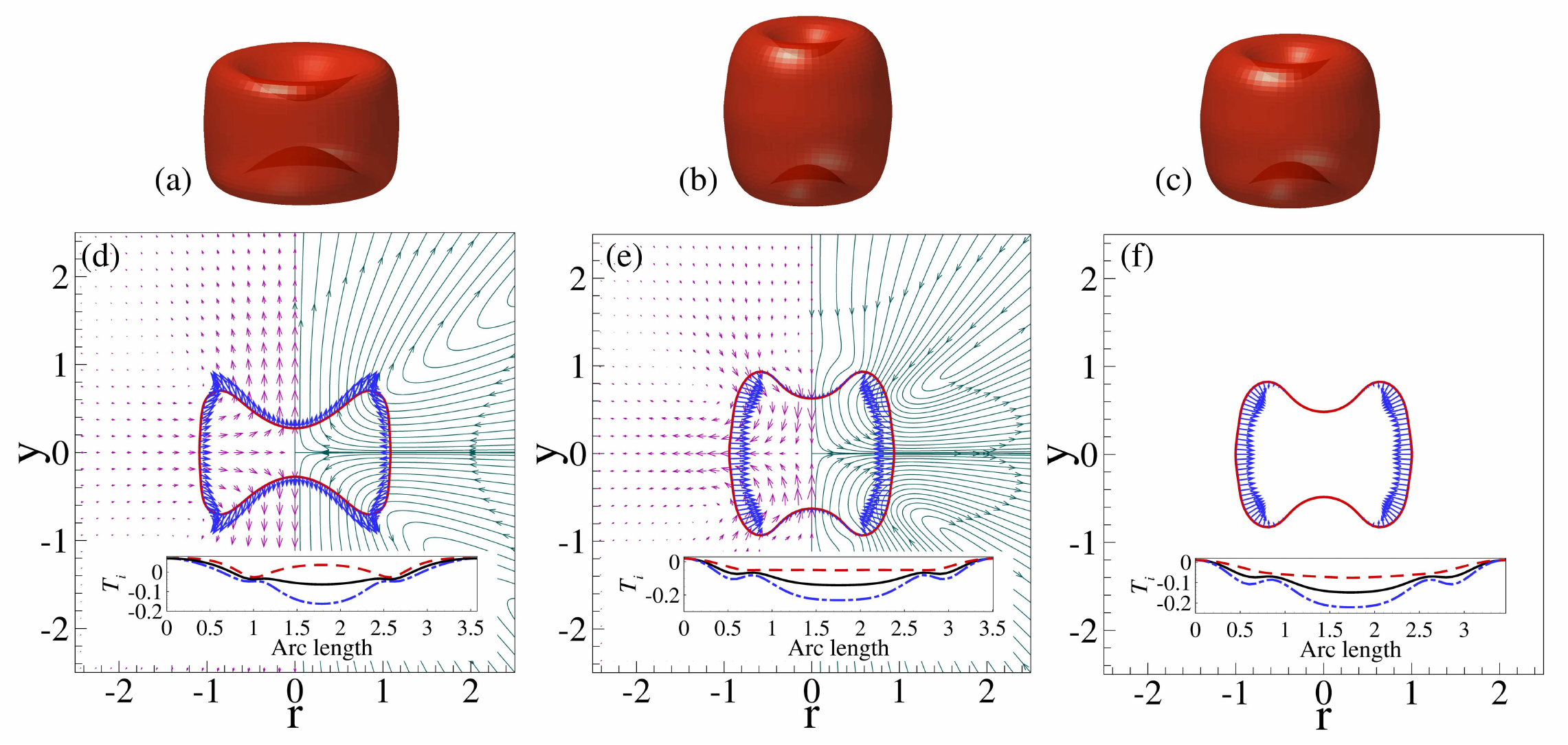}
\caption{During the evolution of a biconcave-discoid capsule, observed shapes at (a) $t=20$, (b) $t=85$ and (c) $t=\infty$ at $Ca_e=0.3$ and $\sigma_r=10$, corresponding streamlines (shown only at right half), velocity profile (relative magnitude of the flow is proportional to the length of the short arrows shown at left half), electric stress distribution (shown by short arrows from the interface) are shown in (d), (e) and (f), respectively. Variation of meridional, $T_s$ (\textcolor{red}{$\pmb{\pmb{--}}$}), azimuthal, $T_\phi$ (\textcolor{blue}{$\pmb{\pmb{-\cdot-}}$}) and mean, $T_m$ (\textcolor{black}{$\pmb{\mi}$}) membrane tensions along the arc length are shown in the insets.}
\label{fig:ca0p3}
\end{figure}

\begin{figure}
\centering
\includegraphics[width=0.95\linewidth]{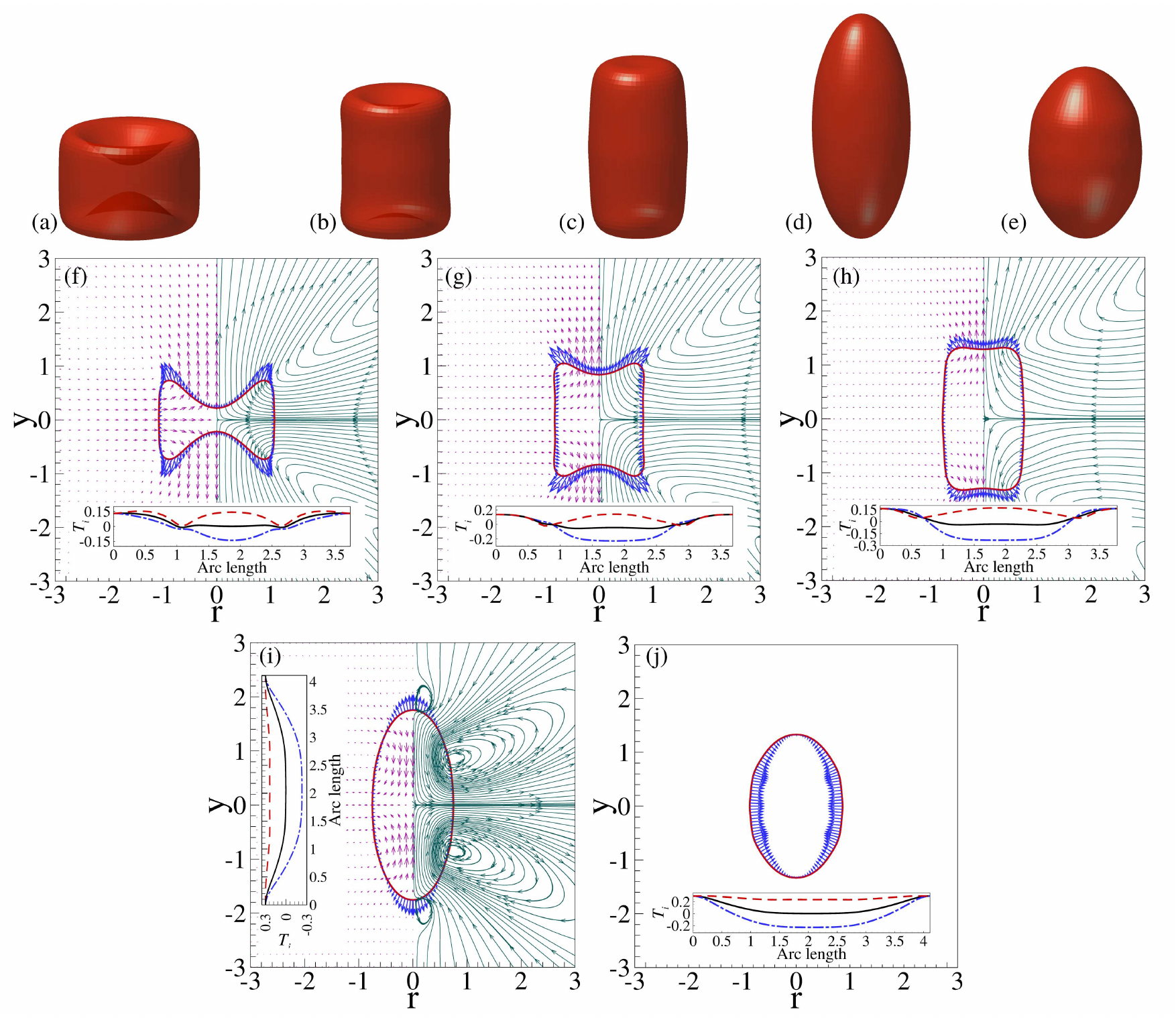}
\caption{During the evolution of a biconcave-discoid capsule, observed shapes at (a) $t=10$, (b) $t=23$, (c) $t=29$, (d) $t=37$ and (e) $t=\infty$ at $Ca_e=0.5$ and $\sigma_r=10$, corresponding streamlines (shown only at right half), velocity profile (relative magnitude of the flow is proportional to the length of the short arrows shown at left half), electric stress distribution (shown by short arrows from the interface) are shown in (f), (g), (h), (i) and (j), respectively. Variation of meridional, $T_s$ (\textcolor{red}{$\pmb{\pmb{--}}$}), azimuthal, $T_\phi$ (\textcolor{blue}{$\pmb{\pmb{-\cdot-}}$}) and mean, $T_m$ (\textcolor{black}{$\pmb{\mi}$}) membrane tensions along the arc length are shown in the insets.}
\label{fig:ca0p5}
\end{figure}

\begin{figure}
\centering
\begin{minipage}{0.45\textwidth}
\centering
\includegraphics[width=1\textwidth]{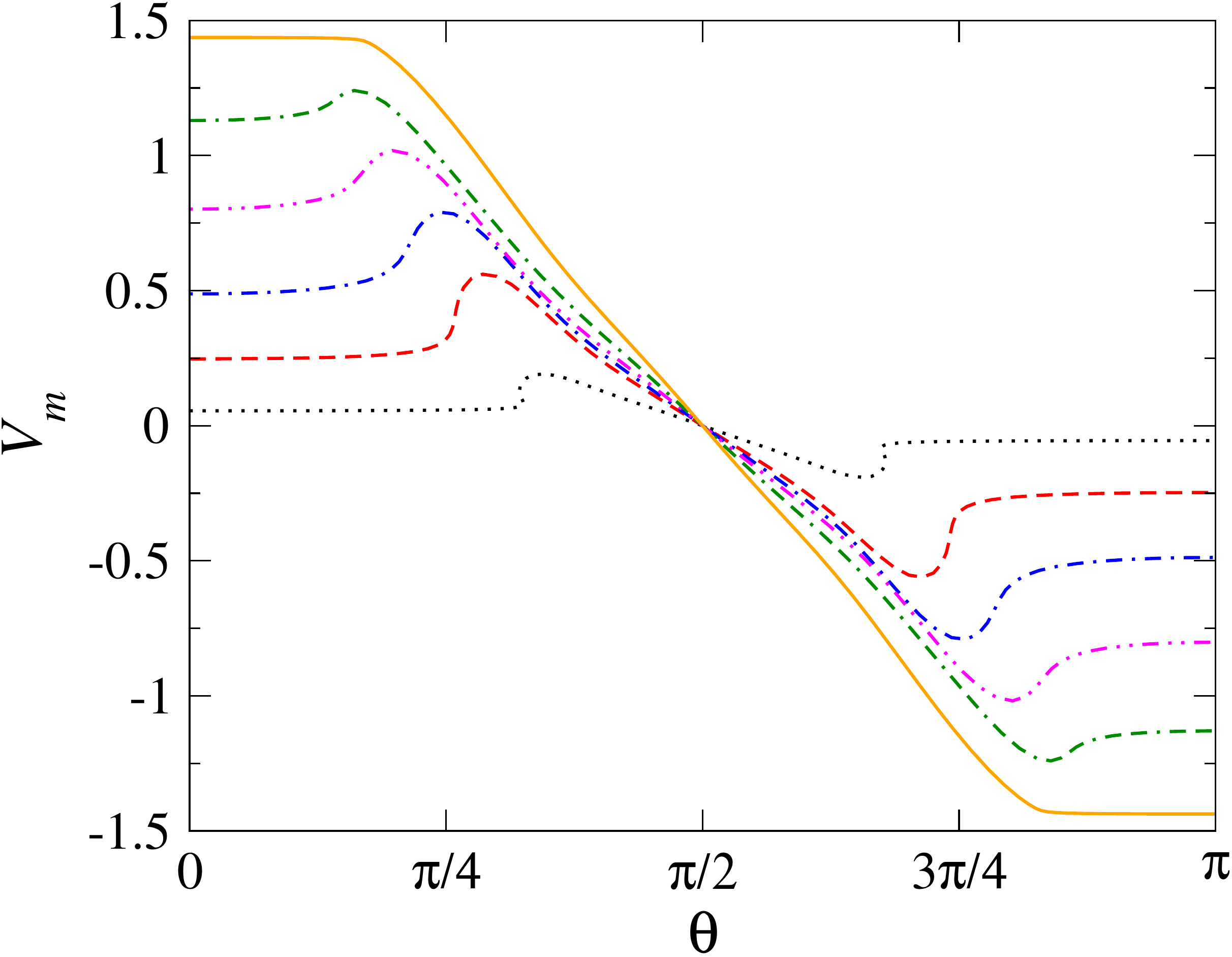}
(a) $Ca=0.3$
\end{minipage}
\begin{minipage}{0.45\textwidth}
\centering
\includegraphics[width=1\textwidth]{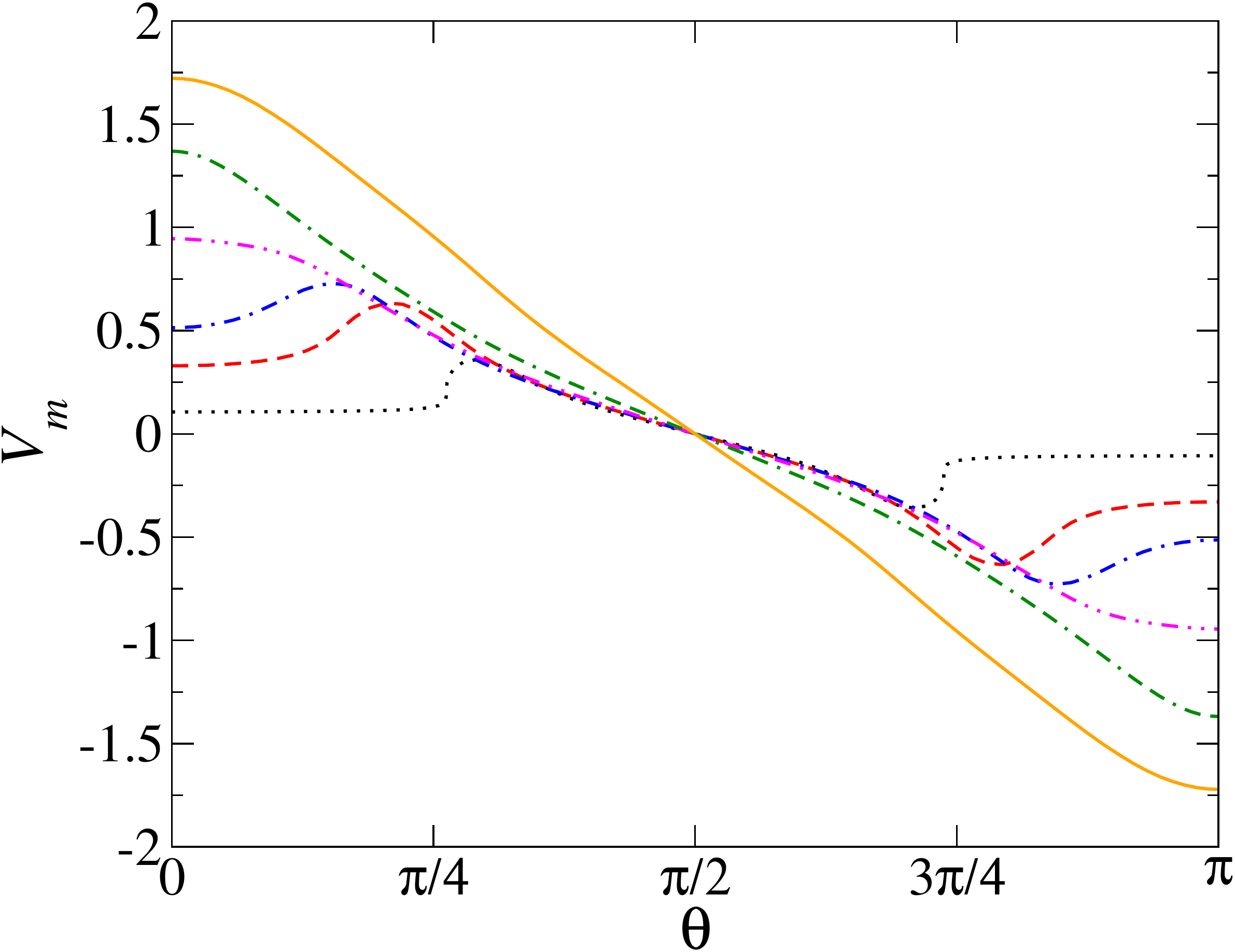}
(b) $Ca=0.5$
\end{minipage}
 \caption{During the shape evolution of a biconcave-discoid capsule, the transmembrane potential at the interface  at (\textcolor{black}{$\pmb{\pmb{\cdots}}$}) $t=5$, (\textcolor{red}{$\pmb{\pmb{--}}$}) $t=20$, (\textcolor{blue}{$\pmb{\pmb{-\cdot-}}$}) $t=35$, (\textcolor{magenta}{$\pmb{\pmb{-\cdot\cdot}}$}) $t=55$, (\textcolor{green}{$\pmb{\pmb{--\cdot}}$}) $t=85$, and (\textcolor{orange}{$\pmb{\mi}$}) $t=\infty$ at (a) $Ca=0.3$, and at (\textcolor{black}{$\pmb{\pmb{\cdots}}$}) $t=10$, (\textcolor{red}{$\pmb{\pmb{--}}$}) $t=23$, (\textcolor{blue}{$\pmb{\pmb{-\cdot-}}$}) $t=29$, (\textcolor{magenta}{$\pmb{\pmb{-\cdot\cdot}}$}) $t=37$, (\textcolor{green}{$\pmb{\pmb{--\cdot}}$}) $t=50$, and (\textcolor{orange}{$\pmb{\mi}$}) $t=\infty$ at (b) $Ca=0.5$.}
 \label{fig:vm}
\end{figure}

When the external fluid is more conducting than the internal fluid $(\sigma_r=0.1)$, at $Ca_e=0.02$, a biconcave-discoid capsule initially undergoes compression at the poles (see Fig.~\ref{fig:el0p1}a and c). In this case, shapes are presented as 2D cross-section in the plane parallel to the axis of symmetry, for better visualization in the small deformation. At $t<t_{cap}$, the polarization vector is in the opposite direction of the applied electric field; therefore, the normal electric stresses are compressive at the poles (see Fig.~\ref{fig:el0p1}c). Later, at $t>t_{cap}$, the membrane becomes charged, and the electric stress becomes compressive with the maxima at the equator (see Fig.~\ref{fig:el0p1}d), drives the poles apart and the capsule reaches a steady shape as shown in Fig.~\ref{fig:el0p1}b. At this low capillary number, a biconcave-discoid capsule does not undergo large deformation due to the developed weak electric stresses at the interface.

At significantly high capillary number ($Ca_e=0.05$), a biconcave-discoid capsule breaks through the merging of the poles (see Fig.~\ref{fig:el0p1}f). At $t<t_{cap}$, the strong compressive electric stresses at the poles and the fluid flow (see Fig.~\ref{fig:el0p1}g) from poles to equator assist the collapse of the capsule (see Fig.~\ref{fig:el0p1}f). Fig.~\ref{fig:el0p1}g is corresponding to the Fig.~\ref{fig:el0p1}e, which is obtained from the valid numerical calculation, whereas the breakup shape (see Fig.~\ref{fig:el0p1}f) obtained due to the merging of poles can be considered as the numerical artifact.

\begin{figure}
\centering
\includegraphics[width=0.95\linewidth]{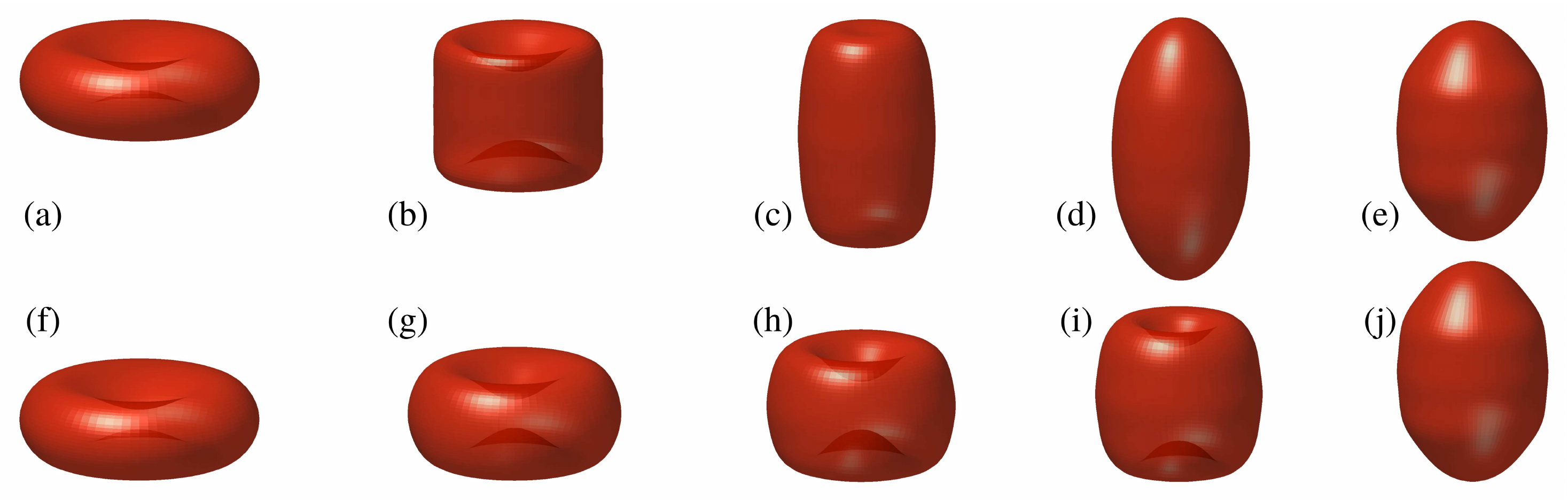}
\caption{Shapes observed at (a) $t=0$, (b) $20$, (c) $42$, (d) $52$ and (e) $\infty$, during the deformation of a biconcave-discoid capsule at $Ca_e=0.4$ for $\sigma_r=10$, considering $k_f=1$ and at (f) $t=0$, (g) $500$, (h) $1500$, (i) $5000$ and (j) $\infty$, considering $k_f=0.01$.}
\label{fig:kf}
\end{figure}

\subsubsection{More conducting internal fluid medium: $\sigma_r=10$} 
When the internal fluid medium is more conducting, the capsule does not collapse; instead, it deforms through a series of  cylindrical shapes to a steady-state shape. At a small capillary number $(Ca_e=0.3)$, a biconcave-discoid capsule deforms through an intermediate cylindrical shape with concave ends (see Fig.~\ref{fig:ca0p3}a). In this case, the polarization vectors align with the applied electric field; therefore, the electric stresses are tensile near the poles and highest at the shoulders (see Fig.~\ref{fig:ca0p3}d). Just after $t>t_{cap}$, as the membrane becomes charged, the tensile electric stresses at the poles disappear, and the compressive electric stresses develop at the equator. In this case, as the normal tensile electric stresses disappear, the intermediate shape (see Fig.~\ref{fig:ca0p3}b) relaxes back (because of the stabilizing elastic traction) to an equilibrium shape (see Fig.~\ref{fig:ca0p3}c). This equilibrium shape is the result of the balance of the elastic traction and compressive electric stress (see Fig.~\ref{fig:ca0p3}f). 
At a high capillary number $(Ca_e=0.5)$, the biconcave-discoid capsule at short times deforms into cylindrical shapes with diminishing concave ends near the poles (see Fig.~\ref{fig:ca0p5}a and b) due to high tensile Maxwell stress at the poles and the shoulders (see Fig.~\ref{fig:ca0p5}f and g). In this case, since the hydrodynamic time scale is small ($t_H=t_e=1$), the deformation progresses rapidly. At $t < t_{cap}$, due to the tensile electric stress (with the maxima near the poles, see Fig.~\ref{fig:ca0p5}h) the biconcave-discoid shape evolves into a cylindrical shape (see Fig.~\ref{fig:ca0p5}c). The tensile normal electric stress at the poles (see Fig.~\ref{fig:ca0p5}i) continues to drive the capsule to evolve into a prolate spheroid shape (see Fig.~\ref{fig:ca0p5}d). Eventually, when the membrane becomes fully charged ($t> t_{cap}$), it retains a steady spheroidal equilibrium shape (see Fig.~\ref{fig:ca0p5}e) wherein the compressive electric stress at the equator (see Fig.~\ref{fig:ca0p5}j) is appropriately balanced by the elastic tensions. 

Thus, similar to the deformation in extensional flow (see Fig.~\ref{fig:c1} and \ref{fig:c10}), a biconcave-discoid capsule deforms into a steady state spheroid (see Fig.~\ref{fig:ca0p5}) when the internal fluid medium is more conducting $(\sigma_r=10)$. In the former case, the deformation is driven by the hydrodynamic pressure generated by the extensional flow around the biconcave-discoid capsule, whereas in the latter case, it is due to the Maxwell electric stress developed at the interface of the capsule. 

During the deformation, the transmembrane potential evolves due to the charging of the membrane as well as the change in the shape of the biconcave-discoid capsule. Fig.~\ref{fig:vm}a and b show the variation of transmembrane potential over the interface for different intermediate shapes during the deformation at $Ca=0.3$ and $Ca=0.5$, respectively. From Fig.~\ref{fig:vm}a it can be observed that the transmembrane potential is always higher at the shoulders until the membrane becomes fully charged, as the shape remains biconcave during the deformation (Fig.~\ref{fig:ca0p3}a,b). For the charged membrane and the steady-state shape (Fig.~\ref{fig:ca0p3}c), the transmembrane potential from the poles to the shoulders are comparable and higher than the other locations at the interface. From Fig.~\ref{fig:vm}b it can be observed that in the deformation at $Ca=0.5$ the transmembrane potential is higher at the shoulders for the cylinders with concave ends (Fig.~\ref{fig:ca0p5}a-c), whereas it is higher the poles for the prolate spheroid (Fig.~\ref{fig:ca0p5}d-e) shapes.

When the hydrodynamic response time and the electric response time are same, $t_H=t_e=1$, i.e., $k_f=1$, near the critical capillary number ($Ca_e=0.4$) for evolving into a prolate spheroid, a biconcave-discoid evolves through the intermediate shapes similar to the case of $Ca_e=0.5$ (see Fig.~\ref{fig:ca0p5}). At this capillary number ($Ca_e=0.4$), the shape evolution is shown in Fig.~\ref{fig:kf}a-e. Fig.~\ref{fig:kf}f-j represent the shape evolution of a biconcave-discoid capsule at $Ca_e=0.4$ and $\sigma_r=10$, considering $k_f=0.01$. In this case, a biconcave-discoid capsule does not respond instantly to the developed electric stresses at the interface. Therefore, at $t<t_{cap}$, the intermediate cylindrical (see Fig.~\ref{fig:kf}c) and highly deformed prolate spheroid (see Fig.~\ref{fig:kf}d) shapes are missing in the deformation considering $k_f=0.01$. Instead, the biconcave-discoid capsule undergoes a slow deformation through few intermediate shapes (see Fig.~\ref{fig:kf}g-i), and eventually attains a prolate spheroid shape (see Fig.~\ref{fig:kf}j) at the equilibrium. Hence, for different hydrodynamic response-times (here, $t_H=1$ and $100$, i.e., $k_f=1$ and $0.01$), even though the final equilibrium shapes a biconcave-discoid capsule attains are same (see Fig.~\ref{fig:kf}e  and Fig.~\ref{fig:kf}j), the pathways for the shape evolution are different. 

\subsection{Dynamics of a RBC in AC electric field}
The electrohydrodynamic deformation of a biconcave-discoid capsule can be extended to explore its suitability for studying a RBC. Physiological parameters for the human RBC, obtained from the literature~\citep{beving1994,haidekker02,poz03,wolf11}, are shown in Table~\ref{tab:parametertable}. Under these  conditions, the non-dimensional parameters are: viscosity ratio $\lambda=5$, ratio of dielectric constants $\epsilon_r=0.6$, conductivity ratio $\sigma_r=1.054$, nondimensional capacitance $\hat{C}_m=282$ and nondimensional conductance $\hat{G}_m=6.75\times 10^{-7}$. Considering scaling parameters as used in the case of deformation of a biconcave-discoid capsule, for the deformation of RBC, the nondimensional hydrodynamic timescale is very large (non-dimensional $t_H=0.625\times 10^{6}$). Therefore, the kinematic factor $k_f=1/t_H\sim O(10^{-6})$ implies that a negligible displacement (eq.~\ref{eq:kin}) of the interface will take place in a considered numerical time-step, $\Delta t\sim O(\tilde{t}_e)$, suggesting that the computation time required for reaching a steady shape could be prohibitively high. 

\begin{table}
  \begin{center}
\def~{\hphantom{0}}
  \begin{tabular}{lcc}
  Parameter & \hspace{0.5in}Value \hspace{0.5in} & Unit\\[3pt]
  $a$ & $5$ & $\mu m$ \\
  $E_s$	& $6.1$& $\mu N/m$\\
  $\mu_e$ & $1.6$& $m Pa s$ at $21^oC$\\
  $\epsilon_e$ & $70$ & - \\
  $\epsilon_i$ & $42$ & - \\
  $\sigma_e$ & $0.74$ & $S/m$ \\
  $\sigma_i$ & $0.78$ & $S/m$ \\
  $C_m$ & $0.035$ & $F/m^2$ \\
  $G_m$ & $0.1$ & $S/m^2$\\
  \end{tabular}
  \caption{Values of parameters for human RBC.}
  \label{tab:parametertable}
  \end{center}
\end{table}

Since the hydrodynamic response time is very high compared to the electric time-scale, the electric field at the interface reaches an equilibrium value (with respect to the transmembrane potential) even before the capsule starts responding to the electric stress developed at the interface. The electric field (eq.~\ref{eq:ponieeqns}) is, therefore, solved with the capacitor time scale, $\tilde{t}_{cap}=t_{cap}\times$ electric time scale $(i.e., \tilde{t}_{cap}=t_{cap}\times \tilde{t}_e=282\times\epsilon_e\epsilon_0/\sigma_e)$, and the transmembrane potential is allowed to reach a steady value at a particular shape of the RBC. Subsequently, the hydrodynamic equation for RBC deformation (eq.~\ref{eq:veleqn}) is solved with the already calculated equilibrium electric stress, considering the hydrodynamic response time $(t_H=\mu_ea/E_s)$ as the scaling parameter for the time. Moreover, as the charge relaxation in both the internal and external fluids are very fast ($\epsilon_e/\sigma_e\ll 1$ and $\epsilon_i/\sigma_i\ll 1$), a simplification of the electric current continuity (eq.~\ref{currentcont}) across the interface is made as
\begin{equation}\label{modccont}
 \sigma_r E_{n,i}=E_{n,e}=\frac{d V_m}{dt}.
\end{equation}

All the variables but time (here $t=\tilde{t}/t_{cap}\times \tilde{t}_e$), for solving the electric potential (eq.~\ref{eq:ponieeqns}) are scaled with the scaling parameters used for the solution of electrostatics in the case of a biconcave-discoid elastic capsule in DC electric field as discussed in the previous section. Thus, $1/(t_{cap}\times \tilde{t}_e)$ is used for scaling the frequency of the applied field. For solving the hydrodynamic boundary integral equation (eq.~\ref{eq:veleqn}), the hydrodynamic time scale is used for scaling. 
 
The hydrodynamic force is balanced by the non-hydrodynamic forces similar to the case of deformation of an elastic capsule in DC electric field (eq.~\ref{eq:fbel}). In this case, a time-averaged electric stress (ignoring the time-periodic stress that has instantaneous dynamics but zero average) is used in the calculation of interfacial velocity (in eq.~\ref{eq:veleqn}). For faster numerical calculations, the viscosity ratio is considered to be unity, $\lambda=1$. The kinematic condition (eq.~\ref{eq:kin}) for the shape evolution from the calculated interfacial velocity, because of $k_f=1$, simplifies to  
 \begin{equation}
  {\bf x}(t+\varDelta t)={\bf x}(t)+{\bf u}(\bf x) \varDelta t.
 \end{equation}

 \begin{figure}
\centering
  \includegraphics[width=0.45\textwidth]{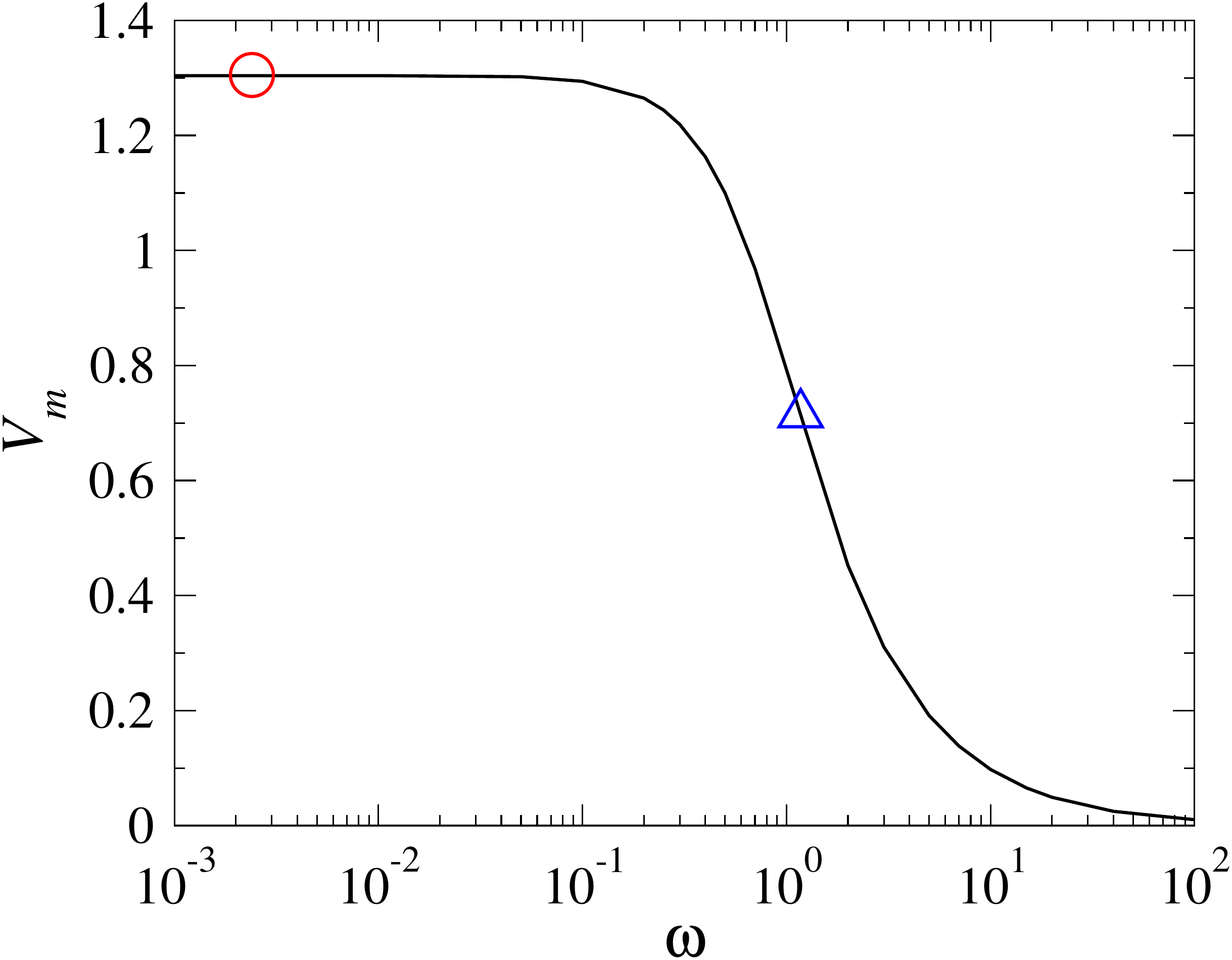}
\caption{Transmembrane potential as a function of frequency. Marker points \rr{$\bigcirc$} and \bb{$\triangle$} represent $t_{cap}^{-1}$ and $t_{MW}^{-1}$, respectively.}
\label{fig:vmvsw}
\end{figure}  
 
The analysis of the deformation of RBC in AC electric field is conducted at a high frequency of the applied field~($\omega=2.5$) (corresponding to $\sim 10 MHz$). This is important in suggested experiments, wherein the transmembrane potential ($V_m$) has to be restricted to a very low value to prevent the possibility of electroporation (see Fig.~\ref{fig:vmvsw})~\cite{joshi12}. The electric capillary number ($Ca_e$) remains the same as used in the case of the deformation of a biconcave-discoid capsule in DC field. In the case of the RBC deformation, $Ca_e=1$ is equivalent to the applied electric field strength of $0.45\ kV/cm$, much below the typical $1\ kV/cm$ or higher used in electroporation studies.

\begin{figure}
\centering
\includegraphics[width=0.95\textwidth]{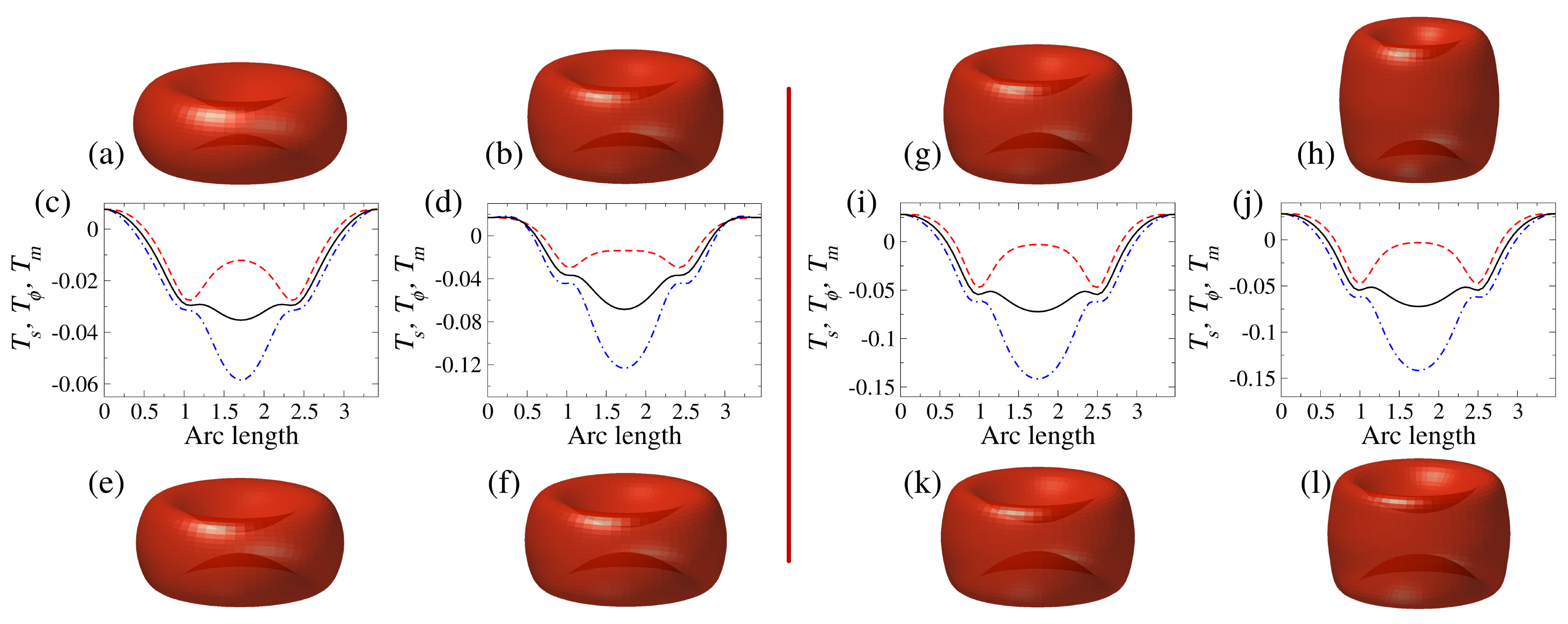}
\caption{Shapes observed at (a) $t=10$ and (b) $t=\infty$ during the deformation of a RBC at $Ca_e=0.5$ and at (g) $t=20$ and (h) $t=\infty$ at $Ca_e=1$ in AC field with $\omega=2.5$, considering $C=1$. Variation of meridional, $T_s$ (\textcolor{red}{$\pmb{\pmb{--}}$}), azimuthal, $T_\phi$ (\textcolor{blue}{$\pmb{\pmb{-\cdot-}}$}) and mean, $T_m$ (\textcolor{black}{$\pmb{\mi}$}) elastic tensions along the arc length for the corresponding cases are shown in (c, d) and (i, j), respectively. Shapes at (e) $t=24$ and (f) $t=\infty$ at $Ca_e=0.5$ and at (k) $t=20$ and (l) $t=\infty$ at $Ca_e=1$, considering $C=10$.}
\label{fig:rbc}
\end{figure}

At $Ca_e=0.5$, the small electric stresses at the interface lead to a small deformation of the RBC (see Fig.~\ref{fig:rbc}a, b). In case of RBC deformation, the transmembrane potential reaches an equilibrium value before the hydrodynamic calculation begins. Since $\omega<t_{cap}^{-1}$, the electric stress is purely compressive with the maxima at the equator (similar to the case of a biconcave-discoid capsule in DC electric field, shown in Fig.~\ref{fig:ca0p3}f). This compressive electric stress is responsible for the deformation of the RBC into a cylinder with concave ends. At larger capillary number ($Ca_e=1$), a substantial increase in the deformation is observed (see Fig.~\ref{fig:rbc}g, h). From Fig.~\ref{fig:rbc}c, d and~\ref{fig:rbc}i, j, it can be observed that the azimuthal, meridional and mean elastic tensions are compressive near the equator and tensile near the poles.  

\begin{figure}
\centering
\begin{minipage}{0.45\textwidth}
\centering
\includegraphics[width=1\textwidth]{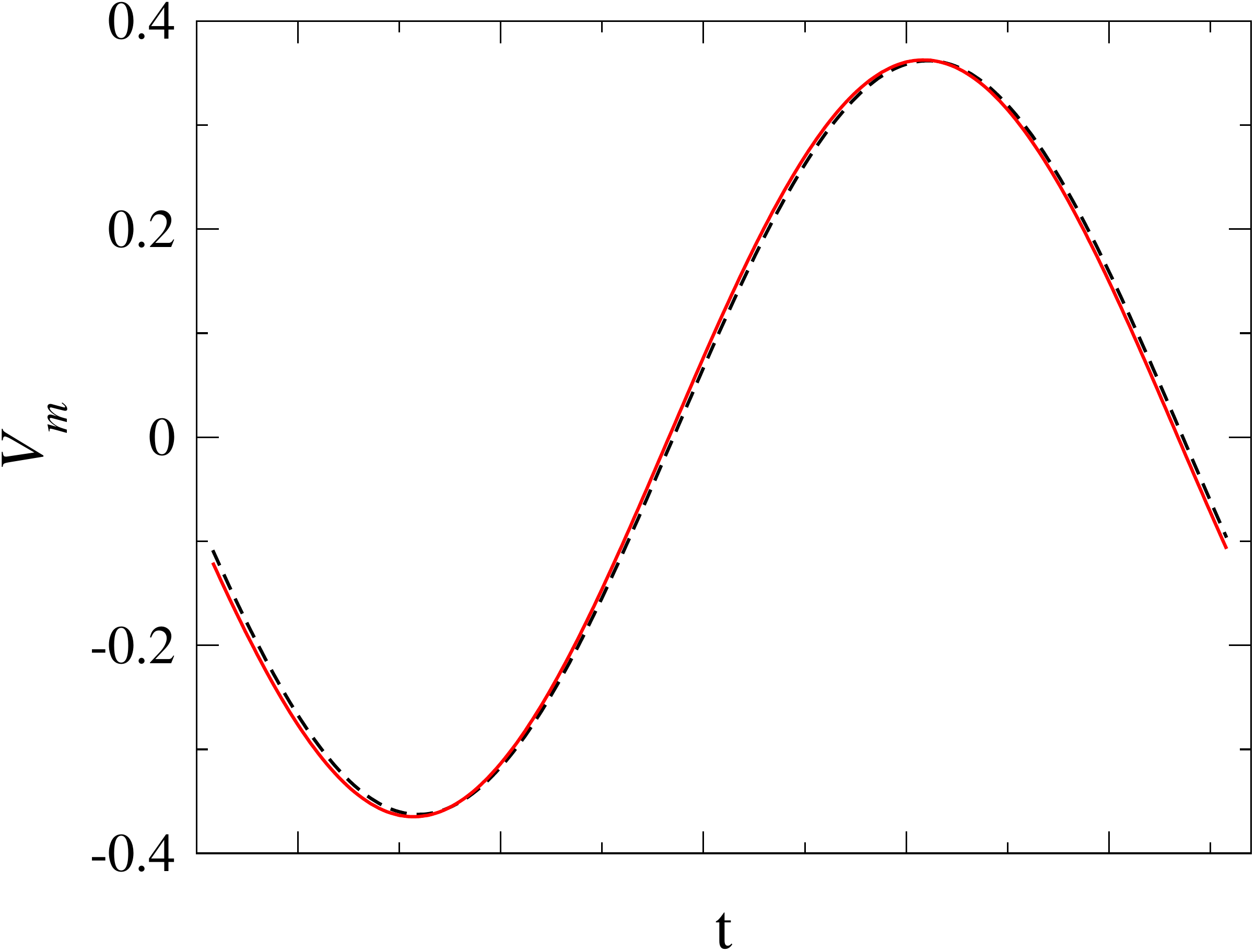}
(a)
\end{minipage}
\begin{minipage}{0.45\textwidth}
\centering
\includegraphics[width=1\textwidth]{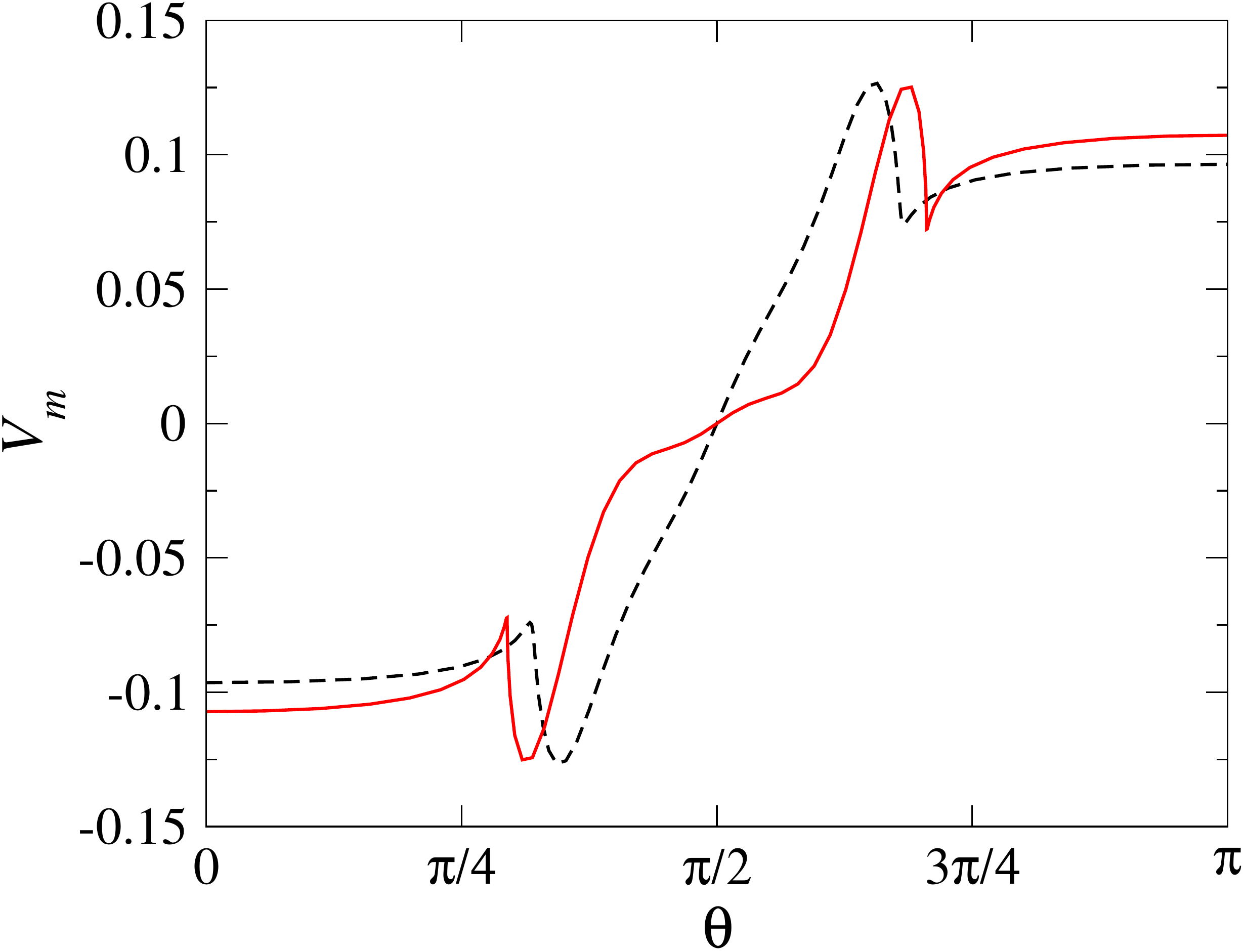}
(b)
\end{minipage}
 \caption{At $Ca=0.5$, (a) the transmembrane potential at the north pole in a single cycle of the applied electric field during the calculation of average electric stress at (\textcolor{black}{$\pmb{\pmb{--}}$}) $t=10$,  and (\textcolor{red}{$\pmb{\mi}$}) $t=\infty$, and (b) the variation of transmembrane potential at the end of the particular cycle of the applied field over the arc length at(\textcolor{black}{$\pmb{\pmb{--}}$}) $t=10$,  and (\textcolor{red}{$\pmb{\mi}$}) $t=\infty$, corresponding to Fig.~\ref{fig:rbc}a and b, respectively.}
 \label{fig:vmrbc}
\end{figure}
\begin{figure}
\centering
\begin{minipage}{0.45\textwidth}
\centering
\includegraphics[width=1\textwidth]{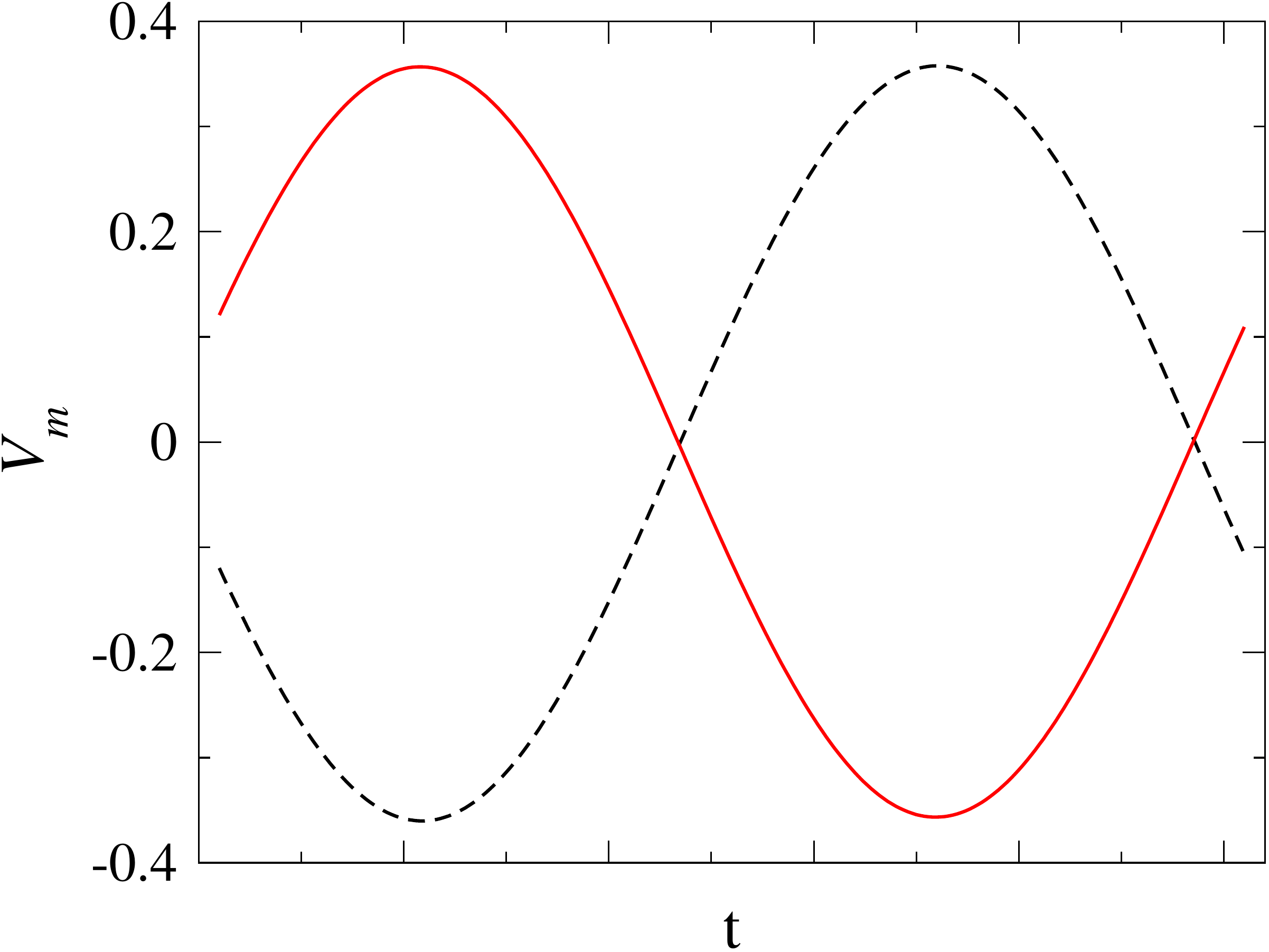}
(a)
\end{minipage}
\begin{minipage}{0.45\textwidth}
\centering
\includegraphics[width=1\textwidth]{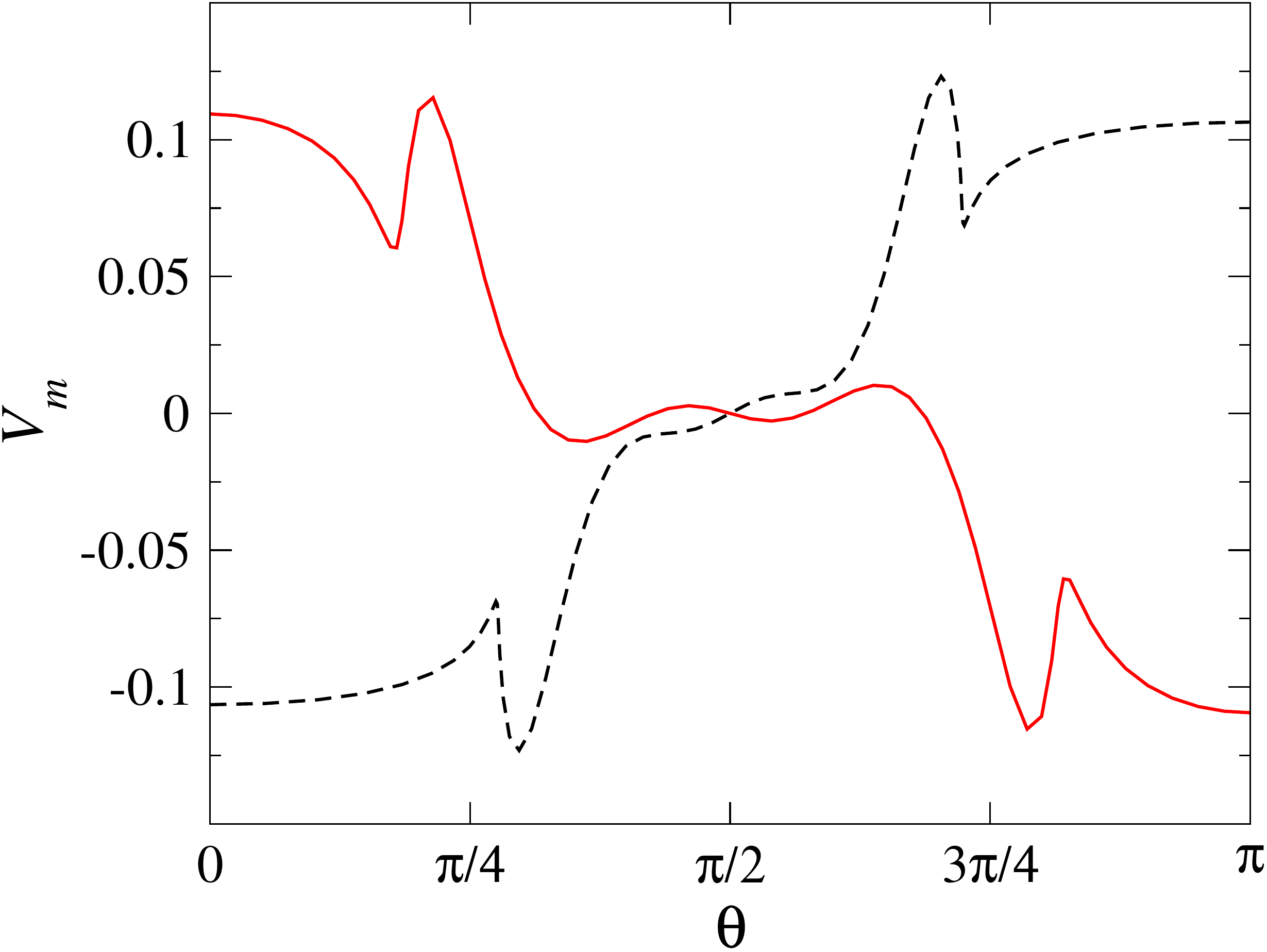}
(b)
\end{minipage}
 \caption{At $Ca=1$, (a) the transmembrane potential at the north pole in a single cycle of the applied electric field during the calculation of average electric stress at (\textcolor{black}{$\pmb{\pmb{--}}$}) $t=20$,  and (\textcolor{red}{$\pmb{\mi}$}) $t=\infty$, and (b) the variation of transmembrane potential at the end of the particular cycle of the applied field over the arc length at(\textcolor{black}{$\pmb{\pmb{--}}$}) $t=20$,  and (\textcolor{red}{$\pmb{\mi}$}) $t=\infty$, corresponding to Fig.~\ref{fig:rbc}g and h, respectively.}
 \label{fig:vmrbc1}
\end{figure}

The consideration of area dilatation parameter $C=1$ allows a substantial change in the area during the deformation of a RBC, i.e., $4.7\%$ and $13.8\%$ at $Ca_e=0.5$ and $Ca_e=1$, respectively. At a higher value of the area dilatation parameter ($C=10$), the constraint on the change in area ($0.82\%$ and $1.75\%$ at $Ca_e=0.5$ and $1$, respectively) restricts the deformation of a RBC. Considering $C=10$, the shapes observed during the deformation of a RBC at $Ca_e=0.5$ and $1$ are shown in Fig.~\ref{fig:rbc}e, f, and~\ref{fig:rbc}k, l, respectively. It can be easily visualized that the deformation of RBC considering $C=10$ is considerably less compared to the deformation with $C=1$.

To understand the mechanical integrity of the cell membrane to the applied electric field, the analysis of the transmembrane potential, during the deformation, is important. Fig.~\ref{fig:vmrbc}a shows that at $Ca=0.5$ the transmembrane potential of the membrane, at the north pole of a RBC during the calculation of the average electric stress in a cycle of the applied electric field, varies with time. Fig.~\ref{fig:vmrbc}b represents the variation of the transmembrane potential of the interface at the end of the cycle. It can be clearly observed from Fig.~\ref{fig:vmrbc}b that the shoulders attain the maximum transmembrane potential suggesting the possibility of electroporation in those regions. Similar behavior of the membrane is suggested in Fig.~\ref{fig:vmrbc1}a and b for the deformation of a RBC at $Ca=1$. 

 \begin{figure}
\centering
  \includegraphics[width=0.4\textwidth]{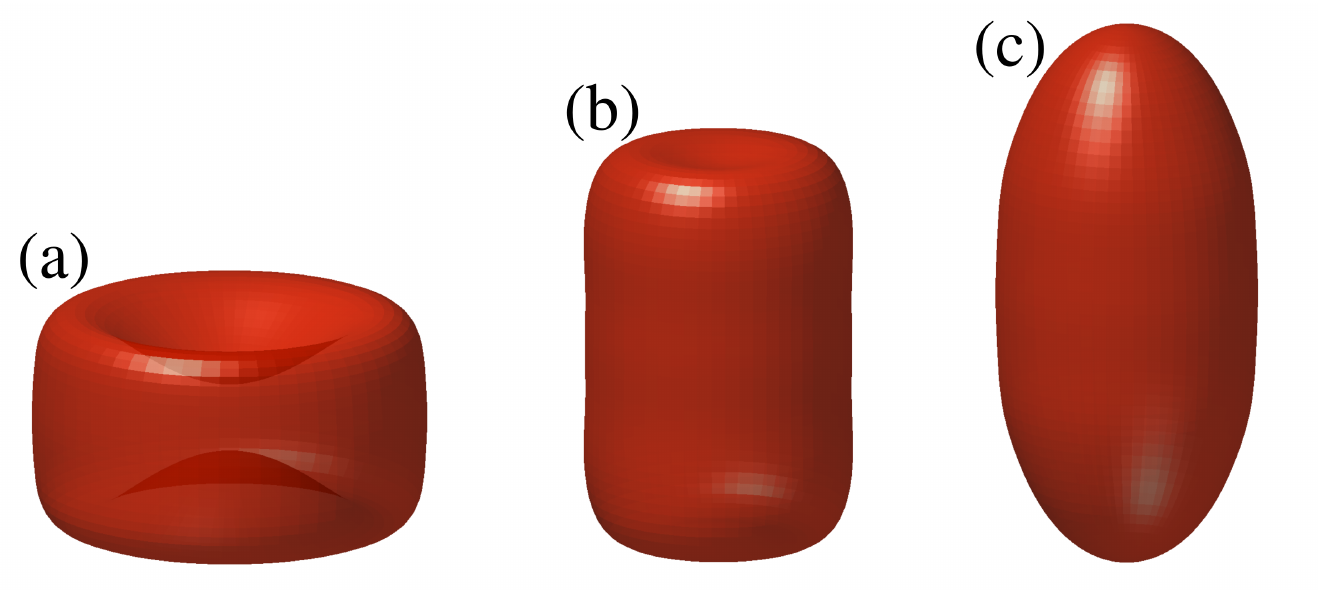}
\caption{Shapes observed at (a) $t=10$, (b) $t=70$ and (b) $t=\infty$ during the deformation of a RBC at $Ca_e=0.5$ in AC field with $\omega=2.5$, considering $C=1$ and $\sigma_r=10$.}
\label{fig:rbcl10}
\end{figure}

\begin{figure}
\centering
\begin{minipage}{0.45\textwidth}
\centering
\includegraphics[width=1\textwidth]{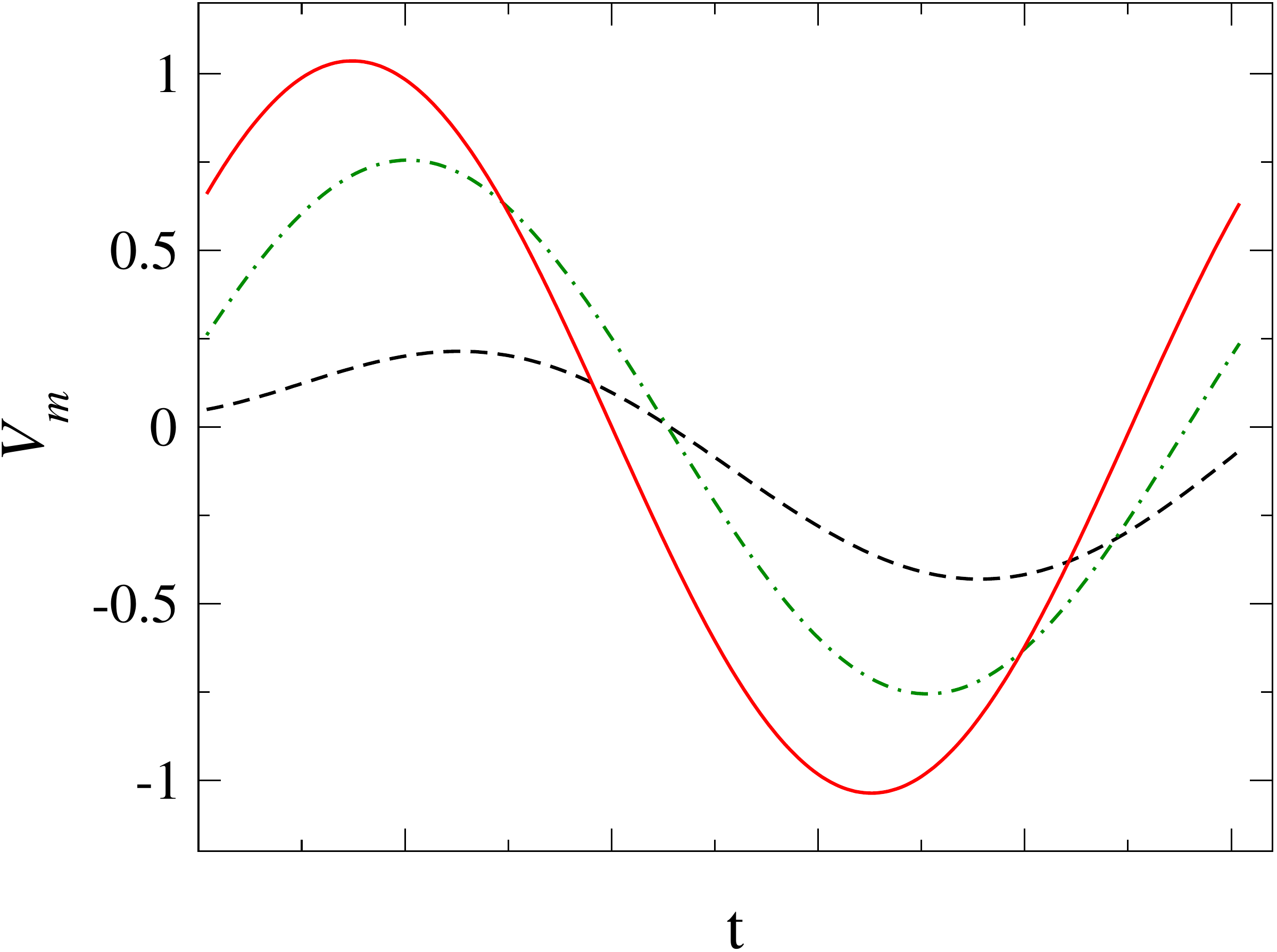}
(a)
\end{minipage}
\begin{minipage}{0.45\textwidth}
\centering
\includegraphics[width=1\textwidth]{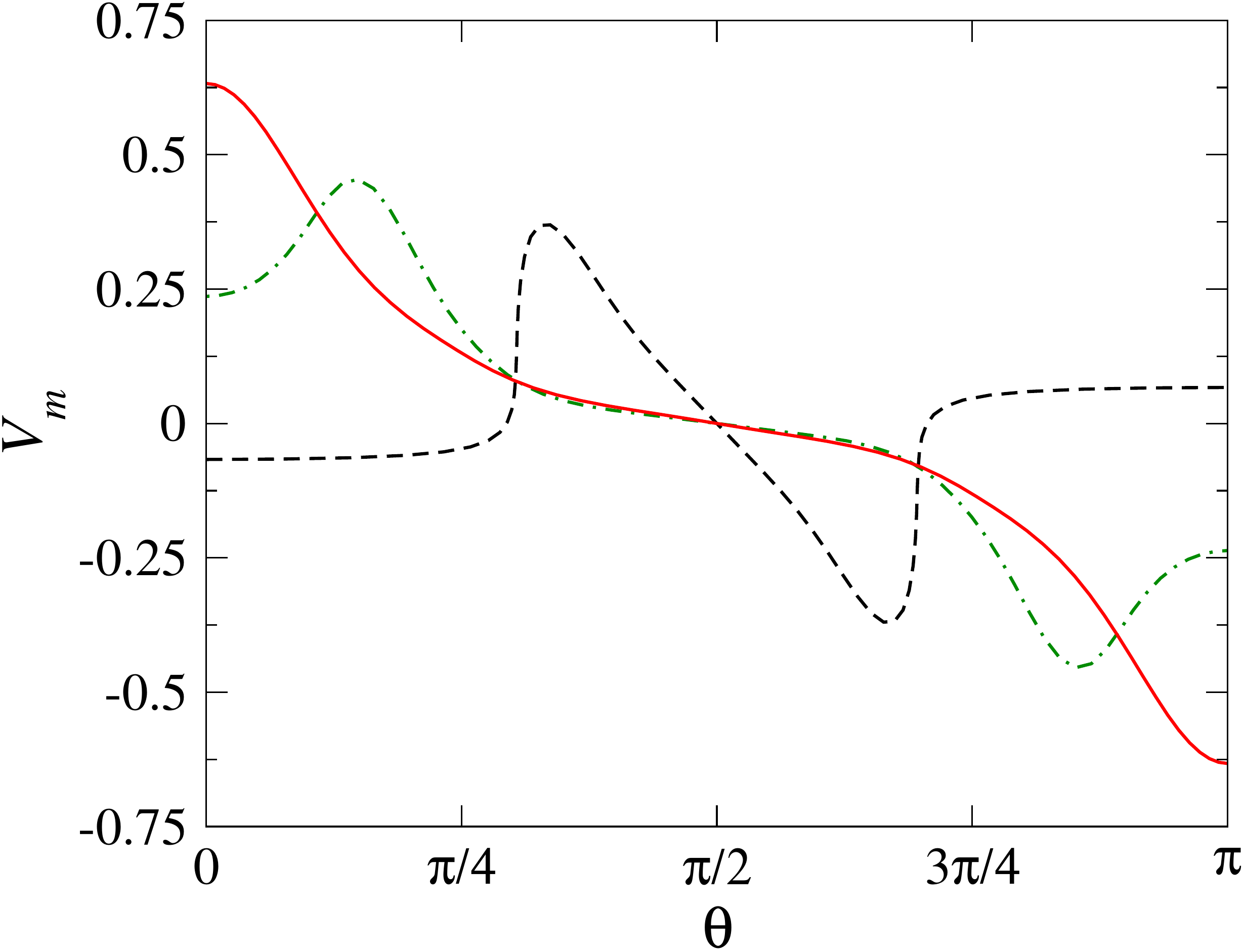}
(b)
\end{minipage}
 \caption{At $Ca=0.5$ and $\sigma_r=10$, (a) the transmembrane potential at the north pole in a single cycle of the applied electric field during the calculation of average electric stress at (\textcolor{black}{$\pmb{\pmb{--}}$}) $t=10$, (\textcolor{green}{$\pmb{\pmb{-\cdot-}}$}) $t=70$, and (\textcolor{red}{$\pmb{\mi}$}) $t=\infty$, and (b) the variation of transmembrane potential at the end of the particular cycle of the applied field over the arc length at(\textcolor{black}{$\pmb{\pmb{--}}$}) $t=10$, (\textcolor{green}{$\pmb{\pmb{-\cdot-}}$}) $t=70$,  and (\textcolor{red}{$\pmb{\mi}$}) $t=\infty$, corresponding to Fig.~\ref{fig:rbcl10}a-c, respectively.}
 \label{fig:vmrbc10}
\end{figure}

In an experimental condition with a low conducting external fluid medium (in this case, $\sigma_r=10$), a RBC undergoes large deformation even at a low capillary number ($Ca_e=0.5$). The strong compressive electric stress causes a RBC to deform through cylindrical shape with biconcave ends (see Fig.~\ref{fig:rbcl10}a) to a cylinder (see Fig.~\ref{fig:rbcl10}b), finally resulting in a prolate spheroid (see Fig.~\ref{fig:rbcl10}c). Unlike the case of a biconcave-discoid capsule (see Fig. 10), a RBC does not show the intermediate large deformation (as observed in Fig.~\ref{fig:ca0p5}d), as the partial charging of the membrane results in consistent tensile electric stress at the poles and compressive at the equator. In this case, since a small value of area dilatation parameter ($C=1$) is used, a significant change in the area is observed. From Fig.~\ref{fig:vmrbc10}a and b, it can be observed that for the intermediate cylinder with concave ends (Fig.~\ref{fig:rbcl10}a) and cylindrical shapes (Fig.~\ref{fig:rbcl10}b) the transmembrane potential is the highest at the shoulders. Whereas, when the RBC evolves into a prolate spheroid at the time-averaged steady-state, the transmembrane potential is maximum at the poles, thereby suggesting the possibility of electroporation at the poles. 

\section{Summary and Conclusions}
In an extensional flow, a RBC undergoes axisymmetric deformation while developing membrane tension. Recalling the dimensional variables for RBC from Table~\ref{tab:parametertable}, the estimated tension in the transition to prolate spheroid is  $\sim O(1) mN/m$ and the corresponding strain rate is $\sim O(10) s^{-1}$. Therefore, considering the suggested tolerance limit of membrane tension in the literature~\citep{evans76}, there is a possibility of a RBC undergoing biconcave-discoid to prolate spheroid transition without the rupture of the cell. In the case of an elastic capsule, with comparatively larger membrane elastic modulus, a high strain rate is required for the biconcave-discoid $\rightarrow$ prolate spheroid transition, resulting in the development of higher membrane tension. 

The electrohydrodynamic steady-state deformation of a spherical capsule is independent of the conductivity ratio~\citep{sudip17}, whereas for a biconcave-discoid capsule the steady-state deformation is strongly dependent on the conductivity ratio. The analysis showed that when $\sigma_r=0.1$, high $Ca_e$ can only lead to the breakup, whereas for $\sigma_r=10$, capsule attains a steady-state deformed shape. Further, from the analysis, it is found that for the kinematic factor ($k_f$, consequently, $t_H$) influences the dynamics of the capsule deformation. For different $k_f$, even though the steady-state electrohydrodynamic deformation is same, but the pathways to reach the steady-state are different. Also, the analysis on the deformation of a RBC in an electric field shows a weak response due to the typical values of the parameters in the physiological condition. In an experimental condition, a lower electrical conductivity of the external fluid media makes the RBC sensitive to the applied AC electric field.

A  biconcave-discoid capsule and RBC undergo axisymmetric deformation in extensional flow due to the developed pressure profile because of the complex geometry. On the other hand, the deformation in an applied uniform electric field is due to the developed Maxwell electric stress at the interface. In extensional flow, the maximum elastic tension develops at the shoulders of the biconcave-discoid shapes, and the tension is maximum at the poles when the biconcave-discoid evolves into a prolate spheroid. Therefore, the possible rupture of the membrane could be at the shoulders or the poles in extensional flow. In an electric field, a similar biconcave-discoid prolate transition is possible with the developed maximum elastic tension at the poles and the maximum transmembrane potential develops at the shoulders of the deformed shape, except for the prolate-spheroids. Therefore, as the elastic tension is very less for deformation of a RBC in AC electric field, the rupture of the membrane can take place through electroporation at the shoulders or at the poles depending upon the location with the maximum value of the transmembrane potential. The measure of maximum elastic tensions in both the cases will help to understand the mechanical stability of the capsule and thereby, help in improvement in designing a capsule for specific applications. 

 \section*{Acknowledgment}
Authors would like to acknowledge Department of Science and Technology, Govt. of India for financial support for this work.

\bibliographystyle{unsrtnat}
\bibliography{rbc}

\begin{thebibliography}{57}
\providecommand{\natexlab}[1]{#1}
\providecommand{\url}[1]{\texttt{#1}}
\expandafter\ifx\csname urlstyle\endcsname\relax
  \providecommand{\doi}[1]{doi: #1}\else
  \providecommand{\doi}{doi: \begingroup \urlstyle{rm}\Url}\fi

\bibitem[Henon et~al.(2014)Henon, Sheard, and Fouras]{henon14}
Yann Henon, Gregory~J. Sheard, and Andreas Fouras.
\newblock Erythrocyte deformation in a microfluidic cross-slot channel.
\newblock \emph{RSC Adv.}, 4:\penalty0 36079--36088, 2014.

\bibitem[Yaginuma et~al.(2013)Yaginuma, Oliveira, Lima, Ishikawa, and
  Yamaguchi]{yaginuma13}
T.~Yaginuma, M.~S.~N. Oliveira, R.~Lima, T.~Ishikawa, and T.~Yamaguchi.
\newblock Human red blood cell behavior under homogeneous extensional flow in a
  hyperbolic-shaped microchannel.
\newblock \emph{Biomicrofluidics}, 7\penalty0 (5):\penalty0 054110, 2013.

\bibitem[Lee et~al.(2009{\natexlab{a}})Lee, Yim, Ahn, and Lee]{Lee09}
Sung~S. Lee, Yoonjae Yim, Kyung~H. Ahn, and Seung~J. Lee.
\newblock Extensional flow-based assessment of red blood cell deformability
  using hyperbolic converging microchannel.
\newblock \emph{Biomedical Microdevices}, 11\penalty0 (5):\penalty0 1021--1027,
  2009{\natexlab{a}}.
\newblock ISSN 1572-8781.

\bibitem[Evans(1980)]{evans80}
E.~A. Evans.
\newblock Minimum energy analysis of membrane deformation applied to pipet
  aspiration and surface adhesion of red blood cells.
\newblock \emph{Biophysical Journal}, 30:\penalty0 265--284, 1980.

\bibitem[Mohandas and Evans(1994)]{mohandas94}
N.~Mohandas and E.~Evans.
\newblock Mechanical properties of the red cell membrane in relation to
  molecular structure and genetic defects.
\newblock \emph{Annual Review of Biophysics and Biomolecular Structure},
  23\penalty0 (1):\penalty0 787--818, 1994.

\bibitem[Pozrikidis(1990)]{poz90}
C.~Pozrikidis.
\newblock The axisymmetric deformation of a red blood cell in uniaxial
  straining stokes flow.
\newblock \emph{Journal of Fluid Mechanics}, 216:\penalty0 231--254, 1990.
\newblock ISSN 1469-7645.

\bibitem[Pozrikidis(2003{\natexlab{a}})]{pozmodel03}
C.~Pozrikidis.
\newblock \emph{Modeling and Simulation of Capsules and Biological Cells}.
\newblock Chapman \& HALL/CRC, 2003{\natexlab{a}}.

\bibitem[Kwak and Pozrikidis(2001)]{kwak01}
Sehoon Kwak and C.~Pozrikidis.
\newblock Effect of membrane bending stiffness on the axisymmetric deformation
  of capsules in uniaxial extensional flow.
\newblock \emph{Physics of Fluids}, 13\penalty0 (5):\penalty0 1234--1242, 2001.
\newblock \doi{10.1063/1.1352629}.

\bibitem[Kozlovskaya et~al.(2014)Kozlovskaya, Alexander, Wang, Kuncewicz, Liu,
  Godin, and Kharlampieva]{veronika14}
Veronika Kozlovskaya, Jenolyn~F. Alexander, Yun Wang, Thomas Kuncewicz, Xuewu
  Liu, Biana Godin, and Eugenia Kharlampieva.
\newblock Internalization of red blood cell-mimicking hydrogel capsules with
  ph-triggered shape responses.
\newblock \emph{ACS Nano}, 8\penalty0 (6):\penalty0 5725--5737, 2014.

\bibitem[She et~al.(2013)She, Li, Shan, Tong, and Gao]{shupeng13}
Shupeng She, Qinqin Li, Bowen Shan, Weijun Tong, and Changyou Gao.
\newblock Fabrication of red-blood-cell-like polyelectrolyte microcapsules and
  their deformation and recovery behavior through a microcapillary.
\newblock \emph{Advanced Materials}, 25\penalty0 (40):\penalty0 5814--5818, 10
  2013.
\newblock ISSN 1521-4095.

\bibitem[She et~al.(2014)She, Yu, Han, Tong, Mao, and Gao]{She14}
Shupeng She, Dahai Yu, Xu~Han, Weijun Tong, Zhengwei Mao, and Changyou Gao.
\newblock Fabrication of biconcave discoidal silica capsules and their uptake
  behavior by smooth muscle cells.
\newblock \emph{Journal of Colloid and Interface Science}, 426:\penalty0 124 --
  130, 2014.
\newblock ISSN 0021-9797.

\bibitem[Merkel et~al.(2011)Merkel, Jones, Herlihy, Kersey, Shields, Napier,
  Luft, Wu, Zamboni, Wang, Bear, and DeSimone]{Merkel11}
Timothy~J. Merkel, Stephen~W. Jones, Kevin~P. Herlihy, Farrell~R. Kersey,
  Adam~R. Shields, Mary Napier, J.~Christopher Luft, Huali Wu, William~C.
  Zamboni, Andrew~Z. Wang, James~E. Bear, and Joseph~M. DeSimone.
\newblock Using mechanobiological mimicry of red blood cells to extend
  circulation times of hydrogel microparticles.
\newblock \emph{Proceedings of the National Academy of Sciences}, 108\penalty0
  (2):\penalty0 586--591, 2011.

\bibitem[Champion et~al.(2007)Champion, Katare, and Mitragotri]{Champion07}
Julie~A. Champion, Yogesh~K. Katare, and Samir Mitragotri.
\newblock Particle shape: A new design parameter for micro- and nanoscale drug
  delivery carriers.
\newblock \emph{Journal of Controlled Release}, 121\penalty0 (1–2):\penalty0
  3 -- 9, 2007.
\newblock ISSN 0168-3659.

\bibitem[Venkataraman et~al.(2011)Venkataraman, Hedrick, Ong, Yang, Ee,
  Hammond, and Yang]{Venkataraman11}
Shrinivas Venkataraman, James~L. Hedrick, Zhan~Yuin Ong, Chuan Yang, Pui
  Lai~Rachel Ee, Paula~T. Hammond, and Yi~Yan Yang.
\newblock The effects of polymeric nanostructure shape on drug delivery.
\newblock \emph{Advanced Drug Delivery Reviews}, 63\penalty0
  (14–15):\penalty0 1228 -- 1246, 2011.
\newblock ISSN 0169-409X.

\bibitem[Lee et~al.(2009{\natexlab{b}})Lee, Ferrari, and Decuzzi]{sei09}
Sei-Young Lee, Mauro Ferrari, and Paolo Decuzzi.
\newblock Shaping nano-/micro-particles for enhanced vascular interaction in
  laminar flows.
\newblock \emph{Nanotechnology}, 20\penalty0 (49):\penalty0 495101,
  2009{\natexlab{b}}.

\bibitem[Decuzzi et~al.(2008)Decuzzi, Pasqualini, Arap, and Ferrari]{Decuzzi08}
Paolo Decuzzi, Renata Pasqualini, Wadih Arap, and Mauro Ferrari.
\newblock Intravascular delivery of particulate systems: Does geometry really
  matter?
\newblock \emph{Pharmaceutical Research}, 26\penalty0 (1):\penalty0 235--243,
  2008.
\newblock ISSN 1573-904X.

\bibitem[Evans et~al.(1976)Evans, Waugh, and Melnik]{evans76}
E.~A. Evans, R.~Waugh, and L.~Melnik.
\newblock Elastic area compressibility modulus of red cell membrane.
\newblock \emph{Biophysical Journal}, 16\penalty0 (2):\penalty0 585--595, 1976.

\bibitem[Yen et~al.(2015)Yen, Chen, Chern, and Lu]{yen15}
Jen-Hong Yen, Sheng-Fu Chen, Ming-Kai Chern, and Po-Chien Lu.
\newblock The effects of extensional stress on red blood cell hemolysis.
\newblock \emph{Biomedical Engineering: Applications, Basis and
  Communications}, 27\penalty0 (05):\penalty0 1550042, 2015.

\bibitem[Chang et~al.(1985)Chang, Takashima, and Asakura]{Chang1985}
S~Chang, S~Takashima, and T.~Asakura.
\newblock Volume and shape changes of human erythrocytes induced by electrical
  fields.
\newblock \emph{J. Bioelectr.}, 4\penalty0 (2):\penalty0 301--316, jan 1985.
\newblock ISSN 1536-8386.

\bibitem[Friend et~al.(1975)Friend, Finch, and Schwan]{Friend75}
A~Friend, E~Finch, and H~Schwan.
\newblock Low frequency electric field induced changes in the shape and
  motility of amoebas.
\newblock \emph{Science}, 187:\penalty0 357--359, jan 1975.
\newblock ISSN 1095-9203.

\bibitem[Cruz and Garc\'{\i}a-Diego(1998)]{cruz98}
J~M Cruz and F~J Garc\'{\i}a-Diego.
\newblock Dielectrophoretic motion of oblate spheroidal particles. measurements
  of motion of red blood cells using the stokes method.
\newblock \emph{Journal of Physics D: Applied Physics}, 31\penalty0
  (14):\penalty0 1745, 1998.

\bibitem[Scheurich et~al.(1980)Scheurich, Zimmermann, Mischel, and
  Lamprecht]{Scheurich1980}
P~Scheurich, U~Zimmermann, M~Mischel, and I~Lamprecht.
\newblock Membrane fusion and deformation of red blood cells by electric
  fields.
\newblock \emph{Z. Naturforsch. C.}, 35\penalty0 (11-12):\penalty0 1081--1085,
  1980.

\bibitem[Engelhardt and Sackmann(1988)]{sackmann88}
H.~Engelhardt and E.~Sackmann.
\newblock On the measurement of shear elastic moduli and viscosities of
  erythrocyte plasma membranes by transient deformation in high frequency
  electric fields.
\newblock \emph{Biophysical Journal}, 54\penalty0 (3):\penalty0 495 -- 508,
  1988.
\newblock ISSN 0006-3495.

\bibitem[Cordasco and Bagchi(2017)]{bagchi17}
Daniel Cordasco and Prosenjit Bagchi.
\newblock On the shape memory of red blood cells.
\newblock \emph{Physics of Fluids}, 29\penalty0 (4):\penalty0 041901, 2017.
\newblock \doi{10.1063/1.4979271}.

\bibitem[Gass et~al.(1991)Gass, Chernomordik, and Margolis]{gass91}
G.V. Gass, L.V. Chernomordik, and L.B. Margolis.
\newblock Local deformation of human red blood cells in high frequency electric
  field.
\newblock \emph{Biochimica et Biophysica Acta (BBA) - Molecular Cell Research},
  1093\penalty0 (2):\penalty0 162 -- 167, 1991.
\newblock ISSN 0167-4889.

\bibitem[Krueger and Thom(1997)]{krueger97}
M.~Krueger and F.~Thom.
\newblock Deformability and stability of erythrocytes in high-frequency
  electric fields down to subzero temperatures.
\newblock \emph{Biophysical Journal}, 73\penalty0 (5):\penalty0 2653 -- 2666,
  1997.
\newblock ISSN 0006-3495.

\bibitem[Sukhorukov et~al.(1998)Sukhorukov, Mussauer, and
  Zimmermann]{Sukhorukov98}
V.L. Sukhorukov, H.~Mussauer, and U.~Zimmermann.
\newblock The effect of electrical deformation forces on the
  electropermeabilization of erythrocyte membranes in low- and
  high-conductivity media.
\newblock \emph{The Journal of Membrane Biology}, 163\penalty0 (3):\penalty0
  235--245, 1998.
\newblock ISSN 1432-1424.

\bibitem[Kononenko and Shimkus(2000)]{Kononenko00}
V.L Kononenko and J.K Shimkus.
\newblock Stationary deformations of erythrocytes by high-frequency electric
  field.
\newblock \emph{Bioelectrochemistry}, 52\penalty0 (2):\penalty0 187 -- 196,
  2000.
\newblock ISSN 1567-5394.

\bibitem[Kononenko and Shimkus(2002)]{kononenko02a}
V.L Kononenko and J.K Shimkus.
\newblock Transient dielectro-deformations of erythrocyte governed by time
  variation of cell ionic state.
\newblock \emph{Bioelectrochemistry}, 55\penalty0 (1):\penalty0 97 -- 100,
  2002.
\newblock ISSN 1567-5394.
\newblock \doi{https://doi.org/10.1016/S1567-5394(01)00129-3}.

\bibitem[Sebasti\'{a}n et~al.(2006)Sebasti\'{a}n, Mu\={n}oz, Sancho, and
  Miranda]{sebastian06}
J~L Sebasti\'{a}n, S~Mu\={n}oz, M~Sancho, and J~M Miranda.
\newblock Analysis of the electric field induced forces in erythrocyte membrane
  pores using a realistic cell model.
\newblock \emph{Physics in Medicine $\&$ Biology}, 51\penalty0 (23):\penalty0
  6213, 2006.

\bibitem[Kononenko(2002)]{kononeko02}
Vadim~L. Kononenko.
\newblock Dielectro-deformations and flicker of erythrocytes: fundamental
  aspects of medical diagnostics applications.
\newblock \emph{Proc. SPIE}, 4707:\penalty0 134--143, 2002.

\bibitem[Thom and Gollek(2006)]{Thom06}
F.~Thom and H.~Gollek.
\newblock Calculation of mechanical properties of human red cells based on
  electrically induced deformation experiments.
\newblock \emph{Journal of Electrostatics}, 64\penalty0 (1):\penalty0 53 -- 61,
  2006.
\newblock ISSN 0304-3886.

\bibitem[Thom(2009)]{Thom09}
Fritz Thom.
\newblock Mechanical properties of the human red blood cell membrane at −15
  °c.
\newblock \emph{Cryobiology}, 59\penalty0 (1):\penalty0 24 -- 27, 2009.
\newblock ISSN 0011-2240.

\bibitem[Du et~al.(2014)Du, Dao, and Suresh]{Du14}
E~Du, Ming Dao, and Subra Suresh.
\newblock Quantitative biomechanics of healthy and diseased human red blood
  cells using dielectrophoresis in a microfluidic system.
\newblock \emph{Extreme Mechanics Letters}, 1:\penalty0 35 -- 41, 2014.
\newblock ISSN 2352-4316.

\bibitem[Ur and Lushbaugh(1968)]{amiram68}
Amiram Ur and C.~C. Lushbaugh.
\newblock Some effects of electrical fields on red blood cells with remarks on
  electronic red cell sizing.
\newblock \emph{British Journal of Haematology}, 15\penalty0 (6):\penalty0
  527--538, 1968.
\newblock ISSN 1365-2141.

\bibitem[Ashe et~al.(1988)Ashe, Bogen, and Takashima]{Ashe1988}
JW~Ashe, DK~Bogen, and S~Takashima.
\newblock Deformation of biological cells by electric fields: Theoretical
  prediction of the deformed shape.
\newblock \emph{Ferroelectrics}, 86:\penalty0 311--324, 1988.

\bibitem[Joshi et~al.(2002)Joshi, Hu, Schoenbach, and Beebe]{Joshi2002}
R~P Joshi, Qin Hu, K~H Schoenbach, and S~J Beebe.
\newblock Simulations of electroporation dynamics and shape deformations in
  biological cells subjected to high voltage pulses.
\newblock \emph{IEEE Transsactions Plasma Sci.}, 30\penalty0 (4):\penalty0
  1536--1546, aug 2002.
\newblock ISSN 0093-3813.

\bibitem[Evans and Fung(1972)]{evans72}
Evan Evans and Yuan-Cheng Fung.
\newblock Improved measurements of the erythrocyte geometry.
\newblock \emph{Microvascular Research}, 4\penalty0 (4):\penalty0 335 -- 347,
  1972.
\newblock ISSN 0026-2862.

\bibitem[Helfrich and Deuling(1974)]{Helfrich1974}
W~Helfrich and H~J Deuling.
\newblock Some theoretical shapes of red blood cells.
\newblock pages 327--329, 1974.

\bibitem[Zhou and Pozrikidis(1995)]{zhou95}
H.~Zhou and C.~Pozrikidis.
\newblock Deformation of liquid capsules with incompressible interfaces in
  simple shear flow.
\newblock \emph{Journal of Fluid Mechanics}, 283:\penalty0 175--200, 001 1995.
\newblock \doi{10.1017/S0022112095002278}.

\bibitem[Skalak et~al.(1973)Skalak, Tozeren, Zarda, and Chien]{skalak73}
R.~Skalak, A.~Tozeren, R.P. Zarda, and S.~Chien.
\newblock Strain energy function of red blood cell membranes.
\newblock \emph{Biophysical Journal}, 13\penalty0 (3):\penalty0 245 -- 264,
  1973.
\newblock ISSN 0006-3495.

\bibitem[Barth\`{e}s-Biesel et~al.(2002)Barth\`{e}s-Biesel, Diaz, and
  Dhenin]{barthesbiesel02}
D.~Barth\`{e}s-Biesel, A.~Diaz, and E.~Dhenin.
\newblock Effect of constitutive laws for two-dimensional membranes on
  flow-induced capsule deformation.
\newblock \emph{Journal of Fluid Mechanics}, 460:\penalty0 211--222, 2002.

\bibitem[Vlahovska et~al.(2009)Vlahovska, Podgorski, and Misbah]{vlahovska09}
Petia~M. Vlahovska, Thomas Podgorski, and Chaouqi Misbah.
\newblock Vesicles and red blood cells in flow: From individual dynamics to
  rheology.
\newblock \emph{Comptes Rendus Physique}, 10\penalty0 (8):\penalty0 775 -- 789,
  2009.
\newblock ISSN 1631-0705.
\newblock \doi{https://doi.org/10.1016/j.crhy.2009.10.001}.

\bibitem[Yazdani et~al.(2011)Yazdani, Kalluri, and Bagchi]{bagchi11}
Alireza Z~K Yazdani, R~Murthy Kalluri, and Prosenjit Bagchi.
\newblock Tank-treading and tumbling frequencies of capsules and red blood
  cells.
\newblock \emph{Physical review. E}, 83\penalty0 (4 Pt 2):\penalty0 046305,
  April 2011.
\newblock ISSN 1539-3755.
\newblock \doi{10.1103/physreve.83.046305}.

\bibitem[Hu et~al.(2014)Hu, Kim, and Lai]{hu14}
W.~F. Hu, Y.~Kim, and M.-C. Lai.
\newblock An immersed boundary method for simulating the dynamics of
  three-dimensional axisymmetric vesicles in navier–stokes flows.
\newblock \emph{Journal of Computational Physics}, 257, Part A:\penalty0 670 --
  686, 2014.
\newblock ISSN 0021-9991.

\bibitem[Trefethen(1996)]{trefethen94}
L.~N. Trefethen.
\newblock \emph{Finite difference and spectral methods for ordinary and partial
  differential equations}.
\newblock Unpublished text:
  http://web.comlab.ox.ac.uk/oucl/work/nick.trefethen/pdetext.html, 1996.

\bibitem[Grosse and Schwan(1992)]{grosse92}
C.~Grosse and H.~P. Schwan.
\newblock Cellular membrane potentials induced by alternating fields.
\newblock \emph{Biophysical Journal}, 63\penalty0 (6):\penalty0 1632 -- 1642,
  1992.
\newblock ISSN 0006-3495.

\bibitem[McConnell et~al.(2015)McConnell, Vlahovska, and Miksis]{mcconnell15sm}
Lane~C. McConnell, Petia~M. Vlahovska, and Michael~J. Miksis.
\newblock Vesicle dynamics in uniform electric fields: squaring and breathing.
\newblock \emph{Soft Matter}, 11:\penalty0 4840--4846, 2015.
\newblock \doi{10.1039/C5SM00585J}.

\bibitem[Rallison and Acrivos(1978)]{rallison78}
J.~M. Rallison and A.~Acrivos.
\newblock A numerical study of the deformation and burst of a viscous drop in
  an extensional flow.
\newblock \emph{Journal of Fluid Mechanics}, 89:\penalty0 191--200, 11 1978.
\newblock ISSN 1469-7645.

\bibitem[Pozrikidis(1992)]{pozrikidis92}
C.~Pozrikidis.
\newblock \emph{Boundary integral and singularity methods for linearized
  viscous flow}.
\newblock Cambridge university press, New York, 1992.

\bibitem[Lac et~al.(2007)Lac, Morel, and Barth\`{e}s-Biesel]{lac07}
Etienne Lac, Arnaud Morel, and Dominique Barth\`{e}s-Biesel.
\newblock Hydrodynamic interaction between two identical capsules in simple
  shear flow.
\newblock \emph{Journal of Fluid Mechanics}, 573:\penalty0 149--169, 2007.
\newblock \doi{10.1017/S0022112006003739}.

\bibitem[Beving et~al.(1994)Beving, Eriksson, Davey, and Kell]{beving1994}
H.~Beving, L.~E.~G. Eriksson, C.~L. Davey, and D.~B. Kell.
\newblock Dielectric properties of human blood and erythrocytes at radio
  frequencies (0.2--10 mhz); dependence on cell volume fraction and medium
  composition.
\newblock \emph{European Biophysics Journal}, 23\penalty0 (3):\penalty0
  207--215, 1994.
\newblock ISSN 1432-1017.

\bibitem[Haidekker et~al.(2002)Haidekker, Tsai, Brady, Stevens, Frangos,
  Theodorakis, and Intaglietta]{haidekker02}
Mark~A. Haidekker, Amy~G. Tsai, Thomas Brady, Hazel~Y. Stevens, John~A.
  Frangos, Emmanuel Theodorakis, and Marcos Intaglietta.
\newblock A novel approach to blood plasma viscosity measurement using
  fluorescent molecular rotors.
\newblock \emph{American Journal of Physiology - Heart and Circulatory
  Physiology}, 282\penalty0 (5):\penalty0 H1609--H1614, 2002.
\newblock ISSN 0363-6135.
\newblock \doi{10.1152/ajpheart.00712.2001}.

\bibitem[Pozrikidis(2003{\natexlab{b}})]{poz03}
C.~Pozrikidis.
\newblock Numerical simulation of the flow-induced deformation of red blood
  cells.
\newblock \emph{Annals of Biomedical Engineering}, 31\penalty0 (10):\penalty0
  1194--1205, 2003{\natexlab{b}}.
\newblock ISSN 1573-9686.

\bibitem[Wolf et~al.(2011)Wolf, Gulich, Lunkenheimer, and Loidl]{wolf11}
M.~Wolf, R.~Gulich, P.~Lunkenheimer, and A.~Loidl.
\newblock Broadband dielectric spectroscopy on human blood.
\newblock \emph{Biochimica et Biophysica Acta (BBA) - General Subjects},
  1810\penalty0 (8):\penalty0 727 -- 740, 2011.
\newblock ISSN 0304-4165.

\bibitem[Joshi and Hu(2012)]{joshi12}
R.~P. Joshi and Q.~Hu.
\newblock Role of electropores on membrane blebbing—a model energy-based
  analysis.
\newblock \emph{Journal of Applied Physics}, 112\penalty0 (6):\penalty0 064703,
  2012.
\newblock \doi{10.1063/1.4754568}.

\bibitem[Das and Thaokar(2017)]{sudip17}
Sudip Das and Rochish Thaokar.
\newblock Large deformation electrohydrodynamics of an elastic capsule in dc
  electric field.
\newblock \emph{ArXiv e-prints}, Aug. 2017.
\newblock https://arxiv.org/abs/1708.08802.

\end{thebibliography}

\newpage
\appendix

\section{Supplementary informations}\label{sec:validation}
\section*{Validation: deformation of oblate spheroids in extensional flow}
The developed numerical code to study the deformation of a biconcave-discoid capsule and RBC is validated with the results reported by~\citet{poz90} for the analysis of its deformation in a uniaxial extensional flow. In both ours as well as~\citet{poz90}'s  analyses, the axisymmetric boundary integral method is used. The stress-free shapes of oblate spheroids are considered as the initial shape of the capsule which is perturbed from the sphere with a second degree Legendre polynomial, and the equivalent volume is maintained as unity as assumed by~\citet{poz90}. To define oblate spheroid shapes, the coefficients of Legendre polynomial are considered, these are $\epsilon=-0.3$ and $\epsilon=-0.5$ for two different cases. Similar scaling of variables as reported by~\citet{poz90} are used in this validation of the numerical boundary integral code.  

The neo-Hookean membrane constitutive equation is a special case of the Mooney-Rivlin constitutive equations and is applicable for isotropic membranes. \citet{poz90} considered a neo-Hookean membrane constitutive law to describe a RBC, where the tension components (see Fig.~\ref{fig:schematic2}d) in the meridional and azimuthal directions are given by
\begin{equation}\label{eq:neo}
 \tilde{T}_{s,\phi}^{MR}=\frac{G_{MR}}{\lambda_s\lambda_\phi}\left(\lambda_s^2-\frac{1}{(\lambda_s\lambda_\phi)^2}\right),
\end{equation}
where $G_{MR}$ is the shear modulus in the Mooney-Rivlin model.
To assert a negligible change in area,~\citet{poz90} imposed the incompressibility condition 
\begin{equation}\label{areainc}
 {\tilde{\bf u}}\cdot {\bf j}+\tilde{\sigma} {\bf t}\cdot\frac{\partial{\tilde{\bf u}}}{\partial s}=0,
\end{equation}
where the notations used by~\citet{poz90} are, $\tilde{\sigma}$ is the radial distance in the cylindrical coordinate system,  ${\tilde{\bf u}}$ is the interfacial velocity, ${\bf j}$ is the local surface unit normal, ${\bf t}$ is the local tangent vector, and $s$ is the arc length. 

In the reported analysis~\citep{poz90}, the elastic modulus is replaced with the nondimensional variable $k=G_{MR}/{\mu e a}$. In our numerical computations, we have used the relation between Skalak modulus ($G_{SK}$)  and the surface Young modulus ($E_s$), given by
\begin{equation}
E_s=2G_{SK}\frac{1+2C}{1+C}, 
\end{equation}
where $C$ is the membrane dilatation parameter~\citep{barthesbiesel02}. For the Skalak model, the nondimensional elastic modulus is $k_s=E_s/{\mu e a}$. Therefore, the nondimensional elastic modulus considered in this analysis is thrice of the nondimensional elasticity considered by~\citet{poz90}, i.e., $k_s=3\times k$. Moreover, the area conservation is not explicitly implemented in the present work, since the parameter $C$  can be independently used to enforce area conservation. 

\begin{figure}
\centering
\begin{minipage}{0.35\textwidth}
\centering
\includegraphics[width=1\textwidth]{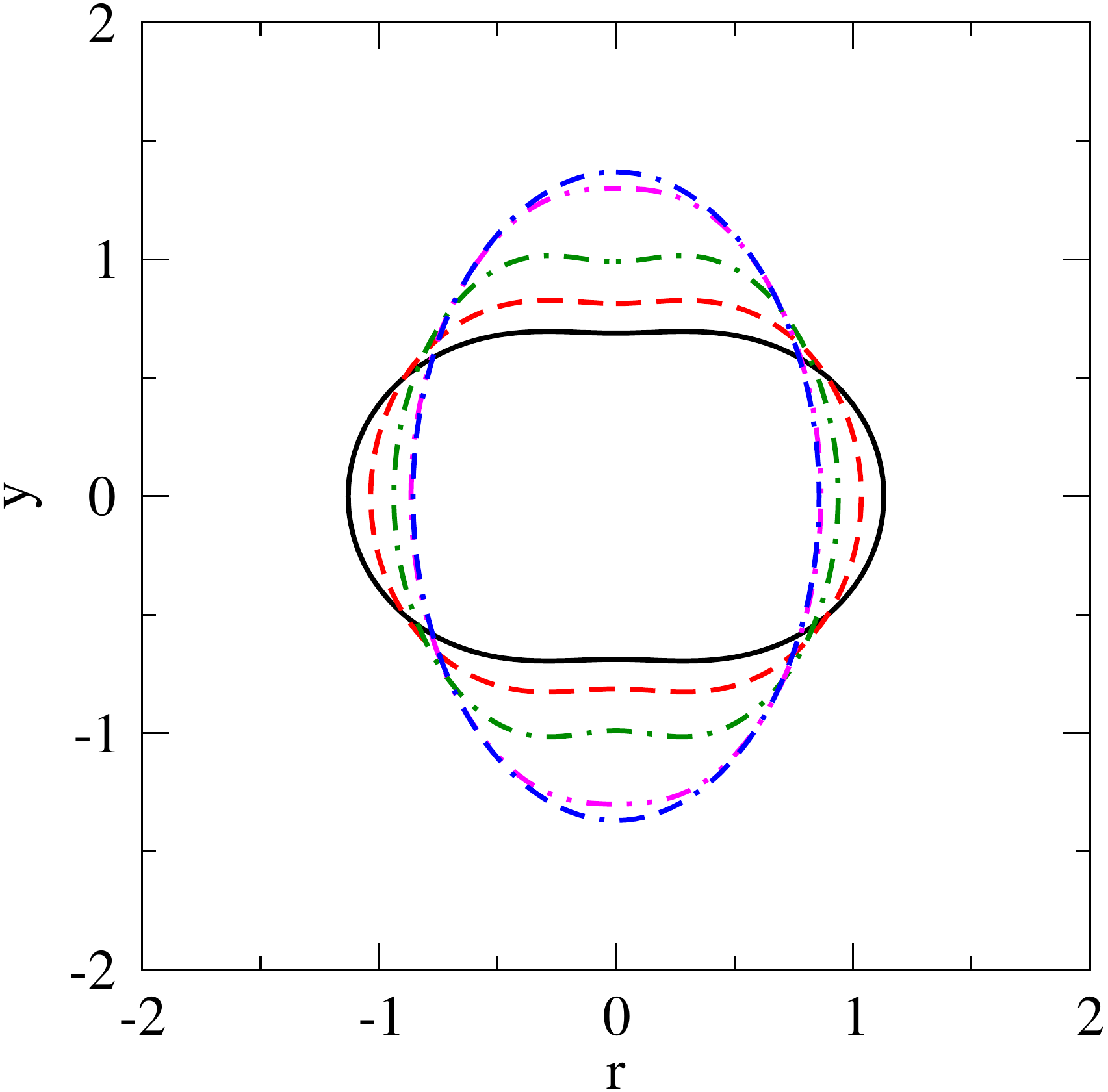}
(a) $C=1$
\end{minipage}
\begin{minipage}{0.35\textwidth}
\centering
\includegraphics[width=1\textwidth]{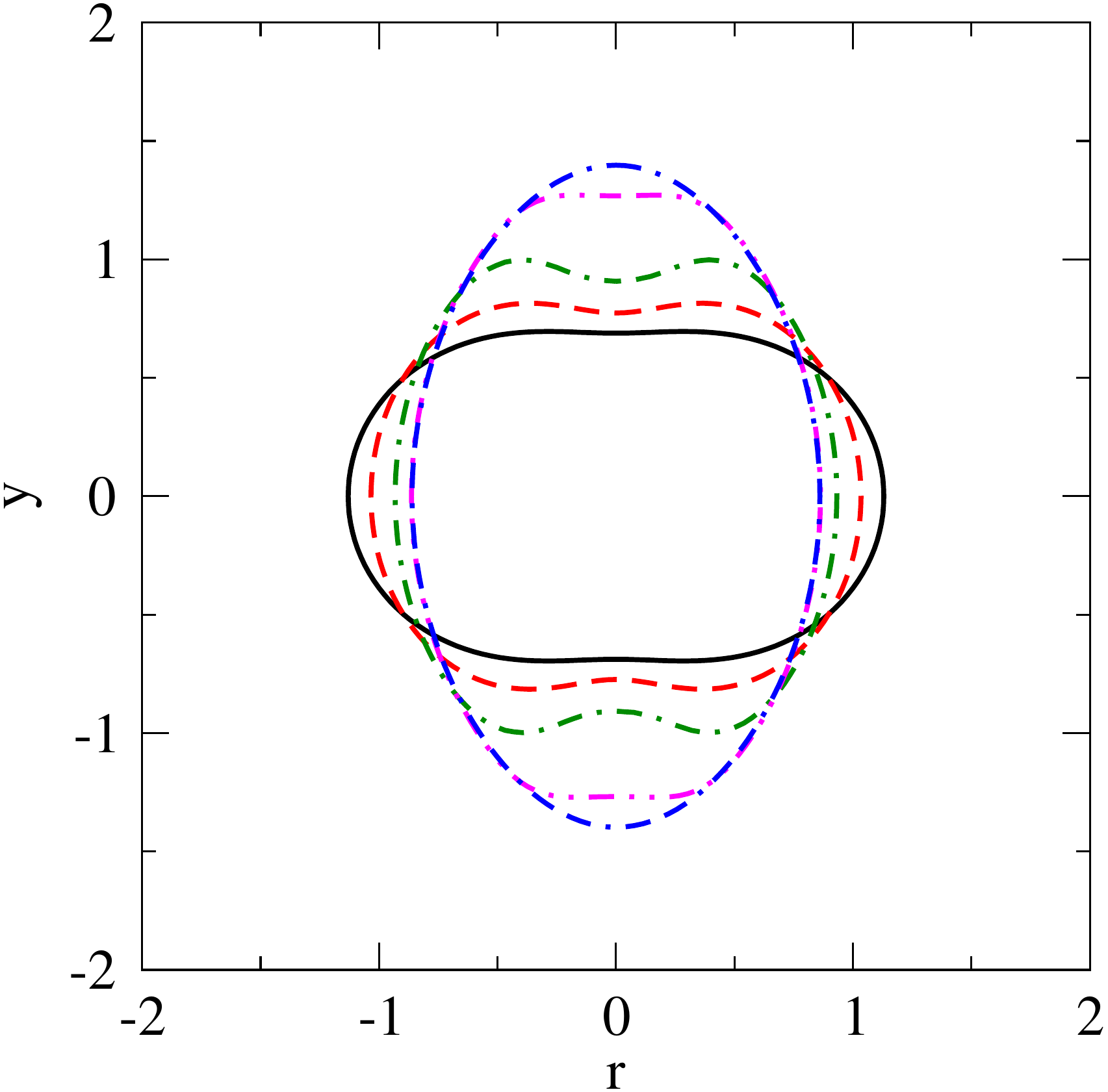}
(b) $C=10$
\end{minipage}
\begin{minipage}{0.35\textwidth}
\centering
\includegraphics[width=1\textwidth]{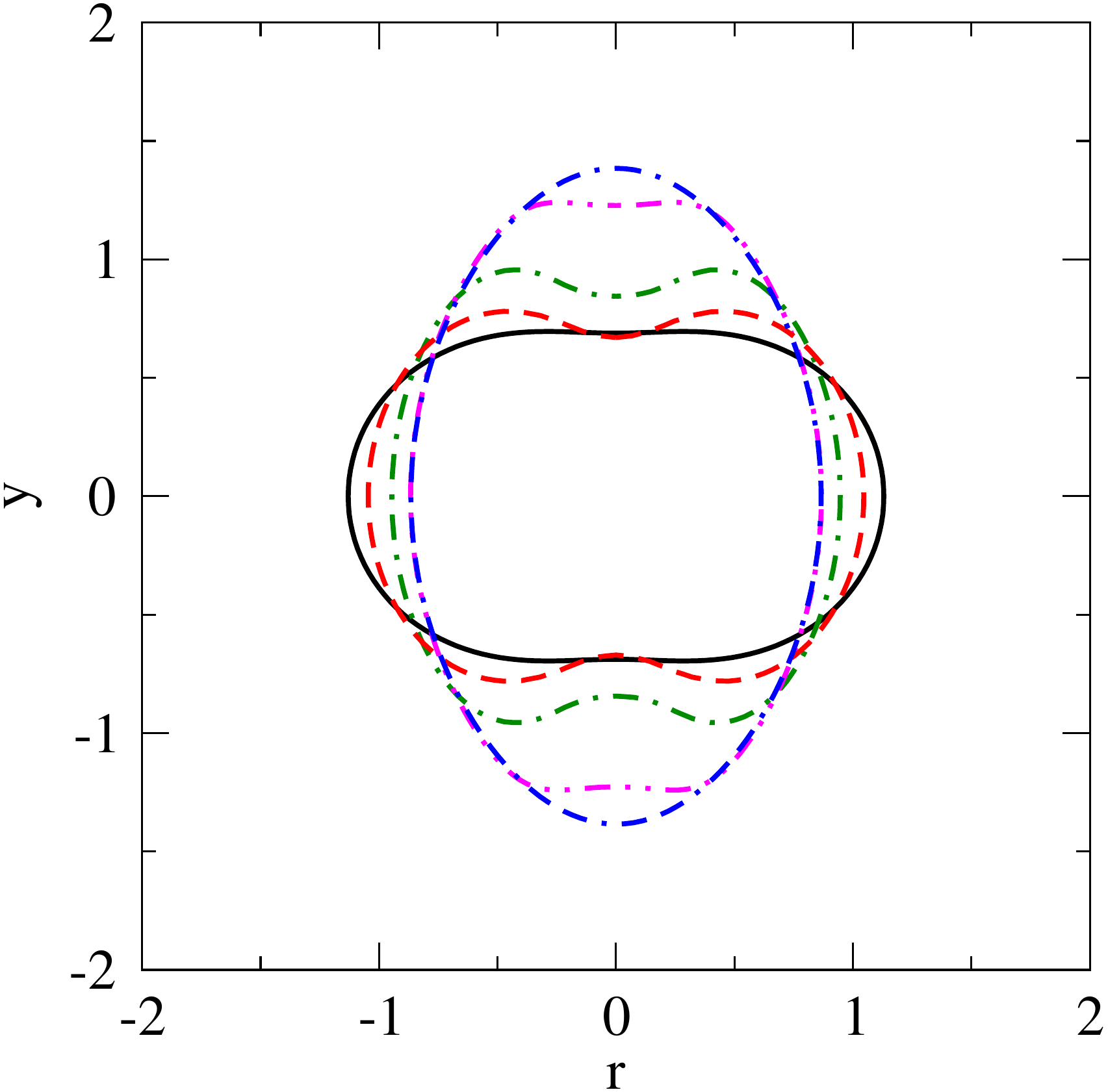}
(c) $C=50$
\end{minipage}
\begin{minipage}{0.35\textwidth}
\centering
\includegraphics[width=1\textwidth]{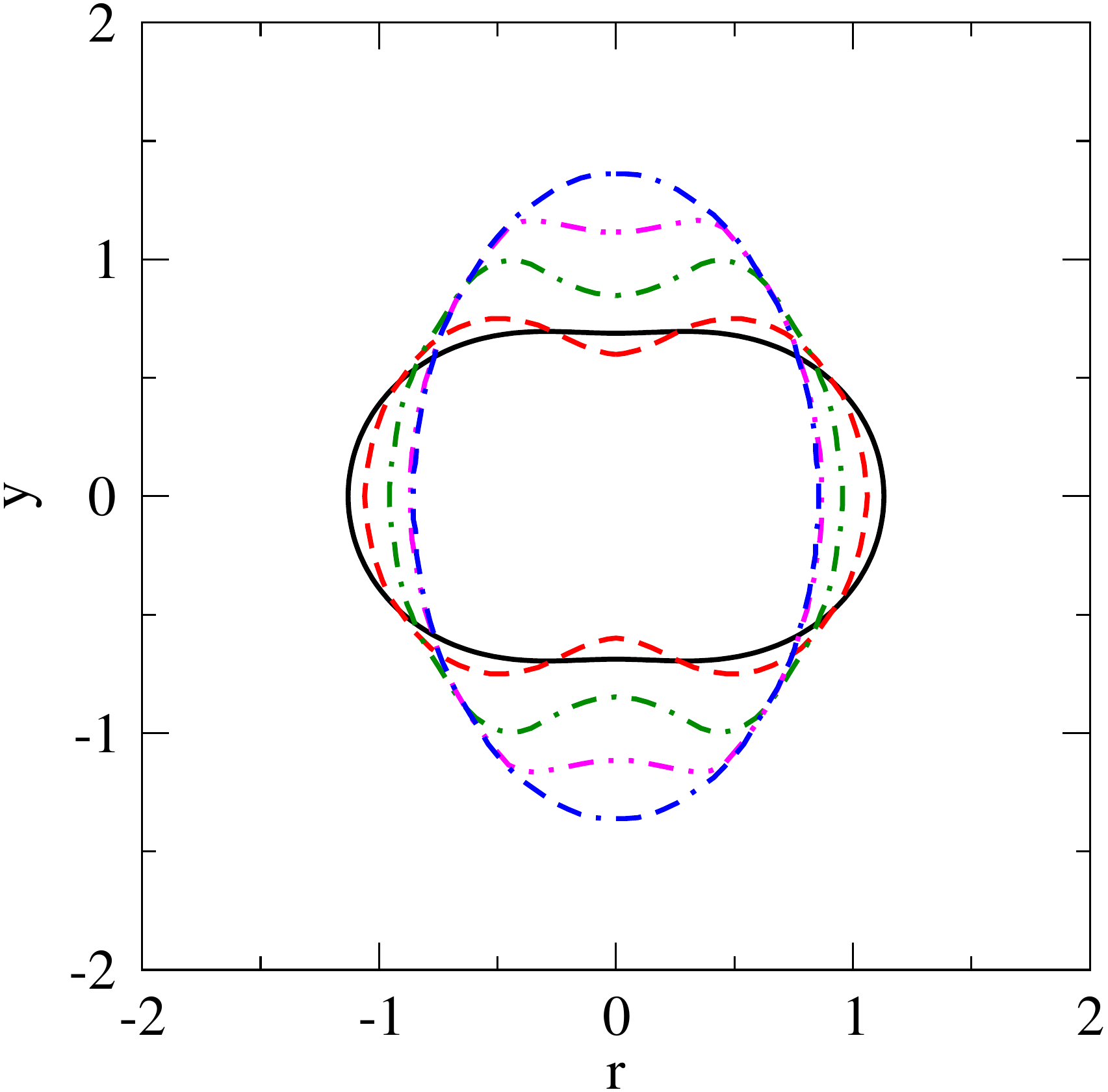}
(d) Reported by~\citeauthor{poz90}~\citep{poz90}
\end{minipage}
 \caption{Comparison of boundary integral simulations at $k_s=30$ and reported~\citep{poz90} evolution at $k=10$ of an oblate spheroid shape perturbed with second degree Legendre mode ($\epsilon=-0.3$). For a particular subfigure shapes with (\textcolor{black}{$\bf{\mi}$}) for $t=0$, ($\textcolor{red}{\pmb{\pmb{--}}}$) for $t=0.095$, (\textcolor{forestgreen}{$\pmb{\pmb{-\cdot -}}$}) for $t=0.235$, (\textcolor{magenta}{$\pmb{\pmb{-\cdot\cdot}}$}) for $t=0.540$ and (\textcolor{blue}{$\pmb{\pmb{--\cdot}}$}) for $t=\infty$. Boundary integral simulation results are reported in subFig. a, b and c, reported~\citep{poz90} shape evolution is shown in d.}
 \label{fig:poz0p3k10}
\end{figure}

\begin{figure}
\centering
\begin{minipage}{0.35\textwidth}
\centering
\includegraphics[width=1\textwidth]{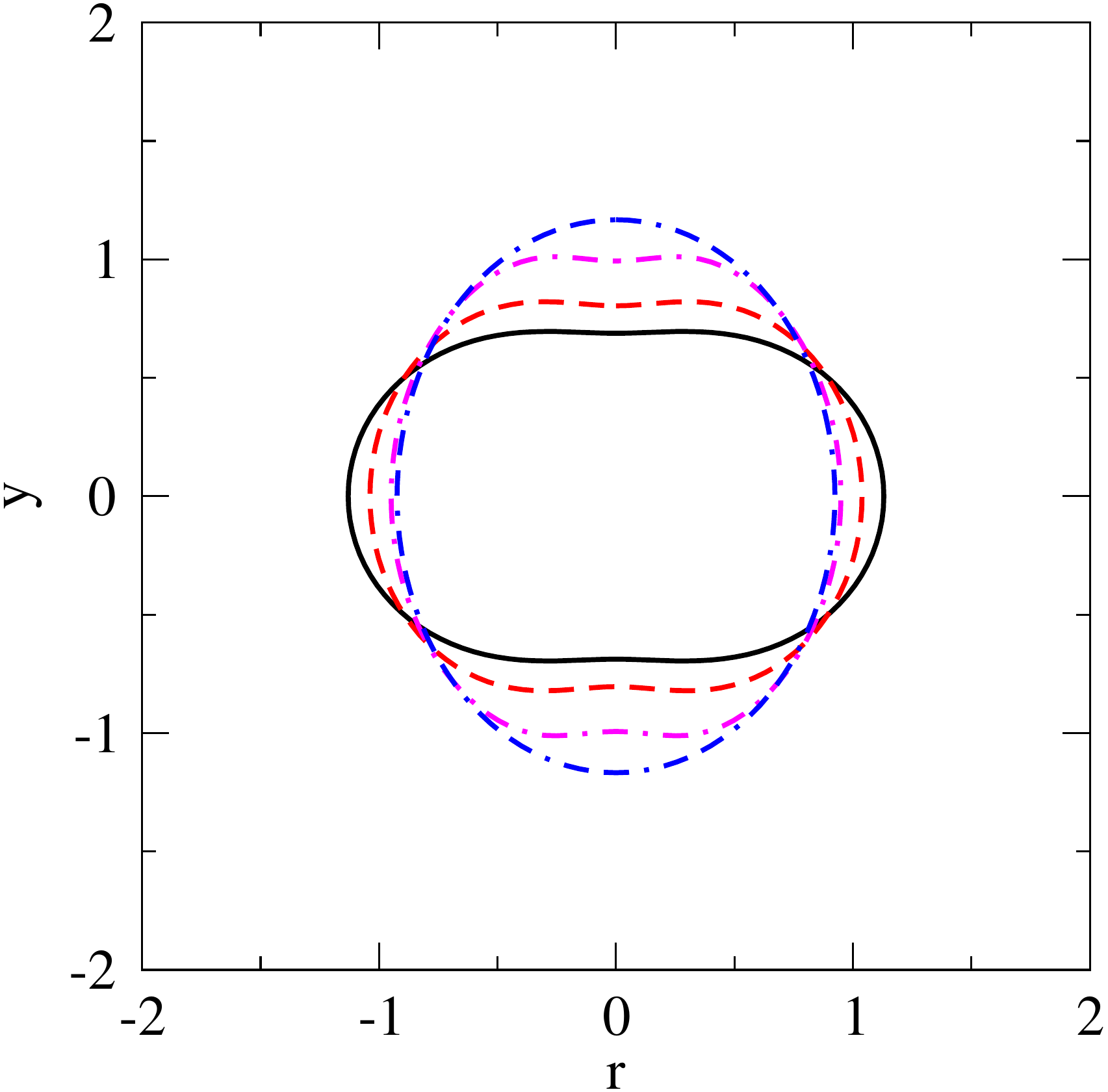}
(a) $C=1$
\end{minipage}
\begin{minipage}{0.35\textwidth}
\centering
\includegraphics[width=1\textwidth]{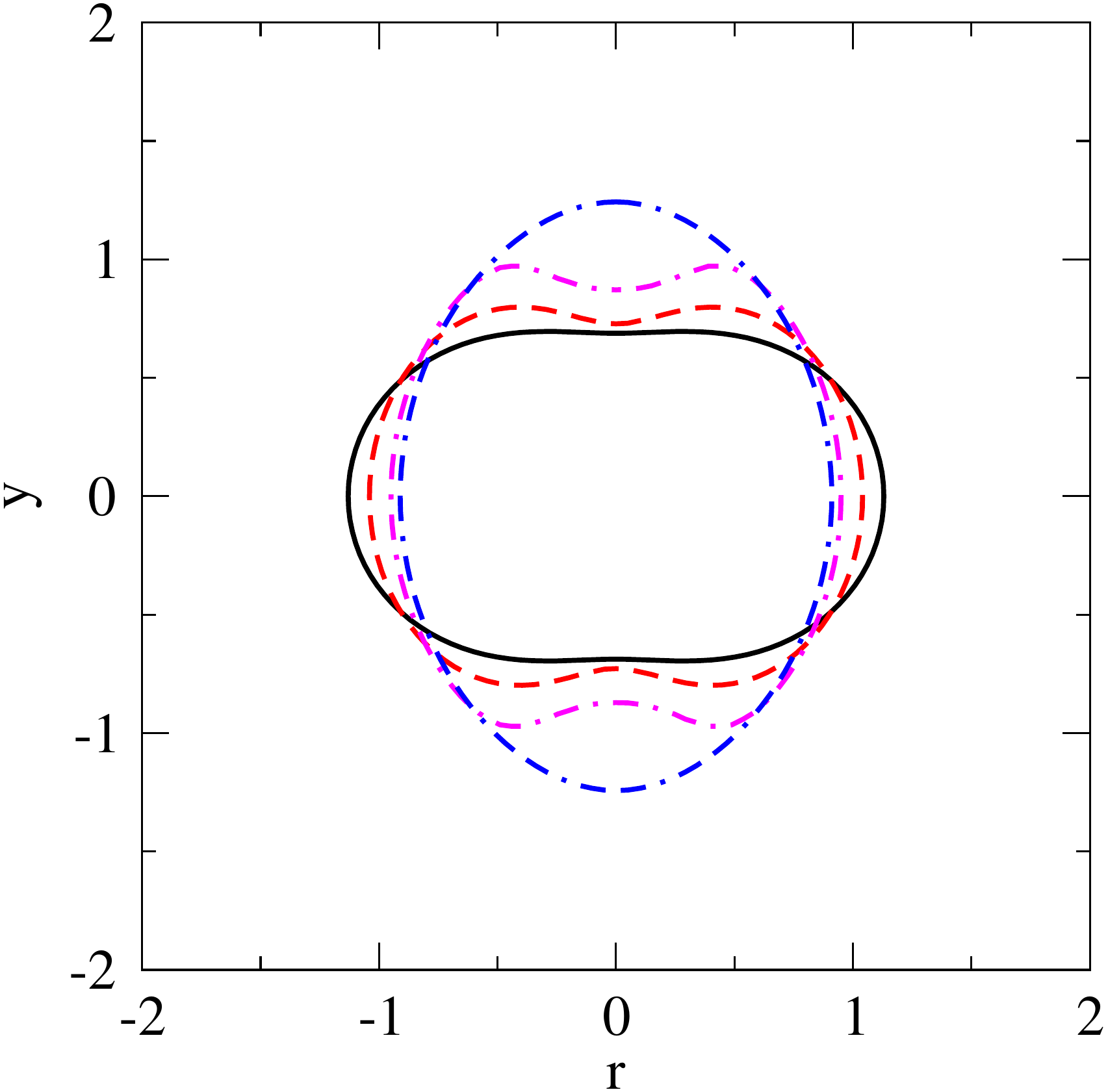}
(b) $C=10$
\end{minipage}
\begin{minipage}{0.35\textwidth}
\centering
\includegraphics[width=1\textwidth]{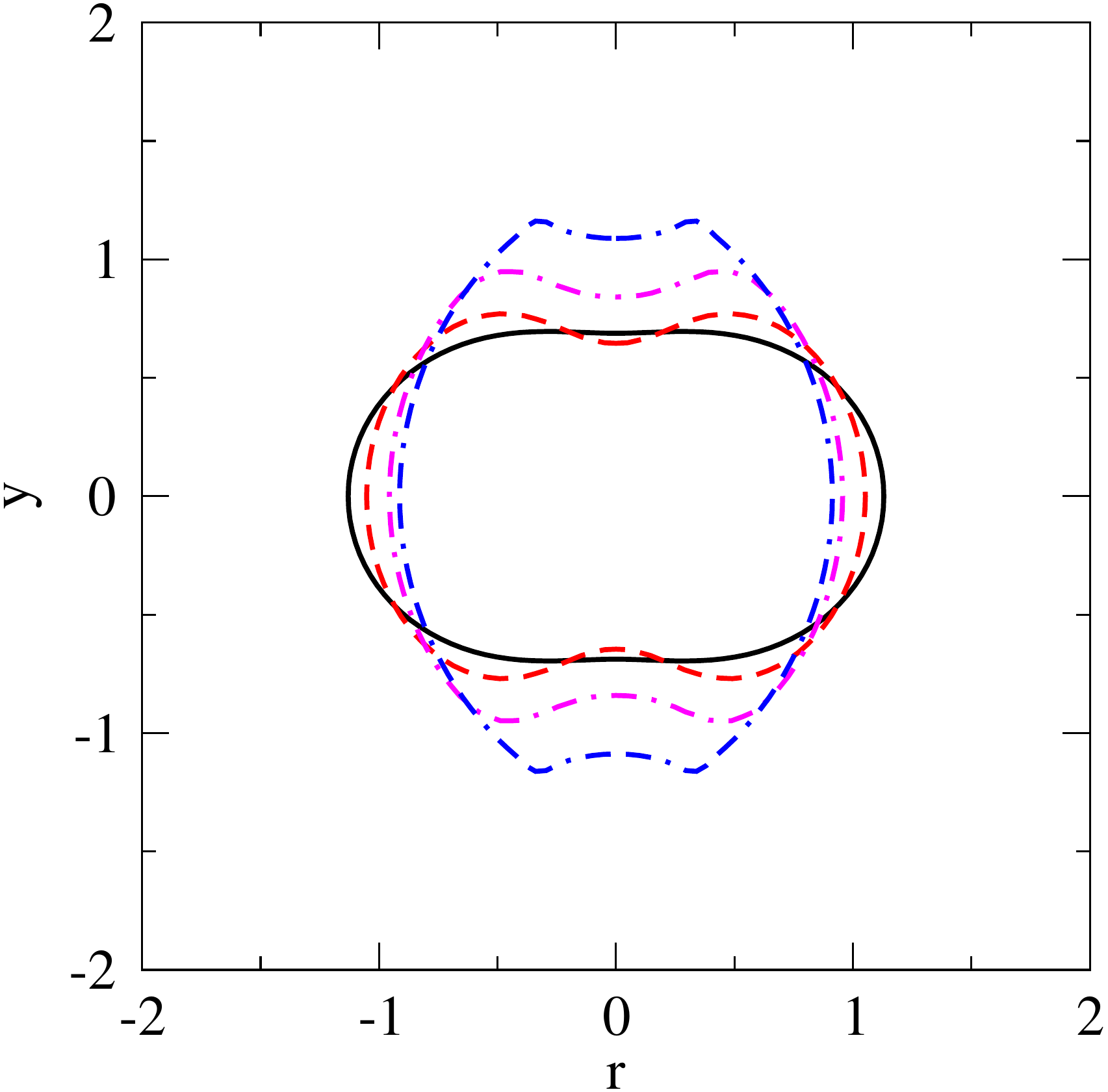}
(c) $C=50$
\end{minipage}
\begin{minipage}{0.35\textwidth}
\centering
\includegraphics[width=1\textwidth]{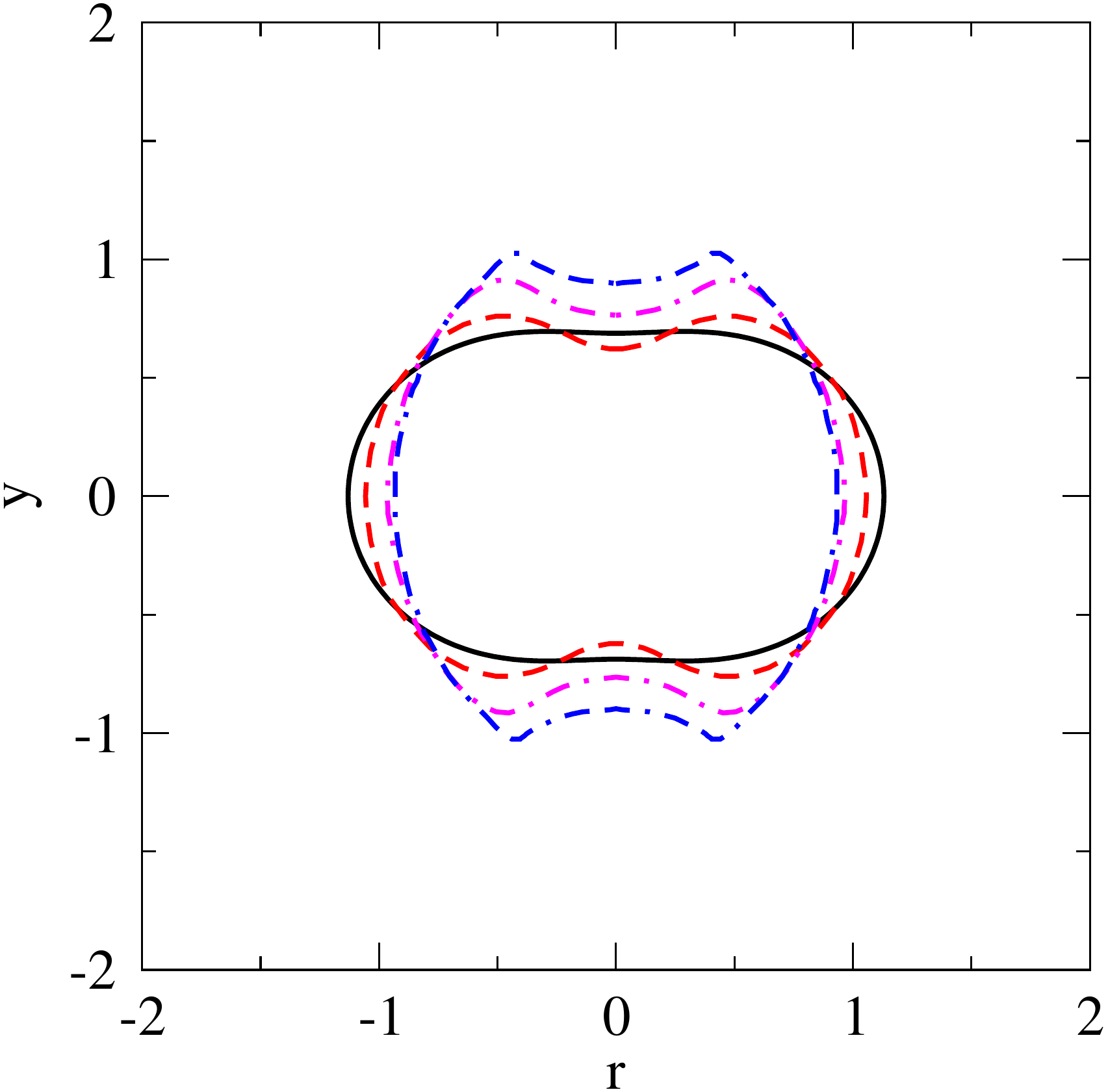}
(d) Reported by~\citeauthor{poz90}~\citep{poz90}
\end{minipage}
 \caption{Comparison of boundary integral simulations at $k_s=60$ and reported~\citep{poz90} evolution at $k=20$ of an oblate spheroid shape perturbed with second degree Legendre mode ($\epsilon=-0.3$). For a particular subfigure shapes with (\textcolor{black}{$\bf{\mi}$}) for $t=0$, ($\textcolor{red}{\pmb{\pmb{--}}}$) for $t=0.1$, (\textcolor{magenta}{$\pmb{\pmb{-\cdot-}}$}) for $t=0.3$ and (\textcolor{blue}{$\pmb{\pmb{--\cdot}}$}) for $t=\infty$. Boundary integral simulation results are reported in subFig. a, b and c, reported~\citep{poz90} shape evolution is shown in d.}
 \label{fig:poz0p3k20}
 \end{figure}

\begin{figure}
\centering
\begin{minipage}{0.35\textwidth}
\centering
\includegraphics[width=1\textwidth]{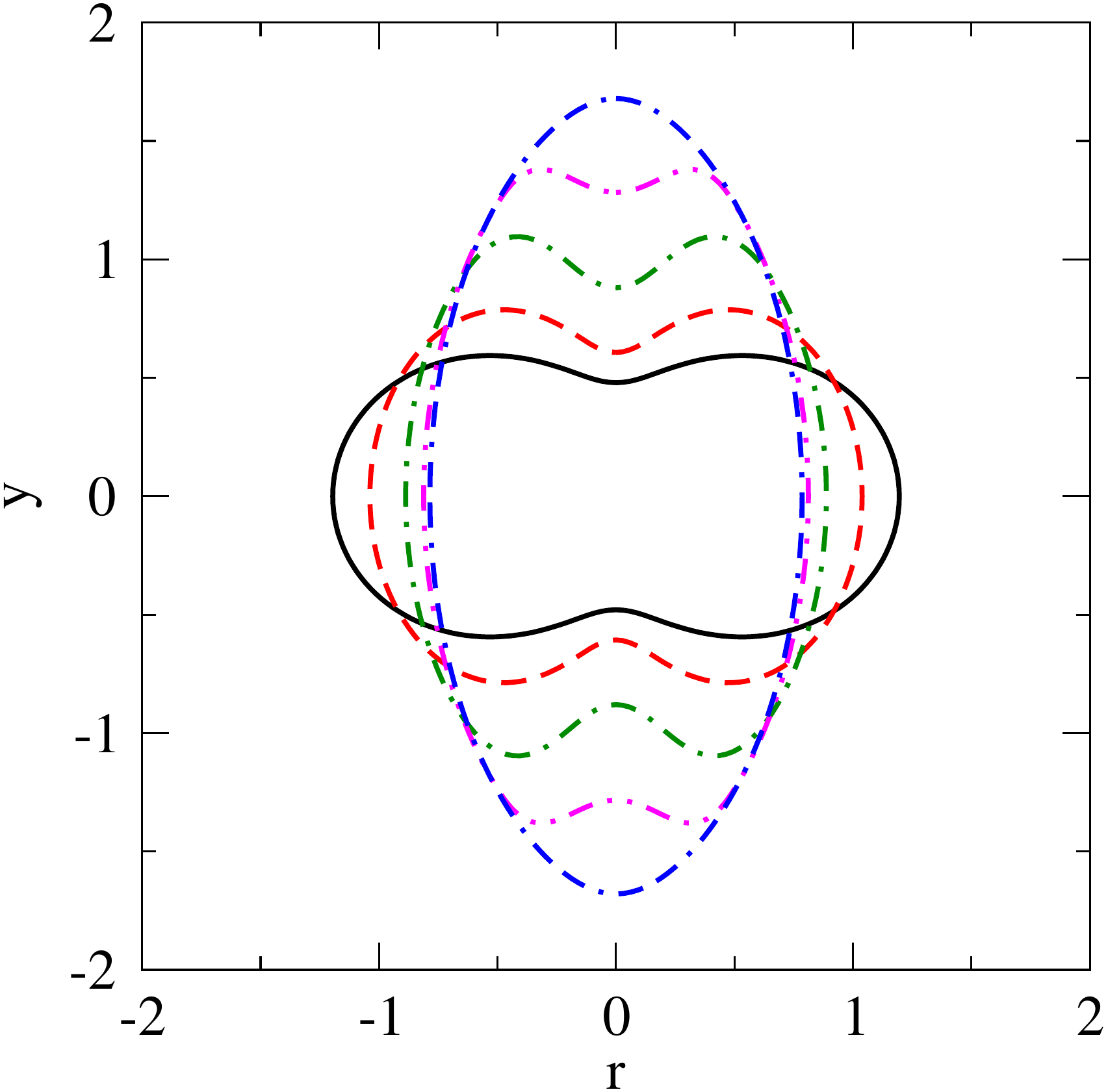}
(a) $C=1$
\end{minipage}
\begin{minipage}{0.35\textwidth}
\centering
\includegraphics[width=1\textwidth]{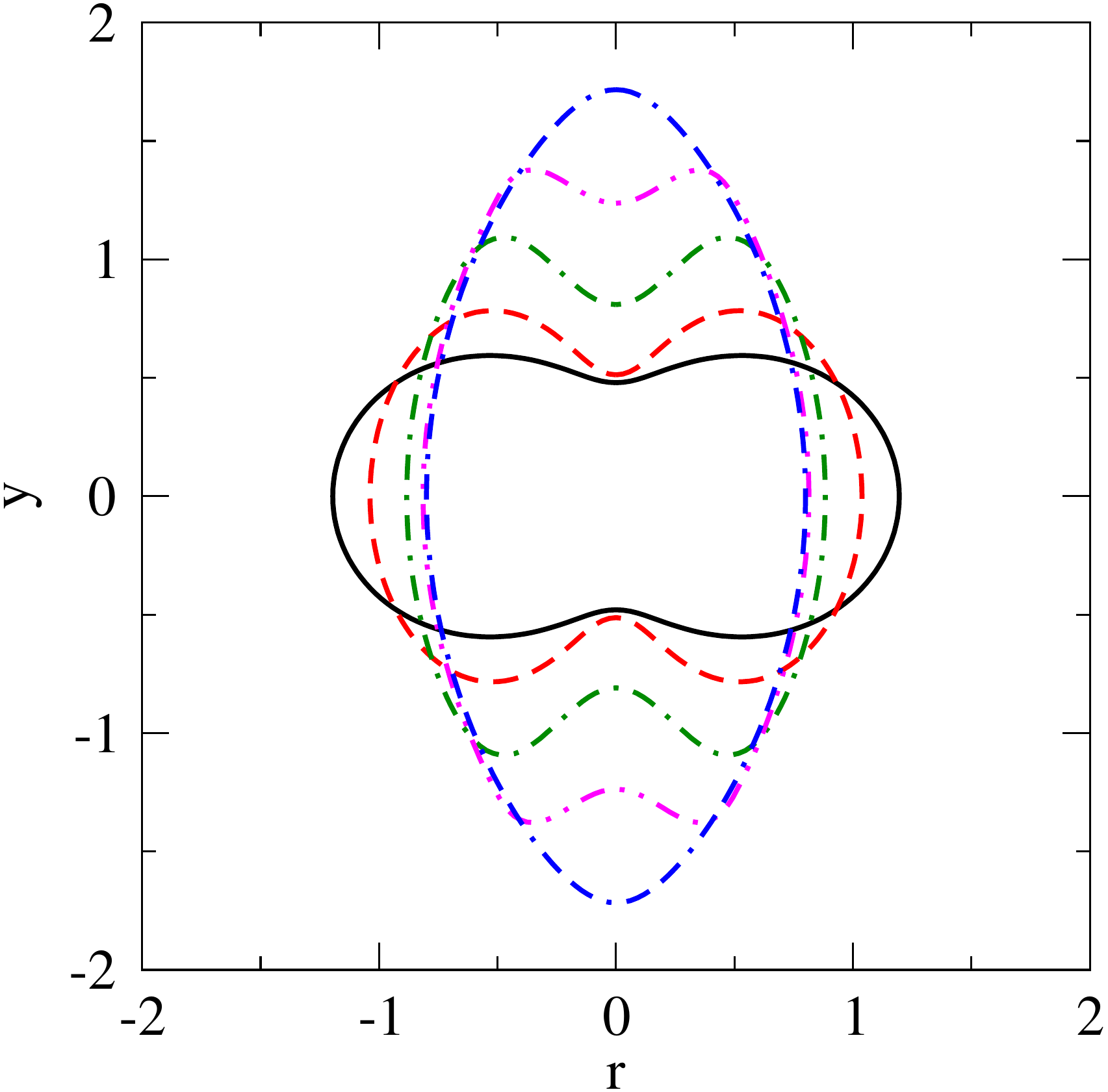}
(b) $C=10$
\end{minipage}
\begin{minipage}{0.35\textwidth}
\centering
\includegraphics[width=1\textwidth]{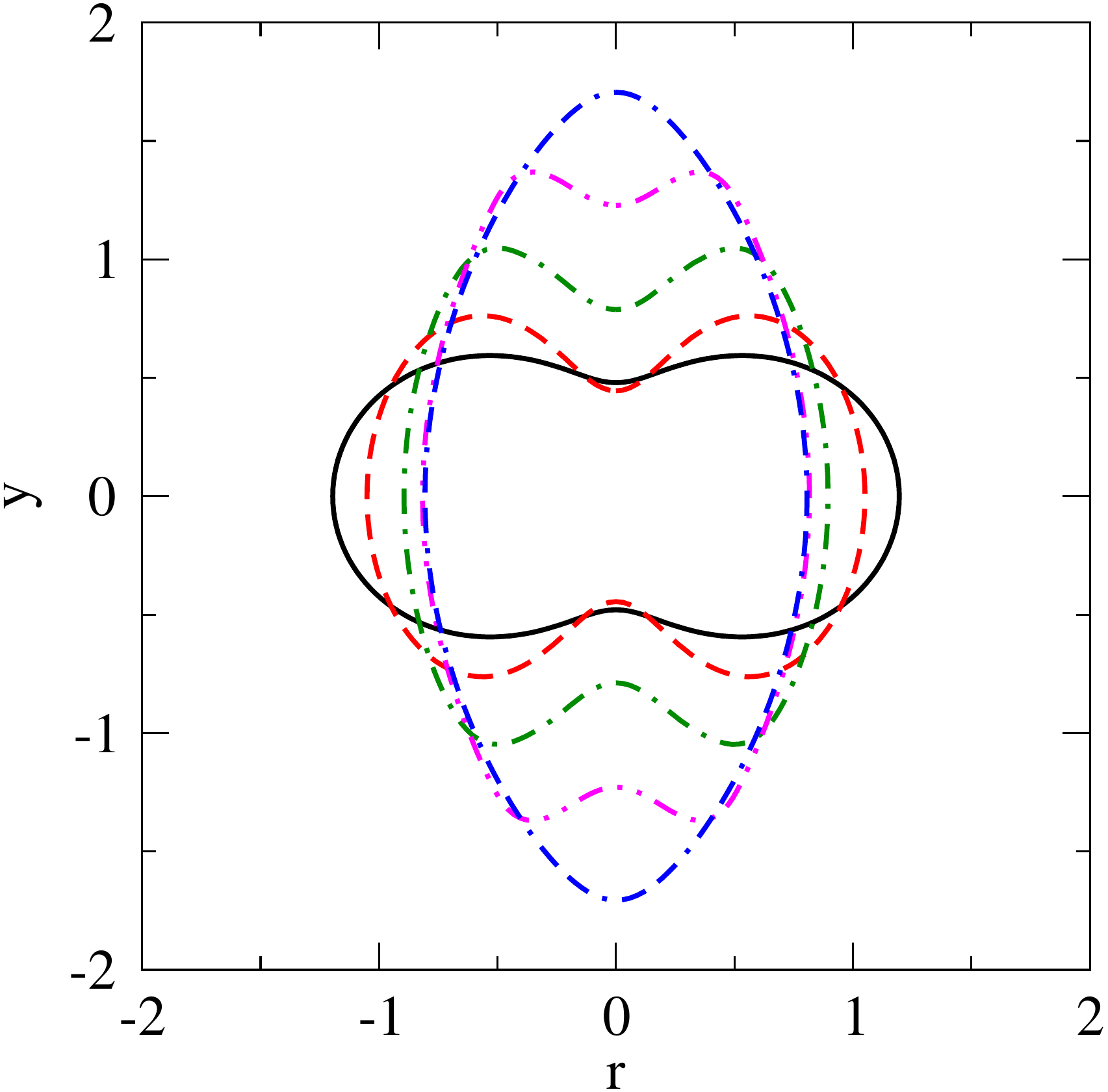}
(c) $C=50$
\end{minipage}
\begin{minipage}{0.35\textwidth}
\centering
\includegraphics[width=1\textwidth]{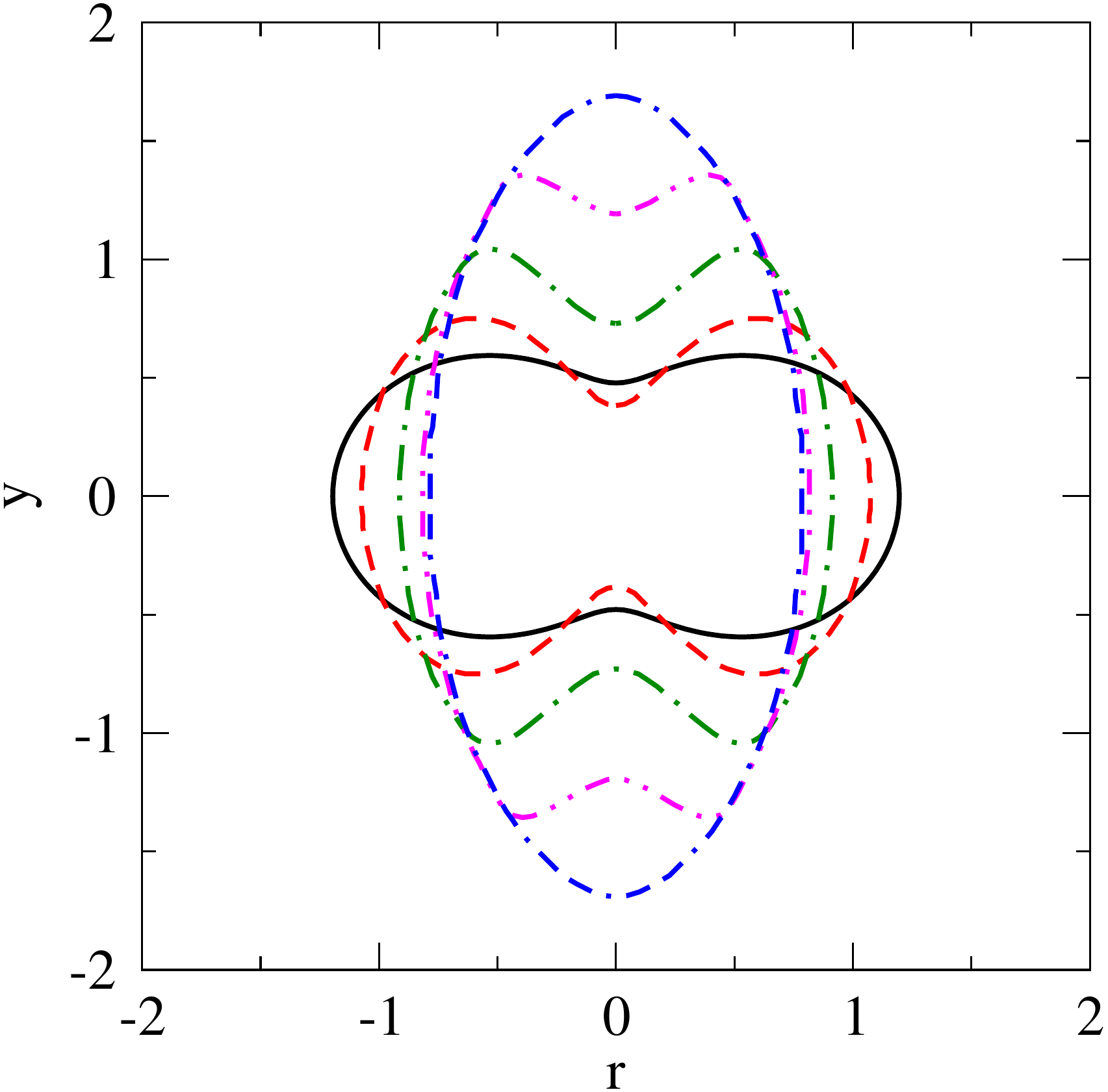}
(d) Reported by~\citeauthor{poz90}~\citep{poz90}
\end{minipage}
 \caption{Comparison of boundary integral simulations at $k_s=15$ and reported~\citep{poz90} evolution at $k=5$ of an oblate spheroid shape perturbed with second degree Legendre mode ($\epsilon=-0.5$). For a particular sub-figure, shapes with (\textcolor{black}{$\bf{\mi}$}) for $t=0$, ($\textcolor{red}{\pmb{\pmb{--}}}$) for $t=0.15$, (\textcolor{forestgreen}{$\pmb{\pmb{-\cdot -}}$}) for $t=0.375$, (\textcolor{magenta}{$\pmb{\pmb{-\cdot\cdot}}$}) for $t=0.682$ and (\textcolor{blue}{$\pmb{\pmb{--\cdot}}$}) for $t=\infty$. Boundary integral simulation results are reported in subFig. a, b and c, reported~\citep{poz90} shape evolution is shown in d.}.
 \label{fig:poz0p5k5}
\end{figure}

In Fig.~\ref{fig:poz0p3k10},~\ref{fig:poz0p3k20}, and~\ref{fig:poz0p5k5}, the comparisons of shape evolution with time are shown for the oblate spheroid shapes with perturbation coefficients of second degree Legendre polynomial $\epsilon=-0.3$, $-0.3$ and $-0.5$, respectively in extensional flow. The reported~\citep{poz90} shape evolution of oblate spheroid with $\epsilon=-0.3$ at $k=10$ and $k=20$ are similar to the obtained boundary integral simulation considering $k_s=30$ and $k_s=60$, respectively. Also, similar dynamics are observed for the deformation of the oblate spheroid with $\epsilon=-0.5$ for reported $k=5$ and calculated with $k_s=15$. The percent changes in the surface area during the deformation of the capsule are shown in Tables~\ref{tab:areachangeem0p3}, \ref{tab:areachangeem0p3a} and \ref{tab:areachangeem0p5} for the boundary integral simulation with $k_s=30$, $k_s=60$ and $k_s=15$, respectively. From the tables \ref{tab:areachangeem0p3}, \ref{tab:areachangeem0p3a} and \ref{tab:areachangeem0p5} and Fig. \ref{fig:poz0p3k10},~\ref{fig:poz0p3k20}, and~\ref{fig:poz0p5k5}, it can be observed that with an increase in the area dilatation parameter, $C$, the change in area decreases and the simulated result with high area dilatation parameter produces similar results as reported by~\citet{poz90}.  

\begin{table}
  \begin{center}
\def~{\hphantom{0}}
  \begin{tabular}{lc}
      $C$ & \hspace{0.5in} Change in area $\%$\\[3pt]
	1	&	2.8\\
      10	&	2.15\\
      50	&	1.05\\
  \end{tabular}
  \caption{Percent change in area at different membrane parameter $C$ for $\epsilon=-0.3$ and $k_s=30$}
  \label{tab:areachangeem0p3}
  \end{center}
\end{table}



 \begin{table}
  \begin{center}
\def~{\hphantom{0}}
  \begin{tabular}{lc}
      $C$ & \hspace{0.5in} Change in area $\%$\\[3pt]
	1	&	2.81\\
      10	&	1.78\\
      50	&	0.62\\
  \end{tabular}
  \caption{Percent change in area at different membrane parameter $C$ for $\epsilon=-0.3$ and $k_s=60$}
  \label{tab:areachangeem0p3a}
  \end{center}
\end{table}


\begin{table}
  \begin{center}
\def~{\hphantom{0}}
  \begin{tabular}{lc}
      $C$ & \hspace{0.5in} Change in area $\%$\\[3pt]
	1	&	4.7\\
      10	&	2.9\\
      50	&	1.22\\
  \end{tabular}
  \caption{Percent change in area at different membrane parameter $C$ for $\epsilon=-0.5$ and $k_s=15$}
  \label{tab:areachangeem0p5}
  \end{center}
\end{table}



\end{document}